\documentstyle[aps]{revtex}
\begin{document}
\draft
\title{{\Huge Statistics of resonance poles, phase shifts and time delays
in  quantum chaotic scattering:}\\
\vspace{1cm}{\Large Random Matrix approach for systems with broken  
time-reversal invariance.}}

\bigskip
\author{Yan  Fyodorov$^{}$\cite{leave} and Hans-J\"{u}rgen Sommers}
\smallskip
\address{Fachbereich Physik, Universit\"at-GH Essen,D-45117 Essen,  
Germany}

\maketitle
\bigskip

\begin{abstract}
Assuming the validity of random matrices for describing the
statistics of a {\it closed} chaotic quantum system, we
study analytically some statistical properties of the S-matrix
characterizing scattering in its {\it open} counterpart.
In the first part of the paper we attempt to expose systematically  
ideas underlying the so-called stochastic (Heidelberg)
 approach to chaotic quantum scattering. Then  we concentrate on systems
with broken time-reversal invariance coupled to continua via M open
channels;$\,\,a=1,2,..,M$. A physical realization of this case  
corresponds to the
chaotic scattering in ballistic microstructures pierced by a strong
enough magnetic flux. By using the supersymmetry method we derive
an explicit expression for the density of S-matrix poles
(resonances) in the complex energy plane. When all scattering
channels are considered to be equivalent our expression describes a
crossover from the $\chi^2$ distribution of resonance widths
(regime of isolated resonances) to a broad power-like distribution
typical for the regime of overlapping resonances. The first moment
is found to reproduce exactly the Moldauer-Simonius relation between
the mean resonance width and the transmission coefficient. Under
the same assumptions we derive an explicit expression for the
parametric correlation function of densities of eigenphases
$\theta_a$ of the S-matrix (taken modulo $2\pi$). We use it to find
the distribution of derivatives  
$\tau_a=\partial{\theta_a}/{\partial E}$ of these eigenphases with  
respect to the energy ("partial delay times" ) as well as with  
respect to an arbitrary
external parameter. We also find the parametric correlations of the
Wigner-Smith time delay $\tau_{w}(E)=
\frac{1}{M}\sum_a\partial{\theta_a}/{\partial E}$  at two
different energies $E-\Omega/2$ and $E+\Omega/2$ as well as at two
different values of the external parameter. The relation between
our results and those following from the semiclassical approach as  
well as the relevance to experiments are briefly discussed.
\end{abstract}
\vfill
\centerline{To be published in the special issue }
\centerline{"Quantum Problems in Condensed Matter Physics"}
\centerline{ of the Journal of Mathematical Physics}
\centerline{version 31.12.96}
\section{Introduction.}
Chaotic scattering has been a subject of a rather intensive research
activity during the last decade, both theoretically (see  
reviews\cite{Smilansky,Gasp,Baldo,chaonan,chaonan1}) and  
experimentally \cite{Marcus,TRS,cav,cav1,Dosmifr,TRScav,Alt}.This  
phenomenon is encountered
in a variety of physical systems ranging from atomic
nuclei\cite{Baldo,chaonuc,Rot},atoms \cite{chaoat,Del,Rydb} and  
molecules \cite{Gasp,Schinke,Lomb} to mesoscopic
ballistic devices \cite{chaonan,Barr} and microwave
cavities\cite{cav,cav1,Dosmifr,TRScav,Alt}.
The most fundamental object characterizing the process of quantum  
scattering is the unitary $S-$matrix relating the amplitudes
of waves incoming onto the system and the amplitudes of scattered
(outgoing) waves. Because of the chaotic nature of the underlying
scattering dynamics the $S-$matrix characteristics behave in an
irregular way when parameters of either incoming waves (e.g. energy)
or of the scattering region (e.g. the form or strength of the
scattering potential, the strength of the magnetic field through the
ballistic microstructure, etc.) are slightly changed. Because of this
fact it seems to be most adequate to describe such a behavior
in terms of some statistical measures: distributions and correlation
functions.

At present, there are two complementary theoretical tools employed to
calculate statistical characteristics of
 open quantum systems exhibiting the phenomenon of chaotic
scattering. These are the semiclassical
\cite{Smilansky,Gasp,chaonan,BS,Eckhardt} and the stochastic
approaches\cite{Weidrev,Melrev}, the relation between both methods
being in some detail discussed in \cite{Lew}. The
semiclassical approach operates with the genuine microscopic
Hamiltonians and allows for treating particular systems with full
account of their specific features. The starting point for such an
approach is a representation of $S$-matrix elements in terms of a sum
over the classical periodic orbits, the method going back to works
by Gutzwiller\cite{Gutzwiller} and Balian and Bloch\cite{Balbl}.
The statistical characteristics are sampled usually over some range  
of energies or changing the system parameters.

It is however known, that the majority of (both closed and open)
 quantum chaotic systems of quite different microscopic nature
shows a great degree of universality in their properties on the  
appropriate energy scale. More precisely, the statistical
characteristics of {\it closed} systems
turn out to be independent of the microscopic
details when sampled on the energy intervals $\delta E$ large in  
comparison
with a mean separation between two adjacent levels $\Delta$, but smaller
then the energy scale $E_c=\hbar/t_e$, with $t_e$ standing for the  
relaxation time necessary for the classically chaotic chaotic system  
to cover uniformly the constant energy shell \cite{AlSi}.
Because of this universality one achieves
the correct description of the properties of such systems
\cite{Bohigas} by exploiting the similarity with ensembles of large  
Gaussian
random matrices $\hat{H}$ of the size $N\times N$
characterized by the following probability
distribution:\begin{equation}\label{GUE}
{\cal P}_{\beta}\propto \exp{-\frac{\beta N}{4}\mbox{Tr}\hat{H}^2}
\end{equation}
where the matrices are considered to be real symmetric ($\beta=1$,  
Gaussian
Orthogonal Ensemble: GOE), Hermitian ($\beta=2$, Gaussian Unitary
Ensemble: GUE) or consisting of real quaternions ($\beta=4$, Gaussian
Symplectic Ensemble: GSE).  Ensembles
with $\beta=1$ ($\beta=2$) serve to describe spectra of closed  
quantum chaotic systems systems with preserved (broken, e.g by  
applied magnetic field or by Aharonov-Bohm magnetic flux)  
time-reversal
invariance, correspondingly. At last, the ensemble corresponding to  
$\beta=4$
describes systems with preserved time-reversal invariance displaying
 strong spin-orbit scattering which should be taken into account.

Properties
of all these ensembles were studied long ago \cite{Mehta,RMT}.
The mean level density $\nu_{sc}(E)=\langle\nu(E)\rangle$
in the limit $N\to \infty$ is given by the
so-called Wigner semicircle law:
\begin{equation}\label{semi}\nu_{sc}(E)=\left\langle\frac{1}{N}\mbox{Tr}\delta(E-\hat{H})\right\rangle=
\frac{1}{2\pi }\sqrt{4-E^2}
\end{equation}
where the angular brackets stand for the ensemble averaging.
The radius of the semicircle is equal (in chosen normalization) to
$E_{sc}=2$, so that the average spacing between eigenvalues is $4/N$
while the local spacing around the point $E$ is  
$\Delta(E)=(N\nu_{sc})^{-1}$.

The mean level density is the simplest quantity characterizing
the spectrum of any system. It is not very informative from a  
physical point of view since it is
insensitive to the fine structure of the spectrum. It is also the same
for all universality classes. Actually, no real
physical system is known where the mean level density follows the
semicircle law, Eq.(\ref{semi}).

In contrast, the two-point spectral correlation function:
\begin{equation}\label{i2}
R_2(\omega)=\Delta^2(E)\left\langle\nu\left(E-\Omega\right)
\nu\left(E+\Omega\right)\right\rangle-1
\end{equation}
where $\omega=2\pi\Omega/\Delta(E)$ is known to be universal, i.e.
independent of the microscopic details and
has the same form both for generic chaotic systems and for
the Gaussian Ensembles of definite symmetry. The same universality
extends to other spectral properties, such as the nearest neighbors
spacing distribution, etc.\cite{Bohigas}.

 Despite the apparent success in the exploitation of random
matrix results for describing spectra of quantum chaotic systems of  
different
nature \cite{Bohigas,RMT} it was a long standing problem to justify  
the validity of such an approach microscopically. Some
 insight was achieved within the semiclassical approach long ago by
Berry\cite{Berry}. Very recently Muzykantsky and
Khmelnitsky\cite{Khm}  and
especially  Andreev et al. \cite{AAA} managed to find a way to  
prove this conjecture by a nontrivial
combination of field-theoretical and semiclassical ideas. In parallel,
traditional semiclassical methods were also significantly improved
\cite{Bogkeat}. These results put applications of random matrices
for the description of universal features of closed chaotic
systems on a firm ground.

Provided the properties of a Hamiltonian ${\cal H}_{in} $ for a  
closed chaotic
system are specified, one can consider its open counterpart
and work out the $S-$ matrix by standard methods in the theory
of quantum scattering\cite{MW,Levine,Livsic,KNO,Pavlov,Mak}. As the  
result, the
scattering matrix is expressed in terms of both the Hamiltonian ${\cal
H}_{in}$ and matrix elements describing the coupling of the
internal motion to "open channels" i.e. the states
of the system asymptotically far from the chaotic  
region.Correspondingly, one can try to extract the statistics of  
$S-$matrix
inherited from the mentioned universal "random matrix"properties of  
${\cal H}_{in}$ \cite{VWZ}.

In principle, it is far from being obvious that the coupling to
continua does not wash out the universal features. The key
observation (made long ago in the context of nuclear physics,see  
e.g. \cite{Fesh})  is that typically there are two well-separated  
time
stages associated with the scattering process: an immediate "prompt"
response (so-called {\it direct} processes) and a delayed, or
equilibrated response associated with the formation of long-living  
states,
or {\it resonances}. In the energy domain direct processes are
described by smooth $S-$matrix characteristics averaged over a  
large energy
interval. Such characteristics must be, of course, highly non-universal
and are determined mainly by system-specific boundary conditions on the
boundary of the scattering region. At the same time, resonance
response happening on much shorter energy scale manifests itself in  
a form of a random signal on top of the smooth averaged
characteristics.  Formation of the long-living resonances
is intimately related to the internal dynamics inside the  
scattering region.  It is natural to expect that the universal
features of the chaotic quantum dynamics will be reflected in the  
universal
statistical characteristics of the $S-$matrix on the "resonance"
energy scale, as long as the characteristic times(e.g. measured by  
inverse widths of the resonances)
 will be much longer than the time scale of the direct response.

To find some adequate description of these universal features
one can substitute the Hamiltonian ${\cal H}_{in}$ by the Gaussian
random matrix, Eq.(\ref{GUE}).  This way was pioneered by  
Verbaarschot, Weidenm\"{u}ller and Zirnbauer
\cite{VWZ} who calculated the correlation function
of $S-$ matrix elements at two different energies
for arbitrary number $M$  of open channels satisfying
$M\ll N$, with $N$ being the total number of resonances.
It was indeed found
that the $N\times M$ matrix elements describing the coupling of the  
internal
region to open channels enter the final result in the form of only  
$M$ simple
combinations, the so-called "transmission coefficients". In full
agreement with the "two distinct time scale" picture,
these coefficients just measure the portion of the flux in
a given channel which is not reflected back immediately,
but penetrates the interaction region and
participates in the formation of the long-living resonances  
\cite{Weidrev,Lew}.   The approach developed in \cite{VWZ}  
(following \cite{Melrev}
we will call it "Heidelberg approach" henceforth)
turned out to be very fruitful
and serves as a case study for all further development
in the field.

One can also try to make use of the expected universality directly on
the level of $S$-matrix without any reference to the system
Hamiltonian. Such a method was developed in great detail in a
series of papers by Mello and co-workers\cite{Mello,Barr,Melrev}.
The probability density for the whole $S-$matrix can be obtained
if one makes the assumption of minimal information content
of such a distribution respecting the requirements of $S-$matrix
 unitarity, analyticity and constraints imposed by absence or presence of
the time-reversal invariance. Provided all the  system-specific  
relevant information is encoded in the value of the average  
$S-$matrix
$\langle \hat{S}\rangle$ the joint probability $P(\hat{S})  
d\mu(\hat{S})$ is given by:
\begin{equation}\label{Pua}
P_{\langle S\rangle}(\hat{S})d\mu(\hat{S})\propto  
\frac{\left[\mbox{det}\left(\hat{I}-
\langle S\rangle\langle S^{\dagger}\rangle\right)\right]^{(\beta
M+2-\beta)/2}}{\mid\mbox{det}\left(\hat{I}-
 S\langle S^{\dagger}\rangle\right)\mid^{(\beta M+2-\beta)}}
d\mu_{\beta}(\hat{S})\end{equation} with the following measure  
$d\mu_{\beta}(\hat{S})$ :
\begin{equation}\label{meas}
d\mu_{\beta}(\hat{S})\propto\prod_{a<b}\mid
e^{i\theta_a}-e^{i\theta_b}\mid^{\beta} \prod_{a=1}^Md\theta_a dU
\end{equation}
where $\theta_a;\quad a=1,...,M$ are eigenphases of the $S-$matrix,
the volume element $dU$ is generated by the corresponding eigenvectors,
$\beta=1,2,4$ as before and $M$ is the dimension
of the $S-$matrix equal to the number of open channels. The   
distribution $P(\hat{S})$ is called the Poisson's kernel.

For the particular case $\langle S\rangle=0$ the distribution  
Eq.(\ref{Pua})
just coincides with the  measure $d\mu_{\beta}(\hat{S})$.
Such distributions were considered long ago by Dyson and known
as the Dyson Circular Ensembles\cite{Mehta}.
They were found to describe very satisfactorily
the $S$-matrix statistics for some realistic models of the chaotic
scattering \cite{Smicol}. The general Poisson's kernel,
Eq.(\ref{Pua}) was verified as well
\cite{Barr,Melrev} and proved to be a very useful tool to predict  
fluctuations of transmissions through ballistic
microstructures.
It is natural to expect that the
same distribution can be actually derived from the Heidelberg
approach. It turned out that the problem is quite involved  
technically, however. In his insightful paper
\cite{Brouwer} Brouwer succeeded to derive the  Poisson's kernel  
distribution assuming that the Hamiltonian ${\cal H}_{in}$ is taken  
from a quite specific Lorentzian ensemble of random
matrices. Since the spectral properties of the latter ensemble
and those of Gaussian matrices, Eq.(\ref{GUE}), are identical
as long as the matrix size $N\to\infty$, one expects that the
equivalence of the two
approaches can be shown for this generic case as well.

If one wishes to study the dependence of the $S-$matrix on external
parameters without explicitly considering the system Hamiltonian,
one should make some additional statistical assumptions beyond
the minimum information approach. One possible way is to simulate
such a dependence by a kind of "Brownian motion" in the corresponding
$S-$ matrix space\cite{FPich}. It turns out, however, that the
Brownian motion picture is in disagreement with the
results obtained starting from the Heidelberg formalism.
Therefore, the Heidelberg approach seems to be the only consistent
stochastic method when we are interested in the parametric
variations of the $S-$matrix characteristics.
 An example of such kind of calculation can be found in\cite{Macedo}.
Another important advantage of the Heidelberg approach as compared
with that by Mello and collaborators is that it operates with
an
{\it energy-dependent} $S-$matrix $S(E)$. As such, it allows to study
not only spectral correlations of different physical quantities
 but, in principle,
also contains information about such features of the $S-$matrix as  
{\it resonances} in the complex energy plane ${\cal E}=E+iY$.

The notion of resonances, representing long-lived intermediate
states of open system to which bound states of its closed counterpart
are converted due to coupling to continua, is one of the most
fundamental concepts in the domain of quantum scattering
\cite{Kukulin}. On a formal
level resonances show up as poles of the scattering matrix occuring
at complex energies ${\cal E}_k=E_k-\frac{i}{2}\Gamma_k$, where
$E_k$ and $\Gamma_k$ are called position and widths of the resonance,
correspondingly.

The general problem of determining the domain of concentration and  
the distribution of poles of the $S-$matrix in the complex plane is  
of
fundamental interest in the general theory of scattering\cite{Zworski}.
Powerful numerical methods are available (e.g. the method of complex
scaling \cite{cosc}) allowing one to extract resonance parameters for
models in atomic and molecular physics.
Whereas the issue of energy level statistics in closed chaotic
systems was addressed in an enormous amount of papers (see
\cite{Bohigas,AlSi} and references therein), statistical
characteristics of resonances are much less studied and attracted
significant attention only recently
\cite{Gasp,Del,Rydb,Burg,Sok,John,Haake,Deso,Dallwig,Lehm2,Dittes,Remacle,chemf1,chemf2,Drozdzh,Weaver,Flam,Hacken}.  

In the case of weak coupling to continua individual resonances
do not overlap: $\left\langle\Gamma\right\rangle\ll \Delta$.
Under these conditions one can use a simple first order perturbation
theory to calculate resonance widths in terms of eigenfunctions
of the closed system (see example in the Sec.III of the present
paper). Quite generally, one finds in such a procedure that the
widths for the chaotic system with $M$ open channels
are distributed according to the so-called $\chi^2$
distribution:
\begin{equation}\label{chi}
\rho(y_s)=\frac{(\nu/2)^{(\nu/2)}}{\Gamma(\nu/2)}y_s^{\nu/2-1}
e^{-\frac{\nu}{2}y_s}
\end{equation}
where $y_s$ stands for the resonance widths normalized to its
mean value: $y_s=\Gamma/\langle\Gamma\rangle$, the parameter$\nu=M$  
($\nu=2M$) for systems with preserved (broken) time
reversal invariance, and  $\Gamma(z)$ in Eq.(\ref{chi}) stands for  
the Euler
Gamma-function. The case $\nu=1$ is known as the Porter-Thomas  
distribution\cite{Porter}. It was shown to be in
agreement with experimental data in neutron-nuclei resonances
\cite{Porter}, the fluorescence excitation spectrum of the
$NO_2$ molecule\cite{NO2}, resonance dissociation of
$HO_2$ molecule\cite{chemf2}, the diamagnetic Rydberg spectrum in
lithium atom\cite{Del} and in microwave cavities\cite{Alt}.
Indirectly that distribution manifests itself in fluctuations
of tunneling conductance through ballistic quantum dots\cite{Alh,Mucc}.

When the coupling to continua increases resonances start to
overlap and the simple perturbative result Eq.(\ref{chi})
loses its validity. Finally, when the coupling to continua exceeds
some critical value, the so-called "trapping phenomenon" occurs:
$M$ very unstable states (broad resonances) are formed, whereas
the rest $N-M$ resonances go back to the real axis, i.e.
become more and more narrow with increasing
coupling, see \cite{Sok,Dittes,Remacle} and the end of Sec.III for  
a more detailed discussion. Such a "reorganization" of the spectrum
is the most pronounced when the number of channels $M$ is of the same
order as the (large) number of resonances $N$. This range of
parameters $M\propto N\gg 1$ always corresponds to the
condition  $\langle\Gamma\rangle\gg\Delta$ which is just the  
opposite limiting case
as compared with the domain of validity of the $\chi^2$ distribution.
 Under this condition
one can calculate the density of resonance poles analytically
\cite{Haake,Lehm2}.
 However, frequently one encounters the case of
few open channels and moderately overlapping resonances  
$\Gamma\sim\Delta$
\cite{Del,Burg,Schinke,chemf2,Barr}. In this situation,
which is in some sense generic, one can neither rely upon the
distribution Eq.(\ref{chi}), nor use the results of
\cite{Haake,Lehm2}. The general distribution of resonance widths
describing a crossover from isolated to overlapping resonances
was found recently by the present authors
for the particular case of an open chaotic system
with broken time-reversal invariance coupled to continua via $M\ll N$
equivalent channels\cite{FSres}. One of the main goals of the present
paper is to give a detailed derivation and subsequent analysis
of that distribution, also for the case of non-equivalent channels.

Apart from the $S-$matrix elements and $S-$matrix poles, the set of
scattering phase shifts $\theta_a$ (defined via the $S-$ matrix  
eigenvalues $\exp{i\theta_a};\quad a=1,2,...,M$ ) are intensively
used to characterize the chaotic scattering, see  
\cite{Smilansky,Smicol,Doron}. Quite recently, their
statistical characteristics were studied numerically in some detail
for chaotic \cite{Selig,Dietz} as well as for disordered
\cite{JPich,Eduardo} systems. The derivatives of phase shifts over
the energy $\tau_a=\partial \theta_a/\partial E$ (we propose to call
them "partial delay times") are particularly interesting and
related to the mean time spent by a quantum particle in the
interaction domain.

The issue of the time scales associated with different stages of the
quantum scattering process (e.g. tunneling, reflection and transmission)
is quite a controversial subject which is under an
intensive discussion for a long time, see  
\cite{Butt,time,delayop,Gasparian}
and references therein. In particular, ambiguities
arise because
there is no a self-adjoint time operator in Hilbert space,
analogous to the position operator; instead, the wave function depends
on time as a parameter.

Relegating all the essential details
and derivations to section II, we just mention here that
if $\psi({\bf x},t)$ denotes a wave packet at time $t$
 for a quantum particle moving in a potential $ U({\bf x})$ (as  
such satisfying the Schr\"{o}dinger equation
 $i\hbar\frac{\partial \psi}
{\partial t}=\left[-\frac{\hbar^2}{2m}\Delta+U({\bf x})\right]\psi$)
 then the real  
number:$$t_{r}\left[\psi\right]=\int_{-\infty}^{\infty}dt\int_{\mid{\bf  
x}\mid\le r}
\mid\psi({\bf x},t)\mid^2 d^3 {\bf x}$$
may be interpreted as the total time spent by this state during its
evolution inside the ball of the radius $r$ centered at the origin
( we assume $\psi({\bf x},t=0)$ normalized to unity). If
$\psi_f({\bf x},t)$ denotes a freely evolving wave packet (i.e.
solution of the same Schr\"{o}dinger equation with $U({\bf x})=0$
and condition: $\psi_f(t=-\infty)\sim\psi(t=-\infty)$), the difference
$\tau(r)=t_{r}[\psi]-t_{r}[\psi_f]$ corresponds to the
time delay inside the same ball due to scattering by the potential
$U({\bf x})$. The {\it global} time delay is defined as
$\tau_d=\tau(r\to\infty)$ and under quite general conditions
(see e.g.\cite{delayop})
can be shown to be equal to the {\it time-independent} expectation  
value\begin{equation}\label{delayop}
\tau_d=\int d{\bf x_1}\int d{\bf x_2} \psi^*({\bf x_1},t)
T_d({\bf x_1},{\bf x_2})\psi({\bf x_2},t)
\end{equation}
where $T_d({\bf x_1},{\bf x_2})$ are matrix elements of a Hermitian
{\it time delay operator} $\hat{T}_d$ in the position representation.
This operator turns out to be commuting with the Hamiltonian and  
intimately
related to the so called Wigner-Smith time delay matrix \cite{Wigner}
defined in terms of the $S-$matrix as :
\begin{equation}\label{wignertime}
\hat{\tau}_w(E)=i\hbar
\frac{\partial \hat{S}^{\dagger}}{\partial E}\hat{S}
\end{equation}
In particular, following  the papers \cite{delayop,Harney}
we show in section II that
the eigenvalues of the operator $\hat{T}_d$ just coincide with the  
eigenvalues of $\hat{\tau}_w(E)$. The quantum mechanical
expectation value of the time delay averaged over different  
channels turns out to be
equal to\cite{Harney,Lyub,Lyubver,Bauer}:
\begin{equation}\label{avdel}
\overline{\tau(E)}=\frac{i\hbar}{M}Tr
\overline{\frac{\partial \hat{S}^{\dagger}}{\partial E}\hat{S}}=
-\frac{i\hbar}{M}\frac{\partial}{\partial E}
\overline{\ln{\mbox{Det}\hat{S(E)}}}=\hbar\frac{1}{M}
\overline{\sum_{a=1}^M
{\partial\theta_a/\partial E}} \end{equation}
where the bar denotes the averaging over the energy spectrum of the
packet. This shows the relation between the phase shift derivatives  
and mean time delay mentioned above.

A quite detailed analysis of the time-delay problem was given
in the context of nuclear physics by Lyuboshits\cite{Lyub,Lyubver}  
and other authors\cite{Bauer}. In particular, for a wave packet of  
arbitrary form Lyuboshits\cite{Lyubver} suggested
a concept of the probability distribution of its time delay. His  
definition is based on the interpretation of the quantity  
$P(t)=\int_{{\bf x}\in V}
\mid\psi({\bf x},t)\mid^2 d^3 {\bf x}$ as the quantum mechanical  
probability to be found within the volume $V$ at instant $t$.
Then the time derivative $\partial P/\partial t$  can be used to define
the distribution of times spent inside the volume $V$.
 A general and illuminating
discussion of the time evolution properties of wave packets in a
generic chaotic systems can be found in \cite{Harney}.

On the other hand, the existence of the Hermitian time-delay operator
$\hat{T}_d$ in Hilbert space suggests an alternative definition of  
the time delay
statistics by the natural requirement that $\overline{\tau^p}=
\langle\Psi(t)\mid \hat{T}^p_d\mid\Psi(t) \rangle$ for any wave packet
 $\Psi(t)$. Then the problem is reduced basically
to study the statistical properties
of the Wigner-Smith time delay matrix $\hat{\tau}_w(E)$.

The chaotic scattering makes the Wigner-Smith time delay matrix
(in particular, the quantity $\frac{1}{M}Tr\hat{\tau}_w$
which is called just Wigner time
delay)  to be
a strongly fluctuating function of the energy $E$ as well as of any  
external
parameter $X$. From this point of view we can speak
about distributions and correlation functions of these quantities.   
Similarly, the distribution of partial delay times can be used to  
characterize variations of time scales associated with the chaotic  
scattering
process.  Various statistical aspects of time
evolution of the chaotic quantum systems were studied earlier
in some details in \cite{Harney,ISS,Lehmann}.

Being an important characteristic of the scattering process,
the statistics of phase shifts and their derivatives deserve to be  
investigated in more
detail. Additional interest to the problem attaches the fact
of relevance of Wigner time delay in condensed matter physics.
Indeed, in a series of papers by B\"{u}ttiker and collaborators the  
Wigner time delay was related to the frequency-dependent response  
of mesoscopic capacitors\cite{But2,Gopar,BB}. A more detailed
discussion of this issue can be found in the last section of the  
present paper.

More general
parametric derivatives of the scattering phase shifts can also be
related to some observable quantities. As a particular example
we mention the relation between the persistent currents and the
derivative of the total phase shift over the magnetic flux derived  
by Akkermans et al.\cite{Akk}.
These authors considered  "open mesoscopic networks":  
two-dimensional systems of conducting loops coupled to infinitely  
long ideal leads (waveguides). The loops can
encircle a flux tube with flux $\phi$. The expectation value of the  
persistent current around the flux
tube in the state $\mid\Psi\rangle$ is $I(\Psi,\phi)=
-\langle \Psi\mid d{\cal H}/d\phi\mid \Psi\rangle$. For the case of  
a closed system
(i.e. when loops are disconnected from the leads)
 each discrete level $E_n(\phi)$ carries the current
$-dE_n/d\phi$. When the system is open , it turns out that the
differential contribution of the
scattering states at energy E to the persistent current can be
expressed in terms of phase shift derivatives as
\cite{Akk}:\begin{equation}\label{Akk}
dI(E,\phi)=\frac{1}{2\pi i}\frac{\partial}{\partial \phi}
\left[\ln\mbox{det}\hat{S}(E,\phi)\right]dE=\frac{1}{2\pi}  
\frac{\partial}{\partial \phi}\left[\sum_{a}\theta(E,\phi)\right]dE
\end{equation}

In the present paper we give quite a detailed analysis
of the statistical properties of scattering phase shifts and their
derivatives for generic chaotic scattering in a system with broken
time-reversal invariance. The extension of our results to other  
symmetry classes as well
as to the crossover regimes will be published elsewhere\cite{FSS}.   
We find it to be most informative to
 concentrate our attention on the so-called $K-$matrix related to  
the scattering
matrix as:
\begin{equation}\label{K}
\hat{S}=\frac{\hat{I}-i\hat{K}}{\hat{I}+i\hat{K}}
\end{equation}
This equation shows that eigenphases $\theta_a(E,X),\quad a=1,...,M$
considered modulo $2\pi$ are determined in a unique way by the
eigenvalues $z_a(E,X)$ of the $K-$matrix , where we have indicated  
explicitly
both the energy dependence and dependence on an external parameter
X. First of all, we calculate explicitly the correlation function
of densities of the eigenvalues of the $K-$matrix at two different  
energies $E\pm \Omega$ and parameter values $X\pm\delta X$.
When $\Omega=\delta X=0$ this correlation function turns out to be
the same as the pair correlation function following from the
Poisson's kernel. This fact confirms the expected equivalence of
the minimum information approach and Hamiltonian approach for
the case of fixed energy, and as such extends the earlier studies
on that subject\cite{Brouwer}.
 From that moment we concentrate on the statistics of delay times
and parametric derivatives of  phase shifts. First, we derive and  
analyze the general expression for the distribution of "partial
delay times" $\tau_a=\partial\theta_a/\partial E$
(here and henceforth we frequently put $\hbar=1$) as well as  
derivatives $\partial\theta_a/\partial X$. This distribution, being  
an interesting characteristic of the
chaotic scattering by itself, also allows us to detect
the qualitative features of the Wigner time delay distribution. In  
particular, for the one-channel system
the partial delay time is exactly the same as the Wigner time  
delay. After that, we derive the parametric correlation function
of the Wigner time delay and show some interesting
correspondences with the results of the semiclassical approach.
A short account of our results was published earlier\cite{FSres,FStime}.

The organization of the paper is as follows. Section II is meant to  
be a kind of introduction to the random matrix method of the  
description of
an open chaotic quantum system. It is based mainly on the original
papers by other authors
\cite{delayop,Harney,Seba1,Seba2}.
 Considering a particular generic example we discuss the main  
ingredients of the model
 and present a quite detailed discussion
of the time-delay operator and other quantities characterizing time
evolution in such systems.

In Section III we derive and analyze the density of the S-matrix poles
in the complex plane. Section IV is devoted to the statistics of
eigenvalues of the $K-$matrix, phase shifts and their derivatives  
and analyses different statistical aspects of the Wigner time delay.
It contains also a kind of semiclassical analysis of the parametric
correlations of Wigner time delays. Concluding remarks and a  
discussion of the potential experimental relevance of the obtained  
results
can be found in the final Section V.
\section{Scattering Problem for Random-Matrix Hamiltonians}
\subsection{General description of the model.}
A model which is most appropriate for incorporating
random matrix ideas for describing the phenomenon of quantum  
chaotic scattering was discussed in great details in the works by
Verbaarschot, Weidenm\"{u}ller and Zirnbauer\cite{VWZ} and Lewenkopf and
Weidenm\"{u}ller \cite{Lew}. A general construction actually goes
back to works by Feshbach \cite{Fesh} and is based on the theory of  
quantum scattering as
formulated in the book by Mahaux and Weidenm\"{u}ller\cite{MW} on
the theory of nuclear reactions, or in the book by Levine\cite{Levine}
 about quantum scattering in molecules. Let us also mention a very  
general and profound analysis of open quantum and classical wave  
systems performed in the book by
$\mbox{Liv\v{s}i\v{c}}$ \cite{Livsic}.
In order to make the present paper self-contained we give below a
list of the main ingredients of the model. Since the general
construction is a rather abstract one, we find it instructive
to illustrate it by presenting the explicit derivation of the
 expression for the scattering matrix and its subsequent analysis
 for a  generic system of much interest: scattering
of a quantum particle moving in a perfect lead of width $d$ in
contact with a chaotic region, simulated by a random matrix  
Hamiltonian. Such a derivation follows that of the papers   
\cite{Seba1,Seba2} (see also the book \cite{Livsic} and the works by  
Pavlov and collaborators \cite{Pavlov}).

When dealing with a generic scattering problem it is natural to
assume that the scattering event is always confined inside a compact
part of the available space which is called the "interaction region".
Outside this region interaction is absent
and fragments exhibit a free motion characterized (apart from the
total energy $E$)
by a set of quantum numbers describing the internal quantum state of
each fragment.
As such, these quantum numbers specify the states in which  
particles can be
found long before and long after the scattering takes place -
the so-called "channels of reaction". For example, in nuclear and  
molecular physics different channels are marked by relative
angular momentum and spins of colliding particles. In the particular
example considered in much detail below the motion of a particle
 along an infinite lead of width $d$ is quantized along the  
transverse direction, the different transverse modes being different  
channels. Assuming that exactly $M$ channels
at given energy $E$ are "open" (i.e. allow for an unbounded motion
of the particles), we associate with the channel region a {\it  
continuous} set of functions
$\mid a,E\rangle;\quad a=1,...,M$ normalized as $\langle a,E_1\mid  
b,E_2\rangle=
\delta_{ab}\delta(E_1-E_2)$. An analogous, but {\it discrete} set of
orthogonal states $\mid n\rangle;\quad n=1,2,...,N$
is associated with the compact interaction region.

In the absence of interaction between the states in channels and
the internal states the Hamiltonian of the system has obviously
the form:
\begin{equation}\label{noint}
{\cal H}_0=\sum_{l,m}\mid l\rangle \left(H_{in}\right)_{l,m}
\langle m \mid+ \sum_a\int dE\mid a,E\rangle E\langle a,E\mid
\end{equation}
where the integration goes over the energy region where the given
channel $a$ is open. The model is simplified by neglecting any direct
coupling between different channels; hence the corresponding term
in the Eq.(\ref{noint}) is diagonal in $a$. The first term describes the
Hamiltonian $\hat{H}_{in}$ of the "closed"
chaotic system possessing $N\gg 1$ bound states, which are
eigenstates of $\hat{H}_{in}$. In the spirit of the
Random Matrix universality conjecture, we simulate this part of
the Hamiltonian by taking $\hat{H}_{in}$ to be a Gaussian random
$N\times N$ matrix. The number $N$ is considered to be
large: $N\gg 1$.
To describe the interaction between channels and bound  
states(converting bound states into resonances) one
adds to the Hamiltonian Eq.(\ref{noint}) the "interaction term":
\begin{equation}\label{intterm}
\hat{V}=
\sum_{l,a}\left(\mid l\rangle\int dE W_{la}\langle a,E\mid
+\mbox{herm.conj.} \right)
\end{equation}In any practical implementation of such a procedure  
one should make
sure that the total Hamiltonian ${\cal H}={\cal H}_0+\hat{V}$
is self-adjoint. This point is not at all trivial (see the example
below). The general way of self-adjoint matching of Hamiltonians  
with internal structure and those describing motion in external  
scattering channels was suggested by Pavlov and developed by Pavlov  
and collaborators \cite{Pavlov}.
A  particularly convenient formulation of the Pavlov's method
suggested by Makarov \cite{Mak}
was applied to the problem of chaotic scattering in the recent  
papers \cite{Seba1,Seba2}.

After the self-adjoint matching is done one can employ standard
 methods in scattering theory (see the mentioned books \cite{MW,Levine})
in order to write down the Lipmann-Schwinger equation
for the in-and outgoing scattered waves and find an explicit  
expression for
the scattering matrix.
\subsection{ From Random Matrix Hamiltonian to Scattering Matrix.}
Instead of demonstrating such a formal derivation within a quite
abstract "projection formalism" (see e.g. \cite{Lew,Harney})
we find it to be
more illuminating to show how to derive the $S-$matrix in an  
alternative way
\cite{Seba1,Seba2}. To do this let us confine ourself to a
particular generic example of a scattering system depicted schematically
in Fig.1.:  a two dimensional cavity  of irregular shape with  
impenetrable walls coupled to
an infinite waveguide (lead) of width $d$.
  Let us mention, that it is one of the favorite models
for the study of generic features of chaotic scattering, both
theoretically (see e.g. the "frying pan" model in \cite{Smicol})
and experimentally \cite{cav1,Dosmifr}.
The propagation of a quantum particle inside the lead is described by
the Schr\"{o}dinger equation:
\begin{equation}\label{schlead}
-\frac{\hbar^2}{2m}\left(\frac{\partial^2}{\partial x^2}+
\frac{\partial^2}{\partial y^2}\right)\Psi(x,y)=\frac{\hbar^2k^2}{2m}
\Psi(x,y);\quad \Psi(x,y=\pm d/2)=0
\end{equation}
whose general solution can be represented as a $M-$component vector:
${\bf \Psi}=\left(\Psi_1(x,y),...,\Psi_M(x,y)\right)^T$ with
components $\Psi_a(x,y)=\psi_a(x)\phi_a(y)$ ,  
where\begin{equation}\label{chanst}
\psi_a(x)=\frac{1}{(2\pi\hbar^2 k_a/m)^{1/2}}
\left[A_ae^{-ik_ax}+B_ae^{ik_ax}\right];\quad \phi_a(y)=(2/d)^{1/2}
\sin{\left[\left(\frac{a\pi}{d}\right)(y+d/2)\right]}
\end{equation}
for $x\ge0;\,\,\mid y\mid\le d/2;\,\, a=1,2,...,M$, with the number  
$M$ of open channels at the
energy $E=\frac{\hbar^2k^2}{2m}$ being equal to the largest integer
less or equal to $\frac{kd}{\pi}$ and the wave vector $k_a$ being
equal to $k_a=\left[k^2-\left(\frac{a\pi}{d}\right)^2\right]^{1/2}$,
so that $-\frac{\partial^2}{\partial x^2}\psi_a=k_a^2\psi_a$.
The running waves are properly normalized to energy
$\delta-$functions; unitarity of the $S-$matrix to be introduced later
is related to the conservation of the probability flux.

The situation of the waveguide disconnected from the cavity
we describe by the boundary conditions: $\partial \psi_a/\partial
x\mid_{x=0}=0$. This means that the particle in each channel is
just reflected back: $A_a=B_a$. The corresponding $S-$matrix
relating the vectors
of incoming ${\bf A}=(A_1,...,A_M)^T$ and outgoing $ {\bf  
B}=(B_1,...,B_M)^T$ amplitudes: ${\bf B}=\hat{S}{\bf A}$
is just the unity $M\times M$ matrix: $\hat{S}=\hat{I}$.
The role of the vector $\mid a,E\rangle$ of general construction
 (see Eqs.(\ref{noint},\ref{intterm})) is played by the vector
${\bf\Psi}$ corresponding to the particular choice of amplitudes of
incoming waves: $A_a=1;A_{b\ne a}=0$.
The Hamiltonian of the particle motion inside the cavity
is simulated by the $N\times N$ random Hermitian matrix $\hat{H}_{in}$.
Correspondingly, the "internal"
wave function is represented by the $N-$component vector ${\bf  
u}=(u_1,...,u_N)^T$.
The wave functions of the scattering system as a whole ("cavity  
attached to the lead") are therefore vectors
${\bf \Phi}=\left(\begin{array}{c}{\bf u}\\{\bf
\Psi}\end{array}\right)$ from the Hilbert space $L^2(R^+,C^M)\oplus
C^N$ supplied with the scalar product:
$$
\left({\bf \Phi}_1,{\bf \Phi}_2\right)={\bf u_1}^{\dagger}{\bf u_2}+
({\bf \Psi}_1,{\bf \Psi}_2);\quad \mbox{where}\quad ({\bf \Psi}_1,
{\bf \Psi}_2)=
\int_{-d/2}^{d/2}dy\int_{0}^{\infty}dx{\bf \Psi}_1^{\dagger}{\bf \Psi}_2
$$

 Let us define the Hamiltonian operator ${\cal H}$ of the system
as a whole acting in that Hilbert space  as:
\begin{equation}\label{hamwhole}
{\cal H}\left(\begin{array}{c}{\bf u}\\{\bf
\Psi}\end{array}\right)=\left(\begin{array}{c}\hat{H}_{in}{\bf u}+
\int_{-d/2}^{d/2}dy\int_0^\infty dx{\cal W}(x,y){\bf \Psi}(x,y)
\\ {\cal V}(x,y){\bf u}+\hat{{\cal H}}_{ch}{\bf\Psi}
\end{array}\right)
\end{equation}
where $\hat{{\cal H}}_{ch}$ is the operator diagonal in the channel  
space:
$$\hat{{\cal H}}_{ch}=
\mbox{diag}\left(\frac{-\hbar^2}{2m}(\partial_x^2+\partial_y^2),...,
\frac{-\hbar^2}{2m}(\partial_x^2+\partial_y^2)\right)$$
and ${\cal W}(x,y)$ and $ {\cal V}(x,y)$ are $N\times M$ and $M\times
N$ rectangular matrices describing a coupling between two parts of
the Hilbert space.
Let us assume for simplicity that the coupling is local along the  
waveguide \cite{Seba1,Seba2}: ${\cal
W}(x,y)= \delta(x) {\cal W}(y)$, so that  
$\int_{-d/2}^{d/2}dy\int_0^\infty dx{\cal W}(x,y){\bf \Psi}(x,y)=
\int_{-d/2}^{d/2}dy{\cal W}(y){\bf  
\Psi}\mid_{x=0}=\hat{w}{\bf\psi}(x=0)$, where  
$w_{ia}=\int_{-d/2}^{d/2}dy{\cal W}_{ia}(y)\phi_a(y)
\quad i=1,...,N,\,a=1,...,M$ and $
\psi(x)=(\psi_1(x),...,\psi_M(x))^T$. On the other hand,
  we have to put ${\cal V}(x,y)\equiv 0$ in order to be consistent  
with the locality of the coupling
and to stay in the space spanned by the vectors ${\bf \Psi}$.

The operator ${\cal H}$ defined in such a way is not, in general, a
self-adjoint one. Indeed,
\begin{equation}\label{sadj1}
\left({\cal H}{\bf \Phi_1},{\bf \Phi_2}\right)=
{\bf u_1}^{\dagger}\hat{H}_{in}^{\dagger}{\bf u}_2+
\left(\hat{{\cal H}}_{ch}{\bf \Psi}_1,{\bf
\Psi}_2\right)+{\bf \psi}_1^{\dagger}(x=0)\hat{w}^{\dagger}{\bf u}_2
\end{equation}
and
$$
\left({\bf \Phi_1},{\cal H}{\bf \Phi_2}\right)=
{\bf u_1}^{\dagger}\hat{H}_{in}{\bf u}_2+
\left({\bf \Psi}_1,\hat{{\cal H}}_{ch}{\bf
\Psi}_2\right)+{\bf u}_1^{\dagger}\hat{w}
{\bf \psi}_2(x=0)
$$
 From the definition of the operator $\hat{{\cal H}}_{ch}$ and
that of the scalar product $({\bf \Psi}_1,{\bf \Psi}_2)$ one can easily
find after using partial integration and the fact
$\hat{H}_{in}^{\dagger}=\hat{H}_{in}$ that:
\begin{eqnarray}\label{sss1}
\left({\cal H}{\bf \Phi}_1,{\bf \Phi}_2\right)-
\left({\bf \Phi}_1,{\cal H}{\bf \Phi}_2\right)=&&\\ \nonumber
& \frac{\hbar^2}{2m}
\left\{\left(\frac{\partial}{\partial
x}\psi^{\dagger}_1\right)\psi_2-
\psi_1^{\dagger}
\left(\frac{\partial}{\partial
x}\psi_2\right)\right\}\large\mid_{x=0}+
{\bf \psi}_1^{\dagger}(x=0)\hat{w}^{\dagger}{\bf u}_2-
{\bf u}_1^{\dagger}\hat{w}
{\bf \psi}_2(x=0)
&
\end{eqnarray}
In order to have a self-adjoint Hamiltonian operator ${\cal H}$
one has to impose some appropriate boundary conditions at the
point $x=0$ ensuring that the expression above is vanishing \cite{fnote}.
 The most obvious (however, not the most
general) choice is:
\begin{equation}\label{boundcon}
\hat{w}^{\dagger}{\bf u}=\frac{\hbar^2}{2m}
\left(\frac{\partial}{\partial x}\psi\right)
\mid_{x=0}
\end{equation}
On the other hand, the solution of the Schr\"{o}dinger equation for
the whole system: ${\cal H}{\bf \Phi}=E{\bf \Phi};\quad
E=\frac{\hbar^2 k^2}{2m} $ (we call these solutions the "scattering
states" and denote them ${\bf \Phi}_E$ henceforth)
leads immediately to the relation:
\begin{equation}\label{rel}
{\bf u}=\left(E-\hat{H}_{in}\right)^{-1}\hat{w}\psi(x=0)
\end{equation}
which together with Eq.(\ref{boundcon})
yields the following equation for the vector $\psi(x)$:
\begin{equation}\label{rel1}
\hat{w}^{\dagger}\left(E-\hat{H}_{in}\right)^{-1}\hat{w}\psi(x=0)=
\frac{\hbar^2}{2m}
\left(\frac{\partial}{\partial x}\psi\right)
\mid_{x=0};\quad \psi(x)=\left(\frac{m}{2\pi\hbar^2}\right)^{1/2}
\left(\begin{array}{c}
\frac{1}{\sqrt{k_1}}\left[A_1e^{-ik_1x}+B_1e^{ik_1x}\right]\\ ...\\
\frac{1}{\sqrt{k_M}}\left[A_Me^{-ik_Mx}+B_Me^{ik_Mx}\right]
\end{array}\right)\end{equation}This equation allows us to find  
easily the unitary scattering matrix:
\begin{equation}\label{smat1}
\hat{S}=\left[\hat{I}-i\hat{K}\right]\times
\left[\hat{I}+i\hat{K}\right]^{-1};\quad \hat{K}=\pi\hat{W}^{\dagger}
\frac{1}{E-\hat{H}_{in}}\hat{W}
\end{equation}
where $\hat{W}=\sqrt{2m/\pi\hbar^2}\hat{w}
\mbox{diag}(k_1^{-1/2},...,k_M^{-1/2})$, $\hat{I}$ is the unity  
matrix of the corresponding dimension.
Often we treat diagonal  
matrices proportional to $\hat{I}$ simply as numbers. Sometimes, we  
indicate the dimension as index, here it is $\hat{I}=\hat{I}_M$.

The expression Eq.(\ref{smat1}) can be also rewritten in another form,
frequently used in applications. To this end we write:
$\left[\hat{I}-i\hat{K}\right]\times
\left[\hat{I}+i\hat{K}\right]^{-1}=\left[\left(\hat{I}+i\hat{K}\right)-
2i\hat{K}\right]\times\left[\hat{I}+i\hat{K}\right]^{-1}=
1-2i\left(\hat{I}+i\hat{K}\right)^{-1}\hat{K}$
and use the identity\begin{eqnarray}\label{eqind}
\left[\hat{I}+i\pi\hat{W}^{\dagger}
\frac{1}{E-\hat{H}_{in}}\hat{W}\right]^{-1}\hat{W}^{\dagger}
\frac{1}{E-\hat{H}_{in}}\hat{W}=\sum_{k=0}^{\infty}(-i\pi)^k
\left[\hat{W}^{\dagger}
\frac{1}{E-\hat{H}_{in}}\hat{W}\right]^{k+1}= \\ \nonumber  
\hat{W}^{\dagger}\left[\hat{I}+i\pi
\frac{1}{E-\hat{H}_{in}}\hat{W}\hat{W}^{\dagger}\right]^{-1}
\frac{1}{E-\hat{H}_{in}}\hat{W}=
\hat{W}^{\dagger}
\frac{1}{E-\hat{H}_{in}+i\pi\hat{W}\hat{W}^{\dagger}}\hat{W}
\end{eqnarray}
which means that the scattering matrix can be written in the
form:
\begin{equation}\label{smat2}
\hat{S}=\hat{I}-2i\pi\hat{W}^{\dagger}
\frac{1}{E-{\cal H}_{ef}}\hat{W}
\end{equation}
where the {\it non Hermitian} effective Hamiltonian ${\cal H}_{ef}$
is given by ${\cal H}_{ef}=\hat{H}_{in}-i\hat{\Gamma}$
and $\hat{\Gamma}=\pi\hat{W}\hat{W}^{\dagger}$.

The expression for the scattering matrix of the form Eq.(\ref{smat2})
appears generally when one describes an open quantum system
 decaying into several open channels, see e.g. \cite{Livsic,KNO}.
In particular, it can be derived from the general Hamiltonian
Eq.(\ref{noint},\ref{intterm}) under the  assumption that the elements
of the matrix $\hat{W}$ are energy-independent.
This was just a starting point in the
approach by Weidenm\"{u}ller and collaborators \cite{VWZ,Lew}.
 In the derivation above the matrix $\hat{W}$ does depend on the
energy $E$ via the parameters  
$k_a=\left[k^2-\left(\frac{a\pi}{d}\right)^2\right]^{1/2}$.
We, however, will be mostly interested in the {\it fluctuation}  
properties of the energy-dependent $S$-matrix.
The typical energy scale of such fluctuations is given by the
mean level spacing $\Delta$ - a typical separation between the adjacent
eigenvalues of the matrix $\hat{H}_{in}$. Far from the thresholds,  
as long as $\Delta$
is negligible in comparison with the difference between the adjacent
threshold energies  
$\frac{\hbar^2}{2m}(k_{M}^2-k_{M-1}^2)=\frac{2M-1}{2m}\left(\frac{\hbar\pi}{d}\right)^2$  
we can safely neglect
the energy dependence of the matrix $\hat{W}$.
In view of $\Delta\propto 1/N$ the latter requirement is always  
satisfied in the limit
$N\gg 1$ which is the only case studied in the present paper.

\subsection{Time Evolution of the Wave Packets: Staying Probability  
and Time Delay Operator for Open Quantum Chaotic Systems}

Before describing the time evolution of wave packets
let us note that
any particular scattering state ${\bf \Phi}_E=
\left({\bf u}(E),{\bf\Psi}(E)\right)^T$ is uniquely specified
by the set of incoming amplitudes ${\bf A}=
\left(A_1(E),...,A_M(E)\right)^T$. Being the eigenfunctions of the
Hermitian Hamiltonian ${\cal H}$ the scattering states must be
orthogonal. Below we verify by
a direct calculation the validity of the orthogonality condition:
\begin{equation}\label{orthog}
\left({\bf \Phi}_{E_1},{\bf \Phi}_{E_2}\right)=
{\bf u}^{\dagger}(E_1){\bf u}(E_2)+
\left({\bf \Psi}_{E_1},{\bf \Psi}_{E_2}\right)=
\delta(E_1-E_2){\bf A}^{\dagger}_1{\bf A}_2\end{equation}
Such a calculation allows one to derive some helpful relations that
are used later on.

By using the identities$$\frac{1}{2\pi}\int_{0}^{\infty}dx  
e^{-iux}=\frac{1}{2}\delta(u)+
\frac{1}{2\pi iu};\quad
\frac{m}{\hbar^2}\frac{1}{\sqrt{k_a(E_1)k_a(E_2)}}\delta\left(
k_a(E_1)-k_a(E_2)\right)=\delta(E_1-E_2)$$
and exploiting the definition of the $S$-matrix: ${\bf
B}(E)=\hat{S}(E){\bf A}(E)$ and its unitarity
one easily finds:
\begin{equation}\label{nor2}
\left({\bf \Psi}_{E_1},{\bf
\Psi}_{E_2}\right)=\delta(E_1-E_2)\left({\bf
A}^{\dagger}_1{\bf A}_2\right)
+\frac{1}{2\pi i(E_1-E_2)}{\bf
A}_1^{\dagger}\left(\hat{S}^{\dagger}(E_1)
\hat{S}(E_2)-\hat{I} \right){\bf A}_2
\end{equation}
Here we assumed that both energies $E_1,E_2$ are far from thresholds and
close to one another, so that effectively we can put
$k_a(E_1)=k_a(E_2)$ in the expression above everywhere (this is  
consistent with
neglecting the energy dependence of $\hat{W}$ as discussed  
earlier), except in the denominator, where $(E_1-E_2)\propto  
\left(k_a^2(E_1)-k_a^2(E_2)\right)$.

Let us now use the relation Eq.(\ref{rel}) rewritten in the
following form:
\begin{equation}\label{boun}
{\bf u}_{E}=\frac{1}{2}\frac{1}{E-\hat{H}_{in}}\hat{W}\left(\hat{I}+
\hat{S}(E)\right){\bf A}
\end{equation}
 Hence:
\begin{equation}\label{boun1}
{\bf u}^{\dagger}_{E_1}{\bf u}_{E_2}=\frac{1}{4}
{\bf A}^{\dagger}_1\left(\hat{I}+\hat{S}^{\dagger}(E_1)\right)
\hat{W}^{\dagger}\frac{1}{E_1-\hat{H}_{in}}
\frac{1}{E_2-\hat{H}_{in}}\hat{W}\left(\hat{I}+
\hat{S}(E_2)\right){\bf A}_2
\end{equation}
Now we use the identity:
\begin{eqnarray}\label{bbv}
\hat{W}^{\dagger}\frac{1}{E_1-\hat{H}_{in}}
\frac{1}{E_2-\hat{H}_{in}}\hat{W}=\frac{1}{E_2-E_1}\hat{W}^{\dagger}\left[
\frac{1}{E_1-\hat{H}_{in}}-\frac{1}{E_2-\hat{H}_{in}}\right]\hat{W}=\\  
\nonumber \frac{1}{\pi(E_2-E_1)}
\left(\hat{K}(E_1)-\hat{K}(E_2)\right)
\end{eqnarray}
where we exploited the definition of the $\hat{K}$ matrix, see  
Eq.(\ref{smat1})  neglecting the energy
dependence of the matrix $\hat{W}$. Relation Eq.(\ref{smat1})  
between $S$-matrix
and $K-$matrix can be written also as $\hat{K}(E)=-i\left(\hat{I}-
\hat{S}(E)\right)\left(\hat{I}+\hat{S}(E)\right)^{-1}=\hat{K}^{\dagger}(E)$.  
Substituting
this relation into Eq.(\ref{bbv}) we use it to reduce Eq.(\ref{boun1})
to the following final form: \begin{equation}\label{ufin}
{\bf u}^{\dagger}_{E_1}{\bf u}_{E_2}=-\frac{1}{2\pi i(E_1-E_2)}
{\bf  
A}^{\dagger}_1\left(\hat{S}^{\dagger}(E_1)\hat{S}(E_2)-\hat{I}\right)
{\bf A}_2
\end{equation}
We see that when combined together the Eqs.(\ref{nor2}) and (\ref{ufin})
produce exactly the orthogonality condition Eq.(\ref{orthog}).
In particular, this orthogonality condition allows us to use the  
scattering states ${\bf \Phi}_E^{(a)}$ corresponding
to the choice of incoming amplitudes $A_a=1,\,A_{b\ne a}=0$
as a convenient basis in the full Hilbert space. Denoting
${\bf \Phi}_E^{(a)}\equiv\mid {\bf \Phi}_E^{(a)}\rangle$ one can
write down the total Hamitonian ${\cal H}$ as :
\begin{equation}\label{Htot} {\cal H}=\int dE \sum_{a=1}^M\mid{\bf  
\Phi}_E^{(a)}\rangle\,
E\,\langle {\bf \Phi}_E^{(a)}\mid\end{equation}

Now we are prepared to answer the following question: given a wave packet
$\mid {\bf \Phi}(t)\rangle=\left({\bf u}(t),{\bf \Psi}(t)\right)^T$
 which evolves according the Schr\"{o}dinger
equation $i\hbar(\partial/\partial t)\mid{\bf \Phi(t)}\rangle=
{\cal H}\mid {\bf\Phi}(t)\rangle$,   how to express in terms of the  
scattering matrix
 the probability for the
corresponding particle to be found inside the "chaotic" domain $x<0$
 at instant $t$.

According to the rules of quantum mechanics this probability
is just given by $P(t)={\bf u}^{\dagger}(t){\bf u}(t)$.
Let us expand the wave packet over the scattering states
$\mid {\bf\Phi}_E\rangle=\sum_{a=1}^M A_a\mid{\bf
\Phi}_E^{(a)}\rangle$   as:
\begin{equation}\label{wp1}
\mid {\bf\Phi}(t)\rangle=\int dE f(E) \mid  
{\bf\Phi}_E\rangle\exp{-\frac{it}{\hbar}E}
;\quad \int dE\mid f(E)\mid^2=1
\end{equation}
where the coefficients $f(E)$ determine the initial form of the
wave packet and we assume: ${\bf A}^{\dagger}{\bf A}=1$ so that
$\langle \Phi(t)\mid\Phi(t) \rangle=1$.
 This immediately gives us the desired expression:
\begin{eqnarray}\label{wp2}
P(t)=\int dE_1dE_2 f^*(E_1)f(E_2){\bf u}^{\dagger}_{E_1}
{\bf u}_{E_2}\exp{-\frac{it}{\hbar}(E_1-E_2)}= & & \\ \nonumber
\int\frac{dE_1dE_2}{2\pi i} \frac{f^*(E_1)f(E_2)}{(E_2-E_1)}
e^{-\frac{it}{\hbar}(E_1-E_2)}\sum_{ab}A_a^{*}A_b
\left[\left(\hat{S}^{\dagger}(E_1)\hat{S}(E_2)\right)_{ab}-\delta_{ab}\right]&  
&\end{eqnarray}
where we made use of the Eq.(\ref{ufin}).
The mean time spent in the interaction region (i.e. the mean time  
delay $\tau_d$)
can be found integrating this expression over the time. This
operation produces a $\delta-$ functional factor $\delta(E_1-E_2)$
in the integrand which finally gives:
\begin{equation}\label{wp2a}
\tau_d=\int_{-\infty}^{\infty}P(t)dt=\sum_{a,b}A_a^*A_b
\overline{\hat{\tau}_{ab}(E)}
\end{equation}
where
\begin{equation}\label{uu}\hat{\tau}(E)=i\hbar\frac{\partial
\hat{S}^{\dagger}(E)}{\partial E}\hat{S}(E)=-i\hbar\hat{S}^{\dagger}(E)
\frac{\partial\hat{S}}{\partial E}
\end{equation}
 is the Wigner-Smith time delay matrix \cite{fnote1}
 and the bar
stands for the energy averaging determined by the wave packet spectrum:
$\overline{ (...)}=\int dE (...) \mid f(E)\mid^2$.
If the particle comes only via a particular channel $a$ the
scattering states $\mid{\bf
\Phi}_E\rangle$ coincide with the basis states $\mid{\bf
\Phi}_E^{(a)}\rangle$ and the corresponding time $\tau_d^{(a)}$
 (which is natural to call the "delay time for the channel $a$") is  
given by the
spectral average of the diagonal element $\hat{\tau}_{aa}(E)$.
Then the delay time averaged over all channels
 is given by the spectral average of the Wigner-Smith time  
delay:$\frac{1}{M}\overline{\mbox{Tr}\hat{\tau}(E)}$.

The time derivative $-dP/dt$ is a current out of the chaotic region.
Assuming that the spectral function $f(E)$ varies with
$E$ on a much larger scale than the mean level spacing $\Delta$
( the latter scale is typical for variations of
$S-$matrix elements) we can
put $f(E_1)\approx f(E_2)$ in the expression Eq.(\ref{wp2})
This results in the following expression:
\begin{equation}\label{uu1}
\frac{d}{dt}P(t)=\delta(t)-p(t);\quad p(t)=\frac{1}{2\pi}\int
d\epsilon e^{i\epsilon t/\hbar}\sum_{ab}A_a^*A_b\left\langle\left(
\hat{S}^{\dagger}(E+\epsilon) \hat{S}(E)\right)_{ab}\right\rangle
\end{equation}

This expression can be interpreted as follows\cite{Lyubver,Harney}.
 In our approximation the part of the Hilbert space corresponding
to the chaotic region is not populated at $t<0$. At $t=0$
the wave packet reaches the chaotic region and populates its
states instantly. This fact is described by the $\delta(t)$ term in the
expression Eq.(\ref{uu1}). Then the function $p(t)$ has a meaning
of the distribution of duration of stay inside the chaotic
region. This was just the reason to call $p(t)$
the distribution of time delays \cite{Lyubver}.
On the other hand, according to conventional rules of quantum mechanics
in order to speak about the probability distribution of some  
observable one should be able to find a
Hermitian operator in Hilbert space generating all the moments
of that observable as expectation values of the integer powers
of this operator. Some important insights in the issue of
constructing such an operator can be found in\cite{delayop}.
To this end let us consider the following
{\it time delay operator} constructed in terms of Wigner-Smith
time-delay matrix Eq.(\ref{uu}) as:
\begin{equation}\label{tdo}
\hat{T}_W=\int dE\sum_{ab}\mid {\bf\Phi}_E^{(a)}\rangle
(\hat{\tau})_{ab}(E) \langle {\bf\Phi}_E^{(b)}\mid dE
\end{equation}
The Hermiticity of this operator follows from that of the  
Wigner-Smith matrix.
 It commutes with the Hamiltonian, Eq.(\ref{Htot}),
 but is not at the same time diagonal
 due to the degeneracy of ${\cal H}$.
It is evident that for any wave packet $\mid{\bf\Phi}(t)\rangle$
the mean time delay given in Eq.(\ref{wp2a}) is just the {\it time
independent} expectation value:\begin{equation}\label{mdl}
\langle t\rangle\equiv\tau_d=
\langle{\bf \Phi}(t)\mid\hat{T}_W\mid{\bf \Phi}(t)\rangle
\end{equation}
Then it is natural to define the higher moments of the time delay
as :
\begin{equation}\label{mdl1}
\langle t^p\rangle=
\langle{\bf \Phi}(t)\mid\hat{\left(T_W\right)^p}\mid{\bf \Phi}(t)\rangle=
\sum_{a,b}A^*_aA_b\overline{\left(\hat{\tau}^p\right)_{ab}}
\end{equation}

This should be contrasted with the moments of the distribution
function $p(t)$:
\begin{equation}\label{mdl2}
\int dt t^p p(t)=(i/\hbar)^p
\sum_{a,b}A^*_aA_b\overline{\left(\frac{\partial^p\hat{S}^{\dagger}(E)}
{\partial E^p}\hat{S}(E)\right)_{ab}}
\end{equation}

We see that only the first moment of this distribution coincides
with that given by
Eq.(\ref{mdl1}), all other being different. This particular
example shows certain ambiguity in definition of delay time
statistics. In the present paper we concentrate on statistics of
Wigner-Smith time delays and related quantities: energy derivatives of
$S-$matrix eigenphases.
\subsection{S-matrix characteristics: poles, eigenphases and delay times}
The expression (\ref{smat2}) forms the basis
for extracting the statistics of scattering poles (resonances), which
are merely the complex eigenvalues of the non-Hermitian effective
Hamiltonian ${\cal
H}_{ef}=\hat{H}_{in}-i\pi\hat{W}\hat{W}^{\dagger}$.
At the same time, the expression (\ref{smat1}) turns out to be a
more convenient starting point for studying statistics of scattering
phase shifts and delay times. Indeed, it is evident that scattering
phase shifts $\theta_a; a=1,2,...,M$ [defined via the S-matrix
eigenvalues $\exp{i\theta_a}$] are determined by the
eigenvalues $z_a(E,X)$ of the matrix $\pi\hat{K}$ in view of the
relation: $\theta_a=-2\arctan{z_a}$. Here we indicated explicitly the
dependence of the eigenvalues $z_a$ on the energy $E$ and an external
parameter $X$ originating from the corresponding dependence of
the Hamiltonian on the parameter: $\hat{H}_{in}= \hat{H}_{in}(X)$.
It is therefore convenient to characterize the statistics of phase shifts
via the spectral density:\begin{equation}\label{denz}
\rho_{E,X}(z)=\displaystyle{\frac{1}{M}
\sum_{a=1}^{M} \delta(z-z_a(E,X))}
\end{equation}
Actually, the relation $\theta_a=-2\arctan{z_a}$
determines the phase shifts modulo $2\pi$ only. It is easy to  
understand that
every time the energy $E$ coincides with one of the eigenvalues
$E_n(X),\quad n=1,2,...,N$ of the
Hermitian Hamiltonian $\hat{H}_{in}(X)$ one (and only one) of the  
scattering
phase shifts crosses the value $2\pi \times\mbox{integer} $.
Indeed, assuming that the eigenvalues $E_n$ are generically
not degenerate one can write the matrix elements of the matrix
$\hat{K}$ in the vicinity of $E=E_n$ as $\hat{K}(E\to E_n)=
\frac{\pi}{E-E_n}W^{*}_{na}W_{nb}$,
where $W_{nb}$ are matrix elements of the coupling matrix $\hat{W}$
in the basis of eigenstates $\mid n\rangle$ of the Hamiltonian  
matrix $\hat{H}_{in}$ and  
$\left(W^{\dagger}\right)_{an}=W^{*}_{na}$.
 We see immediately that $\hat{K}(E\to E_n)$
has only one eigenvalue divergent at $E\to E_n$ which is given by  
$z(E)=\frac{\pi}{E-E_n}\sum_{a}
\mid W_{na}\mid^2$, the corresponding eigenvector
being ${\bf v}=(W_{n1}^*,W_{n2}^*,...,W_{nM}^*)^T$. The phase shift
corresponding to the infinite value of $z(E)$ must be an integer
of $2\pi$. At the same time all other eigenvalues of $\hat{K}(E\to E_n)$
are exactly zero in that approximation with corresponding
eigenvectors belonging to the $M-1$ dimensional space orthogonal to  
${\bf v}$.
This fact just means that $M-1$ corresponding eigenvalues of the  
{\it exact} matrix $\hat{K}(E)$
stay finite in the vicinity of $E_n$.
Introducing the exact density of states for the {\it closed} chaotic
system: $\nu_X(E)=\frac{1}{N}\sum_{n}\delta\left(E-E_n(X)\right)$  
and fixing the
phase shift value at $E=-\infty$ to be zero, we conclude that:
\begin{equation}\label{totsh}
\sum_{a=1}^M\theta_a=2\pi
N\int_{-\infty}^E du \nu_X(u)-2\sum_a\arctan{z_a}\end{equation}
Here $\arctan{z_a}$ means the principal branch: $\mid  
\arctan{z_a}\mid <\pi/2$.
As function of energy $E$ this expression is continuous and  
monotonically increasing with $E$. The first term is proportional to  
the level staircase and
we can forget it modulo $2\pi$.

We will use the relation (\ref{totsh}) later on in order to determine
the correlations of the Wigner delay times
$\tau_w(E)=(\partial/\partial E)\frac{1}{M}\sum_a\theta_a\equiv
-(i/M)(\partial/\partial E)\ln{\mbox{Det}\hat{S}(E)}$, which , of course,
are positive.
For the latter quantity we also can find an independent
representation by noticing that:
\begin{equation}\label{wigtime1}
\ln{\mbox{Det}\hat{S}(E)}=\ln{\frac{\mbox{Det}(\hat{I}-i\hat{K})}
{\mbox{Det}(\hat{I}+i\hat{K})}}=\ln{\frac{\mbox{Det}
(\hat{I}-i\pi\frac{1}{E-\hat{H}_{in}}\hat{W}\hat{W}^{\dagger})}
{\mbox{Det}
(\hat{I}+i\pi\frac{1}{E-\hat{H}_{in}}\hat{W}\hat{W}^{\dagger})}}
\end{equation}
where we have made use of the identity\begin{equation}\label{identity}
\mbox{Det}(\hat{I}-\hat{U}\hat{V})=\mbox{Det}(\hat{I}-\hat{V}\hat{U})
\end{equation}
valid for arbitrary (also rectangular) matrices $\hat{U},\hat{V}$.
 From Eq.(\ref{wigtime1}) we immediately obtain the simple relation:
\begin{equation}\label{wigtime2}
\tau_w(E)=-(2/M)\mbox{Im Tr}
\left(E-\hat{H}_{in}+i\pi\hat{W}\hat{W}^{\dagger}\right)^{-1}=
\frac{2}{M}\sum_{n=1}^N\frac{\Gamma_n/2}{(E-E_n)^2+\Gamma_n^2/4}
\end{equation}
which, in particular, shows an intimate relation between the
statistics of Wigner time delay and that of $S$-matrix poles.

To study $S-$matrix characteristics within the framework of the
stochastic approach, one should specify the properties of the
amplitudes $W_{ia}$, which couple the internal chaotic motion to  
$M$ open channels. For the sake of simplicity one can restrict
the consideration to the case when the $S-$ matrix is diagonal  
after averaging: $\langle
S_{ab}\rangle=\delta_{ab}\langle S_{aa}\rangle$. Such a choice is  
related with the mentioned absence
of direct coupling between the channels\cite{VWZ,Lew} and can be
ensured if one considers {\it fixed}
amplitudes $W_{ia}$ satisfying the so-called
orthogonality relations \cite{VWZ}:
\begin{equation}\label{ortho}
\sum_{i}W_{ia}^*W_{ib}=\frac{1}{\pi}\gamma_a\delta_{ab}
\end{equation}
An alternative way to ensure the diagonality of the average
$S-$matrix is to consider the amplitudes $W_{ia}$ to be independent
Gaussian random variables\cite{Sok,Haake,Lehmann,Lehm2}:
\begin{equation}\label{orrand}
\langle W_{ia}\rangle=0;\quad \langle
W_{ia}^*W_{jb}\rangle=\frac{\gamma_a}{N} \delta_{ab}\delta_{ij}
\end{equation}
One can show, following the papers\cite{Lehmann,Lehm2} that
both choices lead to the same results as long as the number $M$
of open channels is negligible in comparison with the number of
bound states $N$. Since this case is the only considered in the
present paper, we restrict ourselves to the condition
Eq.(\ref{ortho}) henceforth.
Provided the orthogonality condition (\ref{ortho})
is fulfilled, one can show that the diagonal elements of the  
$S-$matrix are
given by the following expression:\begin{equation}\label{sav}
\langle S_{aa}\rangle=\frac{1-\gamma_{a}g(E)}{1+\gamma_{a}g(E)}
\end{equation}
where $g(E)=iE/2+(1-E^2/4)^{1/2}$. We do not give here
 the derivation of Eq.(\ref{sav}) referring the interested
reader to the paper \cite{VWZ} for more details. However, we mention
that for the one-channel case $M=1$  relation Eq.(\ref{sav}) follows
directly from the distribution of the phase shift $\theta$ to be found
in Sec.IV.
 The strength of
coupling to continua is convenient to be characterized via the
transmission coefficients $T_{a}=1-|\langle S_{aa}\rangle|^2$ that are
given for the present case by the following
expression:
\begin{equation}\label{trans}
T_{a}^{-1}=\frac{1}{2}\left[1+\frac{1}{2
\mbox{Re}\, g(E)}(\gamma_{a}+\gamma_a^{-1})\right]
\end{equation}
The quantity $T_{a}$ measures the part of the flux in channel $a$
that spends substantial part of the time in the interaction
region\cite{VWZ,Lew}. This interpretation follows
from the fact that the energy averaged $S-$ matrix (equal to the  
ensemble average $\langle S\rangle$ by ergodicity
requirement) describes
a short time scattering ("direct response", see the
Introduction). Let us also note that frequently we find it to be more
convenient to use the  
parameters\begin{equation}\label{defg}g_a=\frac{2}{T_a}-1
\end{equation}
rather than the "transmission coefficients" $T_a$ .

Naively, one could suspect that the larger is the parameter
$\gamma_a$, the larger is the part of the flux effectively
penetrating the chaotic region.
However, we see that this is not the case: both limits $\gamma_{a}\to 0$
and $\gamma_{a}\to \infty$ equally correspond to the weak effective
coupling regime $T_{a}\ll 1$ whereas the strongest coupling (at
fixed energy $E$ ) corresponds to the value $\gamma_{a}=1$.
  The maximal possible coupling corresponding to the upper bound  
$T_a=1$ is
achieved in the present model for an energy interval in the vicinity
of the center $E=0$.

This feature is a purely quantum effect and is not surprising any  
longer, if one remembers the simplest
textbook example of a quantum particle scattered on a one-dimensional
" potential step": $V(x)=0$ for $x<0$ and $V(x)=V$ for $x>0$.
The transmission coefficient for such a problem is given by:
$T=1-\mid S\mid^2=4kK/(K+k)^2$, where $k=(2mE)^{1/2}/\hbar;\quad
K=[2m(E-V)]^{1/2}/\hbar$ and $E>0$ stands for the energy of  
incoming particles.
Similarly to the case above the transmission is very small both for  
the system "almost closed" classically: $E-V\ll E$ when
 $\gamma_0\equiv (K/k)^{1/2}\ll 1$ as well as for systems "very  
open classically" $V<0,\mid V\mid\gg E $ when $\gamma_0\gg 1$, the  
"perfect" transmission $T=1$ being possible
for the only case $V=0$ when $\gamma_0=1$.
This simple example is of course just to remind us of the effect  
known in radiophysics as
"impedance mismatch": the wave is always reflected back at the  
point of contact of two different waveguides, unless special  
boundary
conditions are ensured.\section{Scattering poles in complex plane:
 distribution of resonance widths.}
\subsection{Resonances as eigenvalues of a non-Hermitian Hamiltonian.}

We are interested in determining the average two-dimensional density
\begin{equation}\label{dendef}
\rho(E,Y)=\left\langle\frac{1}{N}\sum_{j=1}^N\delta^{(2)}(
{\cal E}-{\cal E}_j)\right\rangle\equiv\left\langle\frac{1}{N}\sum_{j=1}^
N\delta(E-E_j)\delta(Y-Y_j)\right\rangle
\end{equation}
 of complex eigenvalues ${\cal E}_j=E_j+iY_j,\quad j=1,2,...,N$
 of a non-Hermitian effective
Hamiltonian ${\cal H}_{ef}=\hat{H}_{in}-i\hat{\Gamma}$,
 see Eq.(\ref{smat2}). According to the general discussion presented
in the Introduction we use a $N\times N$ random matrix $\hat{H}_{in}$
from the Gaussian Unitary Ensemble
to model the Hamiltonian of a closed chaotic system with broken
time-reversal symmetry. The entries $\Gamma_{ij}$ of the matrix
$\hat{\Gamma}$ are expressed in terms of the channel amplitudes  
$W_{ia},\quad
a=1,2,...,M $ as $\Gamma_{ij}=\pi\sum_a W_{ia}W_{ja}^*$.
Before presenting a general theory it is instructive to consider the
important limiting case of an extremely weak coupling when we expect
that the resonances are so narrow that their widths are much smaller
in comparison with the mean separation between the unperturbed
levels. Under these conditions a simple first order perturbation
theory is adequate. Using the notation $\mid n\rangle$ for the
eigenvector of $\hat{H}_{in}$ corresponding to the real eigenvalue
$E_n$ of the closed system: $\hat{H}_{in}\mid n\rangle=E_n\mid
n\rangle$, one can estimate the shift of the eigenvalues into the
complex plane as\begin{equation}\label{pertwid}-Y_n=\langle  
n\mid\hat{\Gamma}\mid n\rangle=
\sum_{k,l=1}^N\langle\alpha_l\mid\hat{\Gamma}\mid\alpha_k\rangle\langle
n\mid \alpha_l\rangle\langle\alpha_k\mid n\rangle
\end{equation}
where $\mid\alpha_k\rangle;\quad k=1,2,...,N$ is an arbitrary chosen
basis of orthonormal vectors.
The matrix $\hat{\Gamma}$ can be easily diagonalized and shown to have
exactly $M$ non-zero eigenvalues $\gamma_a=\pi\sum_iW_{ia}^*W_{ia}$.
Choosing $\mid\alpha_k\rangle$ to be eigenbasis of the matrix
$\hat{\Gamma}$ we therefore have: $Y_n=-\sum_{a=1}^M
\gamma_a \langle n\mid \alpha_a\rangle\langle\alpha_a\mid n\rangle$.
Now we use the well-known fact that different
 components $\langle n\mid \alpha_a\rangle=u_a+iv_a$ of  
eigenvectors of the GUE matrices in an {\it arbitrary} basis can be  
treated as
independent complex variables, their real and imaginary partsbeing  
independently distributed according to the Gaussian law
with the variances $\overline{u_a^2}=\overline{v_a^2}=\frac{1}{2N}$.
This fact allows one to calculate the distribution of $Y_n$ easily.
Considering for simplicity the case of all equivalent channels:
$\gamma_a=\gamma$ for any channel, we get:
\begin{eqnarray}\label{disper}
{\cal P}(Y)=\overline{\delta(Y-Y_n)}=\int_{-\infty}^{\infty}e^{ikY}
\frac{dk}{2\pi}\left[\int dudv\frac{N}{\pi}
\exp{-(N-i\gamma k)(u^2+v^2)}\right]^M=& &
\\  
\nonumber\int_{-\infty}^{\infty}\frac{dk}{2\pi}e^{ikY}\frac{1}{(1-ik\gamma/N)^M}=
\frac{N^M}{\gamma^M\Gamma(M)}\mid  
Y\mid^{M-1}\exp{-\left[\frac{N\mid
Y\mid}{\gamma}\right]}& &.
\end{eqnarray}for $Y<0$ and zero otherwise.

We arrive at the well-known result: the widths of resonances for a  
slightly
open chaotic system is given by the so-called
$\chi^2$ distribution \cite{RMT}. Actually, the same form is
applicable also for M-channel open
systems with preserved time-reversal symmetry,
provided one changes $M\to M/2$. The latter distribution for $M=1$  
is known as the Porter-Thomas distibution.

When the coupling to continua increases some resonances start to
overlap and the simple perturbation theory loses its validity.

 A general method for calculating the eigenvalue density
for non-Hermitian random matrices was
proposed by Sommers et al.\cite{Som}. These authors suggested to
recover the density $\rho(E,Y)$ from the "potential function"
\begin{eqnarray}\label{potdef}
 -\Phi(E,Y)=\frac{1}{N}\left\langle\ln{\mbox{Det}
\left({\cal E}-{\cal H}_{ef}\right)^{\dagger}\left({\cal E}-
{\cal H}_{ef}\right)}\right\rangle=\\ \label{potdef1}
\left\langle\frac{1}{N}\sum_j\ln{\mid{\cal E}-{\cal
E}_j\mid^2}\right\rangle \equiv\left\langle\frac{1}{N}\sum_j\ln
{\left[(E-E_j)^2+(Y-Y_j)^2\right]}\right\rangle
\end{eqnarray}

To show that this is indeed possible it is convenient to regularize
the logarithm in eq.(\ref{potdef1}) first and consider \cite{Haake}:
\begin{equation}\label{reg1}
 -\Phi(E,Y)_{\epsilon}=\left\langle\frac{1}{N}\sum_j\ln
{\left[(E-E_j)^2+(Y-Y_j)^2+\epsilon^2\right]}\right\rangle
\end{equation}
Then one notices that the function:
\begin{equation}\label{regden}
\rho(E,Y)_{\epsilon}=-\frac{1}{4\pi}(\partial_E^2+\partial_Y^2)
 \Phi(E,Y)_{\epsilon}=\frac{1}{N\pi}\sum_j
\frac{\epsilon^2}{\left[(E-E_j)^2+(Y-Y_j)^2+\epsilon^2\right]^2}
\end{equation}
produces the required two-dimensional density, Eq.(\ref{dendef}),
when $\epsilon\to 0$. Indeed, for an arbitrary continuous function
$f(E,Y)$ one has:
\begin{eqnarray}\label{delta}
\lim_{\epsilon\to 0}\frac{1}{\pi}
\int_{-\infty}^{\infty}\int_{-\infty}^{\infty}dEdY
f(E,Y)\frac{\epsilon^2}{\left[(E-E_j)^2+(Y-Y_j)^2+\epsilon^2\right]^2}=\\
\lim_{\epsilon\to 0}
\frac{1}{\pi}\int_{-\infty}^{\infty}\int_{-\infty}^{\infty}dudv
f(\epsilon u+E_j,\epsilon v+Y_j)\frac{1}{\left[u^2+v^2+1\right]^2}
= f(E_j,Y_j)\nonumber
\end{eqnarray}
in agreement with the $\delta$- functional property.

In fact, the expressions Eqs.(\ref{reg1},\ref{regden}) show that
$\rho(E,Y)$ can be considered as a two-dimensional density of
fictitious pointlike "electric charges" $1/N$, the function $
\Phi(E,Y)$ playing the role of the electrostatic potential
 for such a system and eq.(\ref{regden}) being the corresponding
Poisson equation \cite{Haake,Som}.

Actually, it turns out to be more  convenient to use
 a slightly different regularization, as it has been actually done in
\cite{Haake,Som}
\begin{equation}\label{logpot}
 \Phi(E,Y;\kappa)=-\frac{1}{N}\left\langle
\ln{\mbox{Det}\left[\left({\cal E}-{\cal H}_{ef}\right)^{\dagger}
\left({\cal E}-{\cal H}_{ef}\right)
+\kappa^2\hat{I}_N\right]}\right\rangle
\end{equation}
performing the limiting procedure $\kappa\to 0$ at the very end.
For the so-called {\it normal} matrices (whose Hermitian conjugate
${\cal H}^{\dagger}$ commutes with ${\cal H}$) the regularized
potential Eq.(\ref{logpot}) coincides with that defined in the  
Eq.(\ref{reg1}). It is a less trivial fact that one recovers
the two-dimensional density of complex eigenvalues from the  
potential $\Phi(E,Y;\kappa)$ also in a general case of nonnormal
matrices ${\cal H}$. We show in the Appendix \ref{nonnormal} that  
it is indeed the case: the density obtained from
Eq.(\ref{logpot}) by Poisson's equation is positive and goes to a
sum of $\delta-$functions with weight $1/N$ near the eigenvalues of
${\cal H}_{ef}$.

The main technical problem is to perform the averaging of the  
logarithmic potential $\Phi(E,Y;\kappa)$ over the random matrices
from the corresponding Gaussian ensemble, eq.(\ref{GUE}). To  
perform such an averaging
the authors of the papers \cite{Haake,Som}
 employed the famous, but somewhat problematic "replica trick".  
This procedure amounts to averaging the
corresponding determinant raised to an arbitrary positive integer  
power $n$,
the average logarithm being recovered as a result of the limiting  
procedure $n\to 0$. In general, however, the analytical
continuation $n\to 0$ is not unique. In particular, it is known that
the replica trick fails to reproduce correctly
the correlation function of densities of {\it real}
eigenvalues  of large  Hermitian
matrices at two points $E\pm\Omega$ of the spectrum \cite{VZ}.
Rather, it succeeds in giving the correct behavior of that correlation
function at the scale $\Omega$ large in comparison with
the typical separation between neighboring eigenvalues, known
as the mean level spacing $\Delta$.
As is shown below, formally the calculation of the mean eigenvalue
density in the {\it complex plane} is very similar to the calculation
of a correlation function of eigenvalue densities on the {\it real
axis}, with the role of $\Omega$ played by the variable $iY$
measuring the distance from the real axis.
 We immediately conclude, that the replica trick must fail
when we are interested in eigenvalues situated sufficiently close to the
 real axis: $Y\sim \Delta$.

 To this end it is necessary to mention, that the
non-Hermiticity of the matrices considered in the papers
\cite{Haake,Som} was, in a sense, quite {\it strong}: the  
probability for an
eigenvalue to be situated at the distance $Y\sim \Delta\propto 1/N$
vanished in the limit $N\to \infty$. Under those conditions it is not
surprising that the replica trick  succeeded in producing the correct
result, the fact verified both by independent methods: by a variant  
of the supersymmetry method \cite{Lehm2} and by direct numerical  
computations\cite{Haake,Som}.

The situation described above is drastically different from
that we expect to happen at the model under the present
consideration. Indeed, it is known, that when the
number $M$ of open channels is small in comparison with the number
$N$ of relevant resonances, the majority $N-M$ of resonances are rather
"narrow" and the corresponding poles are situated close to the real axis
\cite{Sok,Haake,Dittes,ISS}. Under such a situation one has to
discard the replica trick and to seek for a more reliable procedure.
Fortunately, the authors of the mentioned paper \cite{VZ}
showed how to calculate the two-point correlation function
for {\it real} eigenvalues correctly by exploiting the method
pioneered by Efetov \cite{Efrev} in the theory of disordered solids.
A pedagogical introduction to the method can be found in \cite{my}.
In the present paper we adjust this procedure for finding the
density of scattering poles in the complex plane
for the few-channel case \cite{FSres} (for the many-channel case  
$M\propto N$ this density
has already been calculated by N.Lehmann et al.\cite{Lehm2}).

Instead of working directly with the regularized potential
$\Phi(E,Y,\kappa)$, see Eq.(\ref{logpot}),  in terms of which
the two-dimensional density $\rho(E,Y)$ is expressed  
as:\begin{equation}\label{denstart}
\rho(E,Y)=-\lim_{\kappa\to 0}\frac{1}{4\pi}(\partial_E^2+\partial_Y^2)
 \Phi(E,Y,\kappa)
\end{equation}
we prefer to consider the related function:
\begin{equation} \label{aupot}
 \Phi(E,-i\Omega;\kappa)=-\frac{1}{N}\left\langle
\ln{\mbox{Det}\left[
\left(E+\Omega-\hat{H}_{in}+i\Gamma\right)
\left(E-\Omega-\hat{H}_{in}-i\hat{\Gamma}\right)
+\kappa^2\hat{I}_N\right]}\right\rangle
\end{equation}
It is evident, that the potential
$\Phi(E,Y,\kappa)$ can be obtained from the function $  
\Phi(E,-i\Omega;\kappa)$ by analytical continuation $-i\Omega\to Y$.
As long as $\kappa$ is finite, there is a region extending from
positive to negative $Y$ where the function $\Phi$ is analytic in  
$Y=-i\Omega$.
Actually, this continuation is more convenient to perform directly  
on the level of densities, i.e. first to calculate the auxilliary  
function
\cite{fnote2}:
\begin{equation}\label{auden}
\rho(E,-i\Omega)_{\kappa}=-\frac{1}{4\pi}(\partial_E^2-\partial_{\Omega}^2)
 \Phi(E,-i\Omega,\kappa)
\end{equation}
and to restore the true two-dimensional density $\rho(E,Y)$ letting
 $-i\Omega\to Y$ first and then $\kappa\to 0$:
\begin{equation}\label{restore}
\rho(E,Y)=\lim_{\kappa\to 0}\rho(E,-i\Omega=Y)_{\kappa}.
\end{equation}

To this end let us consider the generating  
function:\begin{equation}\label{genfun}
{\cal Z}(E,\Omega;E_b,\Omega_b;\kappa)=
\frac{\mbox{Det}\left[\left(E+\Omega-\hat{H}_{in}+i\hat{\Gamma}\right)
\left(E-\Omega-\hat{H}_{in}-i\hat{\Gamma}\right)
+\kappa^2\hat{I}_N\right]}
{\mbox{Det}\left[\left(E_b+\Omega_b-\hat{H}_{in}+i\hat{\Gamma}\right)
\left(E_b-\Omega_b-\hat{H}_{in}-i\hat{\Gamma}\right)
+\kappa^2\hat{I}_N\right]}
\end{equation}
in terms of which the function $\rho(E,-i\Omega)_{\kappa}$ is
expressed as follows:
\begin{equation}\label{auden1}
\rho(E,-i\Omega)_{\kappa}=
\frac{1}{4\pi}\left[\left(\frac{\partial}{\partial E}
\lim_{(E_b,\Omega_b)\to (E,\Omega)}\frac{\partial}{\partial E}\right)-
\left(\frac{\partial}{\partial \Omega}
\lim_{(E_b,\Omega_b)\to(E,\Omega)}\frac{\partial}{\partial  
\Omega}\right)\right]
\langle{\cal Z}(E,\Omega;E_b,\Omega_b;\kappa)\rangle,
\end{equation}

The determinant in the denominator of expression Eq.(\ref{genfun})  
can be represented in a form of
a conventional Gaussian integral
 over the components of a complex $2N-$component
vector ${\bf S}=\left(\begin{array}{c}{\bf S}_1\\{\bf  
S}_2\end{array}\right);\quad{\bf  
S}_{p=1,2}=\left(S_1^{(p)},...,S_N^{(p)}\right)^{T},\quad
[d{\bf S}]=\displaystyle{
\prod_{j=1}^N \frac{d \mbox{Re} S_j d\mbox{Im}S_j}{\pi}}$:

\begin{eqnarray}\label{Gaubos} \nonumber
\mbox{Det}^{-1}\left[\left(E_b-\Omega_b-\hat{H}_{in}-i\hat{\Gamma}\right)
\left(E_b+\Omega_b-\hat{H}_{in}+i\hat{\Gamma}\right)
+\kappa^2\hat{I}\right]=\\ \nonumber
\int [d{\bf S}_1][{d\bf S}_2]\exp{\left\{-
{\bf S}^{\dagger}\left[\begin{array}{cc}
-i\left(E_b+\Omega_b-\hat{H}_{in}+i\hat{\Gamma}\right)&
 -\kappa \hat{I}\\ \kappa \hat{I} &
i\left(E_b-\Omega_b-\hat{H}_{in}-i\hat{\Gamma}\right)
\end{array}\right]{\bf S}\right\}}\equiv \\
\int [d{\bf S}_1][d{\bf S}_2]\exp{\left\{
\kappa({\bf S}_1^{\dagger}{\bf S}_2
-{\bf S}_2^{\dagger}{\bf S}_1)+
iE_b({\bf S}_1^{\dagger}{\bf S}_1-
{\bf S}_2^{\dagger}{\bf S}_2)
+i({\bf S}_1^{\dagger},{\bf S}_2^{\dagger})
\left(\begin{array}{cc}-\hat{H}_{in} &  
0\\0&\hat{H}_{in}\end{array}\right)
\left(\begin{array}{c}{\bf S}_1\\
 {\bf S}_2\end{array}\right)\right\}}\times \\ \nonumber
\exp{\left\{i\Omega_b({\bf S}_1^{\dagger}{\bf S}_1+
{\bf S}_2^{\dagger}{\bf S}_2)-
({\bf S}_1^{\dagger},{\bf S}_2^{\dagger})
\left(\begin{array}{cc}\hat{\Gamma} & 0\\0&\hat{\Gamma}\end{array}\right)
\left(\begin{array}{c}{\bf S}_1\\
 {\bf S}_2\end{array}\right)\right\}}
\end{eqnarray}

At this point it is worth  mentioning that all eigenvalues of  
${\cal H}_{ef}=\hat{H}_{in}-i\hat{\Gamma}$
 (scattering poles) must be situated in the lower half of the complex
plane Im${\cal E}\le 0$. Formally it is ensured by eigenvalues of  
the matrix $\hat{\Gamma}$ being real non-negative.
We see that it is due to this fact that the Gaussian integral
above is convergent \cite{fnote3}
(for $\Omega$ real other terms in the exponent
are purely imaginary and do not spoil the convergency; at the end we
may continue analytically).

The following comment is appropriate here.
In principle, one can deal directly with the potential
$\Phi(E,Y,\kappa)$, Eq.(\ref{logpot}) and
 succeed in finding the convergent Gaussian representation for the
generating function everywhere in the complex plane $E+iY$(see  
\cite{FSres,Khor}).
However, the evaluation of the averaged generating
function and subsequent restoration of the eigenvalue density    
turns out to be quite a dounting job. This is the reason why we
decided to deal in  our particular case with a less general, but
more tractable representation Eq.(\ref{Gaubos}) allowing to evaluate
the generating function for two  real parameters $E,\Omega$
and then to continue analytically $\Omega\to iY$ as explained above.

Returning to our problem we represent the determinant in the
numerator of the generating function Eq.(\ref{genfun})
in the form of a Gaussian integral over a
$2N-$ dimensional vector ${\bf \chi}=
\left(\begin{array}{c}{\bf\chi}_1\\
{\bf\chi}_2\end{array} \right)$ whose elements
$\chi_j^{(p)},\quad j=1,2,...,N; p=1,2$ are {\it
anticommuting} (Grassmannian) variables (see reviews \cite{Efrev,my}
for more details):
\begin{eqnarray}\label{Gaugr}\nonumber
\mbox{Det}^{-1}\left[\left(E-\Omega-\hat{H}_{in}-i\hat{\Gamma}\right)
\left(E+\Omega-\hat{H}_{in}+i\hat{\Gamma}\right)
+\kappa^2\hat{I}\right]=\\ (-1)^{N}\mbox{Det}
\left[\begin{array}{cc}
-i\left(E+\Omega-\hat{H}_{in}+i\hat{\Gamma}\right)&
\kappa \hat{I}\\ \kappa
\hat{I}&-i\left(E-\Omega-\hat{H}_{in}-i\hat{\Gamma}\right)\end{array}\right]=
\end{eqnarray}
$$
(-1)^{N}\int [d\chi_1][d\chi_2]\exp{\left\{
-\kappa(\chi_1^{\dagger}\chi_2+\chi_2^{\dagger}\chi_1)+
i\chi_2^{\dagger}\left(E-\Omega-\hat{H}_{in}-i\hat{\Gamma}\right)
\chi_2+i\chi_1^{\dagger}\left(E+\Omega-\hat{H}_{in}+i\hat{\Gamma}\right)
 \chi_1\right\}}
$$
where $[d\chi]=\prod_{k=1}^Nd\chi^{*}_kd\chi_k$.
In contrast to the discussion above, the
 integration over Grassmann variables is always well defined
and one does not encounter the convergency problem.

Obviously, our generating function is the product of two
Gaussian integrals defined in Eqs.(\ref{Gaubos},\ref{Gaugr}).
It is convenient to introduce the notion of a supervector
\begin{equation}\label{suvec}
{\bf \Psi}=\left(\begin{array}{c}{\bf \Psi}_1\\{\bf   
\Psi}_2\end{array}\right)
\quad \mbox{where} \quad {\bf \Psi}_p=\left(\begin{array}{c}{\bf  
S}_p\\\chi_p\end{array}\right);\quad p=1,2.
\end{equation}
 Then one can write the generating
function in the following "supersymmetric" form:
\begin{equation}\label{Gaususy}
{\cal Z}(E,\Omega;E_b,\Omega_b;\kappa)=(-1)^N\int[d{\bf\Psi}]\exp{-({\cal
S}_{\delta}[{\bf \Psi}]+{\cal S}_{Ef}[{\bf \Psi}])}\end{equation}
where
\begin{equation}
\label{Efac}
{\cal S}_{Ef}[{\bf \Psi}]=-i\Omega{\bf \Psi}^{\dagger}
\hat{\Lambda}\hat{L}{\bf \Psi}-iE
{\bf \Psi}^{\dagger}\hat{L}{\bf \Psi}+i{\bf \Psi}^{\dagger}
\left(\hat{H}_{in}\otimes\hat{L}
\right){\bf\Psi}
\end{equation}
and
\begin{equation}\label{Efac1} {\cal  
S}_{\delta}[{\bf\Psi}]=\kappa{\bf\Psi}^{\dagger}\hat{\Sigma}{\bf\Psi}+
i(\Omega-\Omega_b){\bf\Psi}^{\dagger}\hat{K}_b{\bf\Psi}+
i(E-E_b){\bf\Psi}^{\dagger}\hat{L}\hat{K}_b{\bf\Psi}+
{\bf\Psi}^{\dagger}
\left(\hat{\Gamma}\otimes\hat{\Lambda}\hat{L}
\right){\bf\Psi}
\end{equation}

Before presenting the explicit expressions for the supermatrices
$\hat{L},\hat{\Lambda},\hat{K}_b$ and $\hat{\Sigma}$ we would like  
to make
a notational convention on arranging elements of supermatrices.
All these (and subsequently appearing) supermatrices
are assumed to act in the space of supervectors ${\bf\Psi}$ whose  
element arrangement
is defined in Eq.(\ref{suvec}). Correspondingly, we subdivide
each $4\times 4$ supermatrix $\hat{Q}$ into four blocks
$\hat{Q}=\left(\begin{array}{cc}\hat{Q}_{11}&\hat{Q}_{12}\\
\hat{Q}_{21}&\hat{Q}_{22}\end{array}\right)$ in such a way that
${\bf\Psi}^{\dagger}\hat{Q}{\bf\Psi}=\sum_{m,n=1}^{2}
{\bf\Psi}_m^{\dagger}\hat{Q}_{mn}
{\bf\Psi}_n$. each of these $\hat{Q}_{mn}$ blocks is in turn a $2\times2$
supermatrix $\hat{Q}_{mn}=\left(\begin{array}{cc}Q^{(mn)}_{bb}
&Q_{bf}^{(mn)}\\
Q_{fb}^{(mn)}&Q_{ff}^{(mn)}\end{array}\right)$
such that ${\bf\Psi}_m^{\dagger}\hat{Q}_{mn}{\bf\Psi}_n=
{\bf S}^{\dagger}_mQ^{(mn)}_{bb}{\bf S}_n+
{\bf S}^{\dagger}_mQ^{(mn)}_{bf}\chi_n+
\chi_m^{\dagger}Q^{(mn)}_{fb}{\bf S}_n+
\chi_m^{\dagger}Q^{(mn)}_{ff}\chi_n$.
The indices $b,f$ remind us of "bosonic"/ "fermionic"
nature of the commuting/ Grassmannian components of
supervectors, respectively.

 It is necessary to note that in the present paper we use the same  
convention
 for Hermitian conjugation of $2\times 2$ supermatrices as in the  
paper \cite{my}
:$\left(\begin{array}{cc}Q_{bb}
&Q_{bf}\\
Q_{fb}&Q_{ff}\end{array}\right)^{\dagger}=\left(\begin{array}{cc}Q_{bb}^{*}
&Q_{fb}^{*}\\
-Q_{bf}^{*}&Q_{ff}^{*}\end{array}\right)$. This is different
from the convention used in \cite{Efrev,Zirn} and results
in some subsequent differences in parametrizations.

With these conventions the $4\times4$ supermatrices appearing in  
Eq.(\ref{Efac},\ref{Efac1}) are given by the following expressions:
\begin{equation}\label{matdiag}
\hat{L}=\mbox{diag}(1,1,-1,1);\quad
\hat{\Lambda}=\mbox{diag}(1,1,-1,-1); \quad
\hat{K_b}=\mbox{diag}(1,0,1,0);\end{equation}
and $\hat{\Sigma}=\left(\begin{array}{cc}0&I_2
\\ \hat{k}&0\end{array}\right)$, where $I_2=\mbox{diag}(1,1);\quad
\hat{k}=\mbox{diag}(-1,1)$.

\subsection{Ensemble-averaged generating function.}
A quick inspection of Eqs.(\ref{Gaususy},\ref{Efac},\ref{matdiag})
makes it clear that the superintegral
in hand is very similar to that emerging in the problem of
calculation of the pair correlation function of eigenvalues of
Hermitian random matrices at points $E\pm\Omega$
, see e.g. \cite{my}. In fact, the two expressions {\it coincide}
 if one neglects the exponent  ${\cal S}_{\delta}$. The neglected  
exponent
 ${\cal S}_{\delta}$ does not contain random variables and can not  
prevent us from using successfully  the main steps of Efetov's  
standard procedure when evaluating
the average value of the generating function. Below we give a short
description of the main steps of the method;
 all further details can be found in
the review \cite{my} and references therein.

\begin{itemize}
\item{\bf Ensemble averaging.}
One can easily perform the averaging over the Gaussian-distributed  
matrix elements
of $\hat{H}_{in}$ by exploiting the
identity
\begin{equation}\label{ident}
\left\langle\exp{\pm  
i\sum_{ij}(\hat{H}_{in})_{ij}\hat{U}_{ij}}\right\rangle=
\exp{-\frac{1}{2N}\sum_{ij}\hat{U}_{ji}\hat{U}_{ij}}
\end{equation}

In order to write down the result of the ensemble
averaging in a convenient form it is useful to introduce the
supermatrix $\hat{A}$ with elements
\begin{equation}\label{matA}
A^{(mn)}_{pq}=\left(\hat{L}^{1/2}\right)^{(mm)}_{pp}
\sum_{i=1}^{N}(\Psi_i)^p_m\left(\Psi_i^{\dagger}\right)^q_n
\left(\hat{L}^{1/2}\right)^{(nn)}_{qq}
\end{equation}
where indices $p$ and $q$ are equal to $b$ or $f$ and we assumed
the convention: $\Psi^b_i\equiv S_i;\quad \Psi^f_i\equiv \chi_i$.
Now the ensemble-averaged value of the corresponding exponent
in eq.(\ref{Gaususy}) can be written as:
\begin{equation}\label{expave}
\left\langle\exp{\left[-i{\bf\Psi}^{\dagger}\left(\hat{H}_{in}\otimes\hat{L}
\right){\bf\Psi}\right]}\right\rangle=\exp{-\frac{1}{2N}\mbox{Str}\hat{A}^2}
\end{equation}
 where the symbol Str stands
for the graded trace $\mbox{Str}\hat{Q}=\mbox{Tr}\hat{Q}_{bb}
-\mbox{Tr}\hat{Q}_{ff}$.
It is also useful to notice that $\mbox{Str}\hat{A}\hat{Q}=
{\bf\Psi}^{\dagger}\hat{L}^{1/2}\hat{Q}\hat{L}^{1/2}{\bf\Psi}$ for  
an arbitary
supermatrix $\hat{Q}$. In particular,
$\mbox{Str}\hat{A}\hat{\Lambda}=
{\bf\Psi}^{\dagger}\hat{\Lambda}\hat{L}{\bf\Psi}$.

\item{\bf Hubbard-Stratonovich transformation.}

As a result of ensemble averaging the superintegral representing
the generating function ceased to be a Gaussian one. The further progress
is based on the following identity:
\begin{eqnarray}\label{HS}
\exp{\left[-\frac{1}{2N}\mbox{Str}\hat{A}^2+i\Omega\mbox{Str}\hat{A}
\hat{\Lambda}\right]}=\\
 \nonumber\int[d\hat{R}]\exp{\left\{-\frac{N}{2}\mbox{Str}
\hat{R}^2+i\mbox{Str}\hat{R}\hat{A}+N\Omega
\mbox{Str}\hat{R}\hat{\Lambda}\right\}}
\end{eqnarray}
known as the Hubbard-Stratonovich (HS) identity.

 Now we can substitute
this relation back into the averaged generating function, to change the
order of integrations over the supervector $\Psi$ and the supermatrices
$\hat{R}$, and to calculate the corresponding (Gaussian) integral
over $\Psi$ exactly using the identity:
\begin{equation}\label{Gausup}
\int[d\Psi]\exp{(-\Psi^{\dagger}\hat{F}\Psi)}=\mbox{Sdet}^{-1}\hat{F},
\end{equation}
where the notation Sdet stands for the graded determinant:
Sdet$\hat{Q}=\exp{\mbox{Str}\ln{Q}}$. It turns out, however, that  
in order to
have {\it both} $\int[d\Psi]...$ and
 $\int [d\hat{R}]...$ convergent, one has to parametrize
the set of supermatrices $\hat{R}$ in the following non-trivial
fashion suggested in \cite{Efrev,VZ} ( see detailed discussion in
\cite{my}):\begin{equation}\label{parR}
\hat{R}=\hat{T}^{-1}\hat{P}\hat{T};\quad
\hat{P}=\left(\begin{array}{cc} \hat{P}_1-i\delta\hat{I}_2&0\\
0&\hat{P}_2+i\delta\hat{I}_2\end{array}\right);\quad
\hat{P}_m=\left(\begin{array}{cc} p_m&\eta_m^*\\ \eta_m&
iq_m\end{array} \right)
\end{equation}
where $p_m,q_m,\,m=1,2$ are real commuting variables, $\eta,\eta^{*}$
are Grassmannians and the supermatrices $\hat{T}$ belong to the graded
coset space U(1,1/2)/U(1/1)$\times$ U(1/1), and $\delta$
 is positive infinitesimal.

\item {\bf Saddle-point calculation.}

Performing the program specified above one gets the following
representation for the average generating function:
\begin{equation}\label{genfunav}
\langle{\cal Z}(E,\Omega;E_b,\Omega_b;\kappa)\rangle=
\int[d\hat{R}]\exp{\{-\frac{N}{2}\mbox{Str}\hat{R}^2
+N\Omega\mbox{Str}\hat{R}\hat{\Lambda}-\mbox{Str}\ln{\hat{G}}\}}
\end{equation}
where\begin{equation}\label{G}
\hat{G}=\hat{G}_1\otimes  
\hat{I}_N-i\hat{\Gamma}\otimes\hat{\Lambda};\quad \hat{G}_1=
-i\kappa\hat{\Sigma}_L-(E_b-E)\hat{K}_b+
(\Omega-\Omega_b)\hat{K}_b\hat{L}-E\hat{I}_4-\hat{R}\end{equation}
and $\hat{\Sigma}_L=\hat{L}^{-1/2}\hat{\Sigma}\hat{L}^{-1/2}$.
Now one can write:
$$
\mbox{Str}\ln{\hat{G}}=N\mbox{Str}\ln{\hat{G}_1}+\mbox{Str}
\ln{\left[\hat{I}_N-i\hat{\Gamma}\otimes(\Lambda\hat{G}^{-1}_1)\right]}
$$
The second term in this expression can be rewritten  as:
\begin{equation}\label{chanexp}
\mbox{Str}
\ln{\left[\hat{I}_N-i\hat{\Gamma}\otimes(\Lambda\hat{G}^{-1}_1)\right]}=
\sum_{a=1}^M\mbox{Str}\ln{\{\hat{I}_4-i\gamma_a(\Lambda\hat{G}^{-1}_1)\}}
\end{equation}
which easily can be veryfied by expanding the logarithm into
the series, exploiting the orthogonality condition,  
Eq.(\ref{ortho}), in each term of that
expansion and resumming the whole series back \cite{fnote4}.
Up to the present point we did not make use of any approximation
and our calculation was essentially exact. However, we are particularly
 interested in the limiting case of many
 resonances $N\gg 1$ coupled with few open
channels $M\ll N$. In this limit we expect that the resonance  
widths are of the same order
 as the mean separation between adjacent resonances
$\Delta\propto 1/N$. Therefore, we can restrict our attention to  
the case $Y\sim 1/N$, and, correspondingly, consider $\Omega\sim  
1/N$.
 The second fact that should be taken into account
is that we are actually interested in the limit
$\kappa,E_b-E,\Omega_b-\Omega\to 0$ when calculating the generating  
function.
These facts taken together make it clear that it is sufficient
 to expand the logarithm $ \mbox{Str}\ln{\hat{G}_1}$ in the exponent of
Eq.(\ref{genfunav}) with respect to $\kappa,E_b-E,\Omega_b-\Omega$  
and retain (apart from the leading terms) only terms linear
in these variables. At the same time we can just neglect all these  
variables in the term
$\sum_{a=1}^M\mbox{Str}\ln{\{\hat{I}_4-i\gamma_a(\Lambda\hat{G}^{-1}_1\}}$
because of the condition $M\ll N$. As the result, we have:
\begin{equation}\label{gfav}
\langle{\cal Z}(E,\Omega;E_b,\Omega_b;\kappa)\rangle=
\int[d\hat{R}]\exp{\left[-N{\cal L}[\hat{R}]+\delta{\cal L}\right]}
\end{equation}
where\begin{eqnarray}\label{L}
{\cal L}[\hat{R}]=\frac{1}{2}\mbox{Str}\hat{R}^2
+\mbox{Str}\ln{(-E\hat{I}-R)}\\
\delta{\cal  
L}=N\Omega\mbox{Str}\hat{R}\hat{\Lambda}+N\mbox{Str}\left[i\kappa
\hat{\Sigma}_L+(E_b-E)\hat{K}_b+(\Omega_b-\Omega)\hat{K}_b\hat{L}
\right](-E\hat{I}-\hat{R})^{-1}
\end{eqnarray}

The form of the integrand in Eq.(\ref{gfav}) suggests that it can be
effectively calculated by the saddle-point method exploiting the
large parameter $N\gg 1$.The saddle point equation is determined by  
stationarity
of the "action" ${\cal L}[\hat{R}]$ and has the form
$\hat{R}=(-E\hat{I}-R)^{-1}$. At the same time the
discussion above makes it clear that
the terms entering $\delta{\cal L}$ are of the order of unity
when $N\to \infty$ and should be disregarded when seeking for the
saddle-point solution.

Actually, it turns out that there is a whole continuous manifold
of the saddle point solutions $\hat{R}_s$
 equally contributing to the integral
Eq.(\ref{gfav}) in the limit $N\to  
\infty$:$\hat{R}_s=-E/2+i\pi\nu_{sc}\hat{T}^{-1}\hat{\Lambda}\hat{T}$,
where $\nu_{sc}=\nu_{sc}(E)$ stands for the semicircular density,  
Eq.
({\ref{semi}).
Introducing a new set of matrices
$\hat{Q}=-i\hat{T}^{-1}\hat{\Lambda}\hat{T}$ satisfying the
conditions: $\hat{Q}^2=-\hat{I}_4;\quad \mbox{Str}\hat{Q}=0$
we finally write down the averaged generating function as:
\begin{eqnarray}\label{genfin}\nonumber\langle{\cal Z}
(E,\Omega;E_b,\Omega_b;\kappa)\rangle=
\int[d\hat{Q}]\prod_{a=1}^M\mbox{Sdet}^{-1}\left[I+i\frac{1}{2}
\gamma_aE\hat{\Lambda}+
i\pi\nu_{sc}\gamma_a\hat{Q}\hat{\Lambda}\right]\times \\  
\exp{\left\{-N\pi\nu_{sc}\Omega
\mbox{Str}\hat{Q}(\hat{\Lambda}-\hat{K}_b\hat{L})-
N\pi\nu_{sc}\Omega_b\mbox{Str}\hat{Q}\hat{K}_b\hat{L}-\right.}\\  
\nonumber \left.
i\pi\rho N\kappa
\mbox{Str}\hat{Q}\hat{\Sigma}_L+N(E_b-E)[E+
\pi\nu_{sc}\mbox{Str}\hat{Q}\hat{K}_b]
\right\}
\end{eqnarray}}
\end{itemize}

To obtain the required function $\rho(E,-i\Omega)_{\kappa}$
we should substitute this expression into the relation
Eq.(\ref{auden}). Upon doing this one immediately notices that
the first term $\frac{\partial}{\partial E}
\lim_{(E_b,\Omega_b)\to (E,\Omega)}
\frac{\partial}{\partial E}\langle Z\rangle$ produces a
 contribution which is of the order of $N$, whereas the term
$\frac{\partial}{\partial \Omega}
\lim_{(E_b,\Omega_b)\to (E,\Omega)}
\frac{\partial}{\partial \Omega}\langle Z\rangle$
produces a much larger contribution of the order of $N^2$.  
Retaining only leading terms as long as $N\to \infty$ we arrive at
the following expression:
\begin{eqnarray}\label{densus}
\frac{\rho(E,-i\omega)_{\epsilon}}
{\nu_{sc}(E)}=-\frac{1}{2}
\int[d\hat{Q}]
\mbox{Str}(\hat{Q}\hat{\Lambda})
\mbox{Str}\left[\hat{Q}(\hat{\Lambda}-\hat{K_b}\hat{L})\right]\times  
& &\\
\nonumber\prod_{a=1}^M
\mbox{Sdet}^{-1}\left[I+i\frac{1}{2}
\gamma_aE\hat{\Lambda}+
i\pi\nu_{sc}\gamma_a\hat{Q}\hat{\Lambda}\right]\exp{\left[-\frac{\omega}{2}
\mbox{Str}\hat{Q}\hat{\Lambda}-\frac{i\epsilon}{2}
\mbox{Str}\hat{Q}\hat{\Sigma}_L\right]}&&
\end{eqnarray}
where we introduced the  "scaled" variables
 $\omega=2\pi\nu_{sc}(E)N\Omega$ and
 $\epsilon=2\pi\nu_{sc}(E)N\kappa$ and the correspondingly rescaled  
$\rho(E,-i\Omega)_{\kappa}\to\rho(E,-i\omega)_{\epsilon} 2\pi  
N\nu_{sc}(E)$.

\subsection{Distribution of resonance widths: general expression.}

To perform the explicit evaluation of the superintegral on the  
right-hand side of Eq.(\ref{densus}) one has to employ the
parametrization of the manifold of the $Q-$ matrices suggested by
Efetov \cite{Efrev}. To make the presentation self-contained we present
such a parametrization in Appendix B. In the same Appendix we
 evalute also some supertraces, superdeterminants and combinations  
of the matrix elements entering Eq.(\ref{densus})
as well as other superintegrals we use later on. Upon substitution
of these expressions into Eq.(\ref{densus}) one can perform the
Grassmannian integration trivially and obtains:
\begin{eqnarray}\label{derden1}
\frac{\partial}{\partial
\omega}\frac{\rho(E,-i\omega)_{\epsilon}}
{\nu_{sc}(E)}=-\frac{i}{2}
\int_{1}^{\infty} d\lambda_1\int_{-1}^{1}d\lambda_2
\int_{0}^{2\pi}\int_{0}^{2\pi}\frac{d\phi_1d\phi_2}{(2\pi)^2}\times
\\ \nonumber
\exp{\left[i\omega(\lambda_1-\lambda_2)+i\epsilon(\mid\mu_1\mid\sin{\phi_1}+
i\mid\mu_2\mid\cos{\phi_2})\right]}\prod_{a=1}^{M}\left(\frac{g_a+\lambda_2}
{g_a+\lambda_1}\right) F_{\epsilon,E}(\lambda_{1,2},\phi_{1,2})
\end{eqnarray}where
\begin{eqnarray}\label{F}
F_{\epsilon,E}(\lambda_{1,2},\phi_{1,2})=
i\epsilon(\lambda_1-\lambda_2)(\mid\mu_1\mid\sin{\phi_1}-i\mid\mu_2\mid
\cos{\phi_2})
-i\lambda_2\times\\
\begin{array}{c}
\left\{\frac{\epsilon}{2}(\mid\mu_1\mid\sin{\phi_1}+i\mid\mu_2\mid
\cos{\phi_2})+i\epsilon^2
\left[\frac{1}{2}(\mid\mu_1\mid\sin{\phi_1}-
i\mid\mu_2\mid\cos{\phi_2})^2+\right.\right.\\
 \left.\left.\frac{1}{2}\left(
\mid\mu_1\mid^2-\mid\mu_2\mid^2-2i\mid\mu_1\mu_2\mid
\sin{(\phi_1-\phi_2)}\right)\right]\right\}
\end{array}\nonumber
\end{eqnarray}
with $g_a=2/T_a-1;\quad \lambda_1^2=1+\mid\mu_1\mid^2, \lambda_2^2=
1-\mid\mu_2\mid^2$.

Having in mind that actually we have to perform the analytic
continuation $-i\omega\to y$ in the   right-hand side of the  
Eq.(\ref{derden1}) we introduce two
functions $F_{1,2}(\epsilon,-i\omega)$ according to the following
definitions:
\begin{eqnarray}\label{F1}\nonumber
F_1(\epsilon,-i\omega)=\frac{1}{2\pi}
\int_{1}^{\infty} d\lambda_1
\int_{0}^{2\pi}d\phi_1
\exp{\left[i\omega\lambda_1+i\epsilon\mid\mu_1\mid\sin{\phi_1}\right]}
\prod_{a=1}^{M}\left(g_a+\lambda_1\right)^{-1}=\\\int_{0}^{\infty}\prod_{a=1}^{M}\left(dS_a\exp{-S_ag_a}\right)
\int_{1}^{\infty} d\lambda_1 J_0(\epsilon\sqrt{\lambda_1^2-1})
\exp{[-\lambda_1(-i\omega+\sum_aS_a)]}
=\\  
\nonumber\int_{0}^{\infty}\prod_{a=1}^{M}\left(dS_a\exp{-S_ag_a}\right)
\Phi_1(\epsilon,-i\omega+\sum_aS_a);\quad \Phi_1(\epsilon,y)=
\frac{\exp{-\sqrt{\epsilon^2+y^2}}}
{\sqrt{\epsilon^2+y^2}}
\end{eqnarray}
and, similarly
\begin{eqnarray}\label{F2}\nonumber
F_2(\epsilon,-i\omega)=\frac{1}{4\pi}
\int_{-1}^{1} d\lambda_2
\int_{0}^{2\pi}d\phi_2
\exp{\left[-i\omega\lambda_2-\epsilon\mid\mu_2\mid\cos{\phi_2}\right]}
\prod_{a=1}^{M}\left(g_a+\lambda_2\right)=\\\sum_{k=0}^M{\cal  
D}^{(M)}_k\{g\}\frac{1}{2}
\int_{-1}^{1} d\lambda_2\lambda_2^k
I_0(\epsilon\mid\mu_2\mid)\exp{-i\omega\lambda_2}=\\
\nonumber\sum_{k=0}^M{\cal  
D}_k^{(M)}\{g\}\frac{\partial^k}{\partial (-i\omega)^k}
\Phi_2(\epsilon,-i\omega);\quad
\Phi_2(\epsilon,y)=\frac{\sinh{\sqrt{\epsilon^2+y^2}}}
{\sqrt{\epsilon^2+y^2}}
\end{eqnarray}
where
\begin{equation}\label{defd}
{\cal D}^{(M)}_0\{g\}=\prod_{a=1}^Mg_a;\quad
{\cal D}_1^{(M)}\{g\}=\sum_{a=1}^M\prod_{b\ne a}g_b;
\quad {\cal D}_2^{(M)}\{g\}=\sum_{a,b=1}^M
\prod_{c\ne a,b}g_c,\mbox{etc}
\end{equation}
and $J_0(z)$ and $I_0(z)=J_0(iz)$ stand for the Bessel functions.

One can easily satisfy oneself that the right-hand
side of Eq.(\ref{derden1}) can be expressed in terms of the
functions $F_{1}(\epsilon,-i\omega)$ and
 $F_2(\epsilon,-i\omega)$ and their derivatives. This gives us the  
possibility to perform the required analytic continuation easily and  
to restore the two-dimensional density
$\rho(E,y);\quad y=2\pi\nu_{sc}(E)NY$, see Eq.(\ref{restore})
in the form:
\begin{eqnarray}\label{diff}
\frac{\partial}{\partial y}
\left(\frac{\rho(E,y)}{\nu_{sc}(E)}\right)
=-\lim_{\epsilon\to 0}
\left\{\frac{\partial}{\partial y}\left( F_2\epsilon
\frac{\partial F_1}{\partial
\epsilon}-F_1\epsilon\frac{\partial}{\partial\epsilon}F_2\right)
-\frac{\epsilon}{2}\frac{\partial}{\partial\epsilon}
\left[F_1 \frac{\partial F_2}{\partial y}\right]\right.
\\ \nonumber\left.
-\frac{\epsilon^2}{2}\left[\frac{\partial^2F_1}{\partial
\epsilon^2}\frac{\partial F_2}{\partial y}+
F_1\frac{\partial^2}{\partial
\epsilon^2}\frac{\partial F_2}{\partial y}
-4\frac{\partial F_1}{\partial
\epsilon}\frac{\partial}{\partial
\epsilon} \frac{\partial F_2}{\partial y}
+\frac{\partial^2F_1}{\partial y^2}
\frac{\partial F_2}{\partial y}-
F_1\frac{\partial^2}{\partial y^2}\frac{\partial F_2}{\partial y}
+2F_1\frac{\partial F_2}{\partial y}\right]
\right\}
\end{eqnarray}

The limiting transition $\epsilon\to 0$ is performed with the help
of the identities:
\begin{equation}\label{limit}
-\lim_{\epsilon\to 0}
\epsilon\frac{\partial}{\partial\epsilon}\Phi_{1}(\epsilon,y)
=\lim_{\epsilon\to 0}
\epsilon^2\frac{\partial^2}{\partial\epsilon^2}\Phi_{1}(\epsilon,y)=
2\delta(y)
\end{equation}
which, in turn, are consequences of the following representations
for the $\delta-$ function valid for an arbitrary integer $k$:
\begin{equation}\label{delfun}
\delta(y)=C_k\lim_{\epsilon\to 0}\frac{\epsilon^{2k}}
{(\epsilon^2+y^2)^{(2k+1)/2}};\quad C_1=1/2;\,C_2=4/3\,  
\mbox{etc}.\end{equation}

It is useful also to note that if we substitute  $\Phi_2(\epsilon,y)$
for $\Phi_1(\epsilon,y)$ in Eq.(\ref{limit}), this will produce  
zero instead of the $\delta$-functions on the right-hand side.
The same is true also for terms like $\lim_{\epsilon\to 0}
\epsilon^2
\frac{\partial^2}{\partial
y^2}\Phi_{1}(\epsilon,y)$.
Using these observations we easily pick up all nonvanishing  
contributions to Eq.(\ref{diff}). Summing them up we get:
\begin{equation}\label{derden}
\frac{\partial}{\partial y}
\frac{\rho(E,y)}
{\nu_{sc}(E)}=\frac{\partial}{\partial y}\lim_{\epsilon\to 0}
\left[F_2\,\epsilon\frac{\partial F_1}{\partial \epsilon}\right]
\end{equation}
that immediately allows us to restore the density $\rho(E,y)$ in
the form:\begin{equation}\label{posy}
\frac{\rho(E,y)}{\nu_{sc}(E)}=
F_2(\epsilon=0,y)\int_{0}^{\infty}
\left(\prod_{a=1}^M dS_a\right)
\delta(y+\sum_aS_a) \exp{-\sum_ag_a S_a}
\end{equation}

 It is clear, however, that
for any positive $y$ the $\delta$-functional constraint in
Eq.(\ref{posy}) is never satisfied and the right-hand side is
identically zero. This result is of course
just a consequence of the fact of
absence of scattering poles in the upper half plane of complex energies.

 We therefore consider $y<0$ from now on, make
 the substitution $y\to -y$ to the previous equation and consider
$y>0$ after that. The M-fold integral can be further simplified
upon using the integral representation: $\delta(u)=
\frac{1}{2\pi}\int_{-\infty}^{\infty}dk\exp{iku}$.  Finally we
arrive at the following expression:
\begin{equation}\label{mainres}
\frac{\rho(E,y)}{\nu_{sc}(E)}={\cal F}_1\left[\{g\},y\right]
{\cal F}_2\left[\{g\},y\right]
\end{equation}
where
\begin{equation}\label{calf1}
{\cal F}_1\left[\{g\},y\right]=
-\frac{1}{2\pi}
\int_{-\infty}^{\infty}dk e^{-iky}\prod_{a=1}^M\frac{1}{g_a-ik}=
(-1)^M\sum_{a=1}^{M}e^{-yg_a}\prod_{b\ne a}\frac{1}{g_a-g_b}
\end{equation}
and
\begin{equation}\label{calf2}
{\cal F}_2\left[\{g\},y\right]=\frac{1}{2}\int_{-1}^{1}d\lambda  
e^{-y\lambda}
\prod_{a=1}^M(g_a+\lambda)=
\sum_{k=0}^M(-1)^k{\cal D}_k^{(M)}\{g\}
\frac{d^k}{dy^k}\,\,\frac{\sinh{y}}{y}
\end{equation}
and the functions ${\cal D}_k^{(M)}\{g\}$ are defined in the  
Eq.(\ref{defd}).

It is clear that the function $\rho_E(y)=\frac{\rho(E,y)}{\nu_{sc}(E)}$
has the meaning of a distribution of (scaled) resonance widths
for those resonances ${\cal E}_j=E_j+iY_j$
 whose positions $E_j$ fall into a narrow window
$\delta E$ around the point $E$ of the spectrum:
$$\rho_E(y)=
\frac{1}{N_E}\sum_{j=1}^{N_E}\delta(y-2\pi\nu_{sc}(E)NY_j).$$
 Such a window
should contain a lot of individual resonances: $N_E\sim\frac{\delta E}
{\Delta}\gg 1$
in order to make the statistics representative. On the other hand,
it should be small in comparison with the total width of the spectrum
(in our model given by the widths of the semicircle) to ensure
that the local mean level spacing $\Delta(E)$ is constant across  
this window.

The expressions Eqs.(\ref{mainres}-\ref{calf2}) provide us
with the explicit formula for the local-in spectrum density
of scattering poles for a generic open chaotic system with broken  
time-reversal invariance
 and constitutes the main result of the present section.

\subsection{ Properties of the resonance width distribution.}
\begin{itemize}
\item{\bf Weak coupling versus strong coupling:  from $\chi^2$
distribution to power-law behavior.}

When all scattering channels are equivalent, i.e. have equal
transmission coefficients $T_a=T$ (hence, equal $g_a=g$)
the distribution Eq.(\ref{mainres})
 can be represented in a quite simple and elegant
form.
Indeed, for this case we have:
\begin{equation}\label{F1eq}
{\cal F}_1(g,y)=-\frac{1}{2\pi}\int_{-\infty}^{\infty}dke^{-iky}
\frac{1}{(g-ik)^M}=-\frac{1}{\Gamma(M)}y^{M-1}e^{-g y}
\end{equation}
where $\Gamma(M)=(M-1)!$ stands for the Euler Gamma-function.
We also can write the function ${\cal F}_2(g,y)$ in this case as
\begin{equation}
{\cal F}_2(g,y)=\sum_{k=0}^M(-1)^k
\left(\begin{array}{c}M\\k\end{array} \right)
g^{M-k}\frac{d^k}{dy^k}\left(\frac{\sinh{y}}{y}\right)
\end{equation}
where $\left(\begin{array}{c}M\\k\end{array} \right)$
stands for the binomial coefficient.
One immediately sees that the expression for the density  
$\rho_E(y)$can be written as :
\begin{equation}\label{maineq}
\rho_E(y)=\frac{(-1)^M}{\Gamma(M)}y^{M-1}
\frac{d^M}{dy^M}\left(e^{-yg}\frac{\sinh{y}}{y}\right)
\end{equation}

Remembering, that the "weak coupling" limit corresponds to large
values of the parameter $g_a\gg 1$, one immediately notices that  
the distribution of resonance widths is exponentially cut at
$y\gtrsim y_{max}= \max_{a}\{g_a^{-1}\}$.
Thus, in the weak coupling limit $y_{max}\ll 1$ and we can put  
$\frac{\sinh{y}}{y}\approx 1$ everywhere.
This procedure immediately results in the well-known $\chi^2$
distribution for the case of equivalent channels, see Eq.(\ref{maineq}).
The condition $y\ll 1$ just means that the resonances are too narrow
to overlap with each other. It is therefore natural that the  
$\chi^2$ distribution
simply follows from a first order perturbation
theory, see the discussion in the beginning of the present section.

As long as the coupling to continuum becomes stronger, the parameters
$g_a$ decrease towards unity. When one or more transmission
coefficients $T_a$ attain their maximal value $T_a=1$ a drastic
modification of the resonance width distribution occurs. Indeed, it  
is easy
to see that when $T_a$ (and hence the corresponding $g_a$) tends to  
unity,
the function ${\cal F}_1\left[\{g\},y\right]$ behaves proportionally
to $\exp{(-y)}$ at large enough $y$. This decay exactly cancels the
growing exponent $\exp{(y)}$ originating asymptotically from ${\cal
F}_2\left[\{g\},y\right]$. As the result, the distribution
function $\rho_E(y)$ must show a {\it pure powerlaw} decay in its
 tail, see Fig.2.

To determine this power explicitly one should make more accurate
estimates of the asymptotic behavior of both ${\cal F}_1$ and ${\cal
F}_2$. Substituting $\sinh{y}\approx
\frac{1}{2}e^{y}$ in the definition of ${\cal F}_2$
one can write:
\begin{equation}\label{pow}
{\cal F}_2\left[\{g\},y\to \infty \right]=\frac{1}{2}
e^{y}\sum_{p=0}^M\frac{(-1)^p}{y^{p+1}}U_p;\quad
U_p=\sum_{k=0}^M(-1)^k \frac{k!}{(k-l)!}{\cal D}_k^{(M)}
\end{equation}
Using the definitions of the coefficients ${\cal D}_k^{(M)}$,
 see Eq.(\ref{defd}) one finds that
$$
U_0=\prod_{a=1}^M (g_a-1);\quad U_1=-\sum_{a=1}^M\prod_{b\ne
a}(g_b-1);\quad ...\quad U_p=(-1)^pp!\sum_{a_1,...,a_l=1}^M
\prod_{c\ne a_1...a_l}(g_c-1)
$$

We see that the leading power of $y$ in the asymptotic behavior of
the function ${\cal F}_2\left[\{g\},y\right]$ is essentially
determined by the number of parameters $g_a$ which are {\it  
simultaneously}
equal to unity. Let us suppose, for definiteness, that exactly $l$
quantities $g_1,g_2,...,g_l$ are equal to unity, whereas all other
$M-l$ parameters $g_{l<a\le M}$ are larger (and, for simplicity, are all
different). Then $U_0=U_1=...=U_{l-1}=0$ and the leading behavior
of the function ${\cal F}_2\left[\{g\},y\right]$ is given by:
\begin{equation}\label{asyf2}
{\cal F}_2\left[\{g\},y\to\infty\right]=(-1)^ll!\frac{e^y}{2y^{l+1}}
\prod_{a=l+1}^{M}(g_a-1)
\end{equation}Under the same conditions one can determine the  
leading asymptotic behavior
of the function ${\cal F}_1\left[\{g\},y\right]$ by calculating
the integral in eq.(\ref{calf1}). The contribution of the $l-$fold pole
at $k=-i$ gives:
\begin{equation}\label{asyf1}
{\cal F}_1\left[\{g\},y\to\infty\right]=\frac{(-1)^l}
{(l-1)!}y^{l-1}e^{-y}\prod_{a=l+1}^{M}\frac{1}{(g_a-1)}
\end{equation}
This finally results in the desirable asymptotic decay law for the
distribution of resonance widths:
\begin{equation}\label{asden}
\rho_E(y\to\infty)=\frac{l}{2y^2}
\end{equation}
The physical origin of such a tail in the width distribution
is discussed below.
It is interesting to note that such a behavior means that
the positive moments of the width  
distribution$\int_0^{\infty}dyy^k\rho_E(y)$ are apparently divergent  
as long as
$k\geq 1$.

It is also instructive to consider briefly the particular case
of very many equivalent channels: $M\gg 1;\quad g_a=g$ for any channel.
We find that it is
most convenient to rewrite the expression Eq.(\ref{maineq})
in an equivalent form:
\begin{equation}\label{intrep}
\rho_E(y)=\frac{1}{2\Gamma(M)y^2}\int_{y(g-1)}^{y(g+1)}dt
\exp{-(t-M\ln{t})}\end{equation}
and to evaluate the integral by the saddle-point method. The exponent
is maximal in the vicinity of $t_s=M$. When this point
is inside the integration region, i.e. $y(g-1)<t_s<y(g+1)$
we have a nonvanishing contribution to the integral. In the  
opposite situation the density of resonances vanishes exponentially
when $M\gg 1$. Picking up the nonvanishing contribution we  
obtain:\begin{equation}\label{gap}
\rho_E(y)\mid_{M\gg 1}=\left\{\begin{array}{cc}\frac{M}{2y^2},&
\frac{M}{g+1}<y<\frac{M}{g-1}\\ 0,\quad &
y<\frac{M}{g+1}\quad\mbox{or}\quad y>\frac{M}{g-1}\end{array}\right.
\end{equation}

Two conclusions can be drawn from this  expression:
\begin{enumerate}
\item In the limit of large number of channels $M\gg 1$ the
distribution of resonance widths shows a gap: there are no resonances
with widths smaller than $y_m=\frac{M}{g+1}$.
\item  A  region of power-law behavior $\rho_E(y)\propto M/y^2$
exists not only for the critical coupling $g=1$, but also in the vicinity
of the critical point: $g-1\ll g$. However, only for $g=1$ the
power-law domain extends to infinity.
\end{enumerate}

The formation of a gap (a strip in the complex
 energy plane free of resonances) was first noticed in the numerical 
experiments by Moldauer\cite{Molgap} long ago and discussed
in much detail by Sokolov and Zelevinsky\cite{Sok} later on.
Gaspard and collaborators (see references in \cite{Gasp})
 observed such a gap in their
studies of chaotic scattering in the so-called three-disk systems.
Semiclassically, the number of open channels was very large and
comparable with the number of resonances. In the limit $M,N\to  
\infty$ but
$m=M/N$ finite the expression for the resonance width
distribution was obtained and analyzed in
the papers\cite{Haake,Lehm2}. Our expression Eq.(\ref{gap})
 obtained under the conditions $1<<M<<N$ perfectly matches the  
$m<<1$ limiting case of their expression.

\item{\bf Mean resonance width: the Moldauer-Simonius relation.}

Having at our disposal the explicit formula Eq.(\ref{mainres})
we can, in particular, easily calculate the mean resonance
width:
\begin{eqnarray}\label{meanwid}\nonumber
\langle y \rangle=\int_0^{\infty}dy y {\cal F}_1{\cal F}_2=
-\frac{1}{4\pi}\int_{-1}^{1}d\lambda\prod_a(g_a+\lambda)
\int_{-\infty}^{\infty} dk\prod_a\frac{1}{g_a-ik}\int_{0}^{\infty}
dy y e^{-y(\lambda+ik)}=\\
\frac{1}{4\pi}\int_{-1}^1d\lambda\prod_a(g_a+\lambda)\frac{\partial}
{\partial \lambda}\int_{-\infty}^{\infty}dk\frac{1}{\lambda+ik}
\prod_a\frac{1}{g_a-ik}=\\  
\nonumber-\frac{1}{2}\int_{-1}^1d\lambda\prod_a(g_a+\lambda)
\frac{\partial}{\partial \lambda}\prod_a\frac{1}{g_a+\lambda}=
\frac{1}{2}\sum_{a=1}^M\ln\frac{g_a+1}{g_a-1}
\end{eqnarray}
Remembering the relation between $g_a$ and the transmission coefficients
$T_a$, see Eq.(\ref{defg}), and using the definition of the scaled
level width $y=\pi\Gamma/\Delta$ we can represent the last result in
the form of a relation between the mean resonance widths $\langle
\Gamma\rangle$ and the transmission coefficients $T_a$:
\begin{equation}\label{MS1}
\langle\Gamma\rangle=-\frac{\Delta}{2\pi}\sum_{a=1}^{M}\ln{(1-T_a)}
\end{equation}
which can be also rewritten as:
\begin{equation}\label{MS2}
\mid\prod_{a=1}^M\langle S_a \rangle\mid=
\exp{-\frac{\pi\langle\Gamma\rangle}{\Delta}}
\end{equation}
in view of the definition $T_a=1-\mid\langle S_a\rangle\mid^2$.

The latter formula is well-known for a long time
in nuclear physics as the Moldauer-Simonius relation\cite{Mol}.
It was derived for systems with unbroken time-reversal symmetry
by averaging the $S-$matrix over the energy spectrum and using the  
unitarity condition. The fact that we recovered this
relation by ensemble averaging is in good agreement with the well-known
ergodicity of the Gaussian ensembles\cite{French}.
 The logarithmic divergency of $\langle
\Gamma\rangle$ at the critical coupling $T_a=1$ is a direct
consequence of the $1/y^2$ decrease of the probability distribution,
see Eq.(\ref{asden}).
\item{\bf Strong chaos as the origin of the power-law tail of the
resonance width distribution.}
The results presented above suggest that the powerlaw
decrease $1/y^2$ should be typical for chaotic systems strongly coupled
to continua and is one of the clear manifestations of the strong
overlap between individual resonances.
It is therefore natural to try to understand the origin of such a
tail qualitatively in terms of the underlying chaotic dynamics.

To this end it is interesting to mention that a little different
powerlaw distribution of resonance widths, that of the form
$\rho(y)\propto y^{-3/2}$, was observed in numerical studies of a
quantum chaotic system with (quasi)
one-dimensional "diffusive" dynamics in the case of
strong coupling to continua\cite{Shep}. The authors suggested a
transparent qualitative explanation of this effect based on the fact
that the resonance width is proportional to the
inverse life-time for a wavepacket injected into the system. The latter
 is determined by the time of
the classical diffusion: $t_{dif}=D/L^2$, where $L$ is the distance
from a given point to the closest (strongly absorbing) boundary
and $D$ is a classical diffusion constant. For a semi-infinite
sample this reasoning immediately gives a powerlaw width distribution
$\rho(\Gamma)\propto \frac{dL}{d\Gamma}\propto\Gamma^{-3/2}$.

Let us remind that the present model is based on the use of the  
Gaussian random matrices. Physically, it
corresponds to the case of strongly chaotic  classical
dynamics for the closed system\cite{Bohigas,AAA}. For such systems
there is a typical time scale $\delta t$ determined by an inverse
Lyapunov exponent $ \lambda^{-1}$ after which the system
effectively loses a memory about
 its initial conditions and can be found in any
part of the available phase space on the energy shell with equal
probability. Such systems are known as the ergodic ones. Let us
show that it is just that type of classical dynamics which is responsible
for the powerlaw tail $1/y^2$ of the width distribution.

To understand this fact let us consider as a particular, but generic
example: a particle moving with a velocity $v$
inside an irregular-shaped two-dimensional
cavity of area $A$ and circumference $C\propto A^{1/2}$. The  
chaoticity is considered to be so strong that complete "loss of
memory" occurs after few collisions with walls so that $\delta t$ can be
estimated as $\delta t\propto C/v\propto A^{1/2}/v$. Let us make a  
small opening of the width
$d<<C$ in the walls so that the particle can escape from the cavity  
whenever it hits the
opening. Subdividing the observation time in intervals of the order
of  $\delta t$ we conclude that the probability
to escape during one interval is just $p_0=d/C\propto dA^{-1/2}<<1$
in view of the ergodicity and escape events during the subsequent
intervals can be treated as independent (memory loss).
Then the probability to stay inside the cavity for a large time $t_e$
and then to leave within the interval $[t_e,t_e+dt]$
can be estimated as $P(t_e)dt=\frac{p_0dt}{\delta  
t}\exp{-p_0t_e/\delta t}$.

Considering our system semiclassically, we associate the wavelength  
$\lambda_d=\frac{\hbar}{mv}$ with our particle and can
estimate the number of the quantum mechanical states available
inside the closed cavity as $N\propto \frac{A}{\lambda_d^2}$.
Since the energy $E=mv^2/2$, the corresponding mean level spacing
being of the order of $\Delta=E/N$ is proportional to  
$mv^2\lambda_d^2/A$.
Interpreting the inverse escape time $\hbar t_e^{-1}$ as the resonance
width $\Gamma$
and measuring it in units of $\Delta$: $y=\Gamma/\Delta$,
one can find the distribution of $y$ to be given by:
$P(y)=\frac{M_{sc}}{y^2}\exp{-\frac{M_{sc}}{y}}\approx  
\frac{M_{sc}}{y^2}$
for $y\gg M_{sc}$, where $M_{sc}=\hbar p_0/(\Delta\delta t)\propto  
\frac{\hbar d}{A^{1/2}}
\frac{A}{mv^2\lambda_d^2}\frac{v}{A^{1/2}}=d/\lambda_d$,
which coincides with the quasiclassical estimate for the number
of open channels for the present problem.

We conclude that the semiclassical arguments faithfully reproduce  
the same powerlaw tail
of the resonance width distribution as that obtained from our random
matrix model. Therefore, we expect such a tail to be a universal
characteristic of the chaotic quantum scattering problem independent
on the specific details of the underlying classical dynamics
being sufficiently chaotic to ensure an exponential escape from the  
compact scattering region.
The following comment is appropriate here.
 Dealing with realistic models of open chaotic systems containing  
no random parameters one always performs
 statistics over an interval of energies $\delta E $
on a real axis
 containing many resonances: $\delta E \gg \Delta$,
 but being small enough for a systematic variation of
 the smoothed level density $\overline{\nu(E)}$ to be neglected:
 $\delta E\ll \overline{\nu}/\frac{d \overline{\nu}}{d E}$
(c.f definition of the quantity $\rho_E(y)$).
 One may expect that {\it universal} features of such statistics
 are adequately reproduced within the framework of the stochastic
 approach, but  only on the level of "local-in-spectrum"
characteristics calculated at {\it fixed} value of E.
 Indeed, any spectral
averaging in the stochastic model performed on a scale
 comparable with the radius of the semicircle unavoidingly mixes up  
data corresponding to very different values of the
 transmission coefficients, the procedure washing out
any relevant physical information. In particular, it seems
 quite meaningless to consider quantities like the "globally"  
averaged  resonance width  
$\Gamma_{gl}=\frac{1}{N}\sum^{N}_{k=1}\Gamma_k$,
where the summation goes over {\it all} $N$ resonances.
 In our model this quantity can be trivially found from the sum rule:
 $N\Gamma_{gl}=-2\mbox{Tr Im}
{\cal H}_{ef}=2\sum_a\gamma_{a}$.
 and can not be related to any particular
transmission coefficient.
 This fact, however, should not be misinterpreted as impossibility  
to have universal statistics of $S-$matrix poles within
the stochastic approach as discussed in\cite{Lew}.
  Rather, the quantity $\Gamma_{gl}$ can be found via
 the direct integration of the  universal {\it local} expression
 $\langle\Gamma\rangle$ , Eq.(\ref{MS1})
 over the energy $E$, upon  substituting there the energydependent  
values $T_a(E),\Delta(E)$ from Eq.(\ref{trans}).
Indeed, the following integral can be easily evaluated \cite{RG}:
\begin{eqnarray}\label{avwid}
\lim_{N\to\infty}\sum_{k}\Gamma_k=N\int_{-2}^{2}\langle
\Gamma(E)\rangle\nu(E)dE=\\ \nonumber
-\frac{1}{2\pi}\sum_a\int_{-2}^2dE\ln{\left(\frac{1-2\gamma_a\sqrt{1-E^2/4}
+\gamma_a^2}{1+2\gamma_a\sqrt{1-E^2/4}
+\gamma_a^2}\right)}=\sum_{a=1}^M\left[\gamma_a+\gamma_a^{-1}-
|\gamma_a-\gamma_a^{-1}|\right],
\end{eqnarray}
 resulting in the expected expression
$2\sum_a\gamma_{a}$ as long as all $\gamma_a\le 1$.

We see that the result of the integration is always finite for any
$\gamma_a$, thus concealing a specific role of the critical  
coupling $\gamma_{a}=1$
 when resonances with  {\it divergent} local mean width
 occur sufficiently close to the center of the spectrum.

Of course, taken literally this divergency has sense
 only in the limit of infinite number of resonances $N\to\infty$.
For any finite $N$ all resonance widths are finite and in any case  
can not
exceed the upper bound $2\sum_a\gamma_a$. Basically, it is related
to the fact that
the distributions Eqs.(\ref{mainres},\ref{maineq}) cease to be valid for
the domain of very broad resonances having widths $\Gamma\sim 1$
(correspondingly, $y\sim N$).
 Alternatively, one may say that for large, but finite $N$
the Moldauer-Simonius relation
is to be modified in a narrow domain $\delta E\propto 1/N$ in the
 vicinity of the energy $E=0$, see \cite{ISS}.

Expression Eq.\ref{avwid} can be also used to describe an interesting
 phenomenon happening when some coupling constants
$\gamma_a$ ( e.g. for the channels $a=1,2,...,M_1\le M$)
 exceed the critical value $\gamma=1$ .
 Under this condition the result of integration in Eq.(\ref{avwid})  
is less than the exact sum
rule value $2\sum_a^M \gamma_a$ by the quantity $\delta \Gamma=
2\sum_{a}^{M_1}(\gamma_a-\gamma_a^{-1})>0$. This deficit reflects
the existence of $M_1$ "broad resonances" of the widths  
$\Gamma_a=2\left(\gamma_a-
\gamma_a^{-1}\right)\gg \Delta\propto 1/N;\quad a=1,...,M_1$.
Such resonances can not be described
by the distributions Eqs.(\ref{mainres},\ref{maineq}) which cover  
only the
resonances satisfying $y=\pi\Gamma/\Delta<\infty$ in the limit $N\to
\infty$. When coupling constants $\gamma_a$ increase further the
group of broad resonances moves away from the real axis, accumulating
in the limit $\gamma_a\gg 1$ the lion's share of the total widths  
$\Gamma_{gl}$. The remaining $N-M_1$
resonances become progressively more and more narrow and their  
widths are well described by the distribution Eq.(\ref{mainres}).

The effect of reorganization of the resonances into two essentially
different groups due to strong coupling to continua ( the "trapping
 phenomenon" ) was studied in some details in the  
papers\cite{Sok,Haake,Dittes,Remacle}.

\end{itemize}
\section{Statistics of scattering phase shifts and time delays}
In the present section we are going to study in much detail
the statistics of individual scattering phase shifts $\theta_a$
and their derivatives, both over the energy $E$ and over an arbitrary
external parameter $X$.
As we already mentioned in Sec.II (see Eq.\ref{denz}),
 it is convenient to characterize
the phase shift properties via the spectral density:
$\rho_{E,X}(z)=
\frac{1}{M}\sum_{a=1}^{M} \delta(z-z_a(E,X))$
of eigenvalues of the matrix $\hat{K}_X(E)=\pi\hat{W}^{\dagger}
\left(E-\hat{H}_{in}(X)\right)^{-1}\hat{W}$. Our object of primary  
interest is
the correlation function\begin{equation}\label{sm1}
{\cal K}_{E,\Omega,X}(z_1,z_2)= \langle
\rho_{E-\Omega/2,-X/2}(z_1)\rho_{E+\Omega/2,X/2}(z_2)\rangle- \langle
\rho_{E-\Omega/2,-X/2}(z_1)\rangle\langle\rho_{E+\Omega/2,X/2}(z_2)
\rangle
\end{equation}
knowledge of which, in particular, allows one to study statistics
of "partial delay times"
$\tau_a=\frac{\partial\theta_a}{\partial E}$ and also the
corresponding parametric derivatives.

Before addressing the issue of the spectral density correlations,
it is instructive to consider in some detail the calculation of
the average spectral density $\langle \rho_{E,0}(z)\rangle$ for the
few channel case. This quantity is
less informative than the correlation function Eq.(\ref{sm1}),
but that simple calculation serves as a reference point for more
interesting cases. Let us mention, that in the limit $M\propto N\gg
1$ the phase shift density was found earlier by Lehmann and Sommers  
\cite{unpub}.

\subsection{Averaged spectral density of $K-$ matrix}

The averaged density can be easily found provided the following
functions are known
\begin{equation}\label{sm2}
f^{\pm}_{E,X}(z)=\left\langle
\mbox{Tr}\frac{1}{z\pm i\epsilon-\hat{K}_{X}(E)}\right\rangle
\end{equation}
in view of the obvious relation: $\rho_{E,X}(z)=\frac{1}{\pi M}
\lim_{\epsilon\to 0}\mbox{Im}f^{-}_{E,X}(z)$. We restrict our  
attention in
the present context by $f^{-}_{E,X=0}(z)$, omitting all the  
indices$\pm,E,X$ for the sake of brevity.

 The function $f(z)$ can be formally written as:
\begin{equation}\label{sm3}
f(z)=\lim_{J\to 0}\frac{\partial}{\partial  
J}\langle\ln{Z(J)}\rangle;\quad
Z(J)=\frac{\mbox{Det}\left((z+J)\hat{I}_M-\hat{K}\right)}
{\mbox{Det}\left(z\hat{I}_M-\hat{K}\right)}
\end{equation}
Here and below we imply $z\equiv z-i\epsilon$ for the sake of
brevity, implying $f(z)$ to be analytic in the lower $z$ half-plane.

Due to the normalisation condition $Z(J=0)=1$ one can write
$f(z)=\lim_{J\to 0}\frac{\partial}{\partial J}\langle Z(J)\rangle$.
In order to perform the ensemble average of the generating function
$Z(J)$ in a standard way
one should first get rid of the following unpleasant feature:
the random matrix $\hat{H}_{in}$ enters the expression for the
generating function only via the matrix $\hat{K}$. To this end, we  
can use the identity Eq.(\ref{identity}) and write
down the determinant in the denominator of the generating function
as:
\begin{eqnarray}\label{sm4}
z^M\mbox{Det}\left(\hat{I}_M-z^{-1}\pi\hat{W}^{\dagger}
(E-\hat{H}_{in})^{-1}\hat{W}\right)=z^M\mbox{Det}\left(\hat{I}_N-z^{-1}\pi
(E-\hat{H}_{in})^{-1}\hat{W}\hat{W}^{\dagger}\right)=\\ \nonumber
z^M\mbox{Det}^{-1}\left(E-\hat{H}_{in}\right)
\mbox{Det}\left(E-\hat{H}_{in}-\pi z^{-1}
\hat{W}\hat{W}^{\dagger}\right)
\end{eqnarray}
 After performing a similar manipulation with the numerator of the
generating function we can write:
\begin{equation}\label{sm5}
f(z)=\lim_{J\to 0}
\frac{\partial}{\partial J}\left[\left(\frac{z+J}{z}\right)^M
{\cal F}(J)\right]; \quad {\cal
F}(J)=\left\langle\frac{\mbox{Det}[E-H_{eff}(z+J)]}
{\mbox{Det}[E-H_{eff}(z)]}\right\rangle
\end{equation}where we introduced the notation
$H_{eff}(z)=\hat{H}_{in}+\frac{\pi}{z}WW^{\dagger}$.

Now we can use a standard procedure and represent the determinants in
the denominator/numerator of the preceeding equation by Gaussian
integrals over $N$ commuting/anticommuting variables. After
introducing $2N-$component supervector ${\bf\Psi}=
\left(\begin{array}{c}\bf{S},\\ \chi\end{array}\right)$, we have:
\begin{equation}\label{sm6}\frac{\mbox{Det}[E-H_{eff}(z+J)]}
{\mbox{Det}[E-H_{eff}(z)]}=\int[d{\bf\Psi}]
\exp{\left\{-i{\bf\Psi}^{\dagger}\left[
(E-\hat{H}_{in})-\hat{\Gamma}\otimes\hat{U}\right]{\bf\Psi}\right\}}
\end{equation}
where the supermatrix $\hat{U}=\mbox{diag}(z^{-1},(z+J)^{-1})$
and $\hat{\Gamma}=\pi\hat{W}\hat{W}^{\dagger}$, as before.

Now one trivially performs the averaging over the ensemble, see  
eq.(\ref{ident}) and decouples the emerging "quartic term"
in the exponent with help of the Hubbard-Stratonovich transformation.
In the present simple case such a decoupling is possible when
one uses the set of $2\times 2$ matrices $\hat{P}$ defined like in  
Eq.(\ref{parR})
as the integration manifold. Changing the order of integrations,
performing the integration over $\Psi$ explicitly, and copying steps
used to derive Eq.(\ref{chanexp}), one obtains:
\begin{equation}\label{sm7}
\left\langle\frac{\mbox{Det}[E-H_{eff}(z+J)]}
{\mbox{Det}[E-H_{eff}(z)]}\right\rangle=\int d\hat{P}e^{-N{\cal
L}(\hat{P})} \prod_{a=1}^{M}\mbox{Sdet}^{-1}\left[\hat{I}_2-\gamma_a
\hat{U}(E-\hat{P})^{-1}\right]
\end{equation}
where$$
{\cal  
L}(\hat{P})=\frac{1}{2}\mbox{Str}\hat{P}^2+\mbox{Str}\ln{(E-\hat{P})}
$$
Now it is convenient to use that: $\frac{(z+J)^M}{z^M}=
\mbox{Sdet}^M\hat{U}$. When combined with the preceeding expression
it gives:
\begin{equation}\label{sm8}
f(z)=\lim_{J\to 0}
\frac{\partial}{\partial J}\int d\hat{P}e^{-N{\cal
L}(\hat{P})} \prod_{a=1}^{M}\mbox{Sdet}^{-1}\left[\hat{U}^{-1}-\gamma_a
(E-\hat{P})^{-1}\right]
\end{equation}
The integral over $\hat{P}$ is calculated in the limit $N\gg M$
by the saddle-point method, with unique diagonal  
saddle-point$\hat{P}_s=(E/2+i\pi\nu_{sc})\hat{I}_2$ accessible by
allowed contour deformation for $\mid E\mid\le 2$.
This immediately yields:
\begin{equation}\label{sm9}
f(z)=\lim_{J\to 0}
\frac{\partial}{\partial J}\prod_{a=1}^M\left(
\frac{(z+J)-\gamma_a(E/2+i\pi\nu_{sc})}
{z-\gamma_a(E/2+i\pi\nu_{sc})}
\right)=\sum_{a=1}^M\frac{1}{z-\gamma_a(E/2+i\pi\nu_{sc})}
\end{equation}
analytic in the lower half-plane, from where we find that the mean  
density of eigenvalues for the matrix
$\hat{K}$ is given by a sum of Lorentzians (with $z$ real):
\begin{equation}\label{sm10}
\rho_E(z)=\frac{1}{M}\sum_{a=1}^M\frac{\nu_{sc}\gamma_a}
{\left(\pi\nu_{sc}\gamma_a\right)^2+\left(z-\frac{\gamma_a  
E}{2}\right)^2}\end{equation}

For the particular case of one open channel $M=1$ the Lorentzian
form of the average spectral densiy was first found by
Mello\cite{Melrev}. Actually, in that particular case
one can check the expression Eq.(\ref{sav}) for the averaged $S$-matrix
using Eq.(\ref{sm10}). Indeed, for $M=1$ the
$S-$matrix is reduced to the only number $S=
\exp{(-2i\arctan{z})}\equiv 1-2iz/(1+iz)$. Therefore
\begin{equation}\label{sm11}
\langle S\rangle=1-2i \int_{-\infty}^{\infty}dz\rho_E(z)\frac{z}{1+iz}
\end{equation}
The integrand has the only pole $z_{-}=E\gamma/2-i\pi\nu_{sc}\gamma$
in the lower half plane $\mbox{Im} z<0$, and the corresponding
residue immediately gives:
\begin{equation}\label{sm12}
\langle S\rangle=1-2i(-2\pi i)\frac{\nu_{sc}\gamma}
{z_{-}-E\gamma/2-i\pi\nu_{sc}\gamma}\frac{z_{-}}{1+iz_{-}}=
\frac{1-\frac{\gamma}{2}(iE+\sqrt{4-E^2})}
{1+\frac{\gamma}{2}(iE+\sqrt{4-E^2})}
\end{equation}
in complete agreement with Eq.(\ref{sav}).

\section{Pair correlation function of $K$-matrix spectral densities}

Let us now turn our attention to the calculation of the pair  
correlation function
${\cal K}_{E,\Omega,X}(z_1,z_2)$, Eq.(\ref{sm1}). To this end let  
us introduce the function
\begin{equation}\label{sm13}
f(z_1,z_2)=\left\langle
\mbox{Tr}\frac{1}{z_1-i\epsilon-\hat{K}_{-X/2}(E-\Omega/2)}
\mbox{Tr}\frac{1}{z_2+i\epsilon-\hat{K}_{X/2}(E+\Omega/2)}\right\rangle
\end{equation}
related to the correlation function in Eq.(\ref{sm1})
as:
\begin{equation}\label{sm14}
{\cal K}_{E,\Omega,X}(z_1,z_2)=\frac{1}{2\pi^2
M^2}\mbox{Re}f_{c}(z_1,z_2); \quad
f_{c}(z_1,z_2)=f(z_1,z_2)-f^{-}(z_1) f^{+}(z_2)
\end{equation}

Performing with each of the two traces of resolvents in the
Eq.(\ref{sm13}) the same manipulations as presented in  
Eqs.(\ref{sm3}-\ref{sm5})
one obviously obtains the following representation:
\begin{equation}\label{sm15}
\begin{array}{c}f(z_1,z_2)=\displaystyle{
\frac{\partial^2}{\partial J_1\partial
J_2}\left[\left(\frac{Z_J^{(1)}Z_J^{(2)}}{Z_{J=0}^
{(1)}Z_{J=0}^{(2)}}\right)^M
{\cal F}(J_1,J_2)\right]\mid_{J_1=J_2=0}} \\\\\displaystyle{ {\cal
F}(J_1,J_2)=\left\langle\frac{\mbox{
Det}[E-\Omega/2-H_{eff}(-X/2;Z_J^{(1)})]
\mbox{Det}[E+\Omega/2-H_{eff}(X/2;Z_J^{(2)})]}
{\mbox{Det}[E-\Omega/2-H_{eff}(-X/2;Z_{J=0}^{(1)})]
\mbox{Det}[E+\Omega/2-H_{eff}(X/2;Z_{J=0}^
{(2)})]}\right\rangle}
\end{array}
\end{equation}
 where we introduced the notations:
$Z_J^{(p)}=z_p+i(-1)^p\epsilon+J_p;\quad p=1,2$ and
$H_{eff}(X;Z_J^{(p)})=\hat{H}_{in}+\frac{X}{N^{1/2}}\hat{H}_{in}^{(1)}
+\frac{\pi}{Z_J^{(p)}}WW^{\dagger}$.

This expression is quite close in its form to the generating function
Eq(\ref{genfun})
appearing in the calculation of resonance widths distributions and  
we can use a similar
representation for it in terms of the Gaussian(super) integrals  
(cf.Eq.(\ref{Gaususy})) :
\begin{eqnarray}\label{sm16}
{\cal F}(J_1,J_2)=(-1)^N
\int[d\Psi]\exp{\left\{-iE\Psi^{\dagger}\hat{L}\Psi-i\frac{\Omega}{2}
\Psi^{\dagger}\hat{L}\hat{\Lambda}\Psi
+i\Psi^{\dagger}
\hat{\Gamma}\otimes\left(\hat{L}\hat{U}\right)
\Psi\right\}}\times\\ \nonumber
\left\langle\exp{\left\{i\Psi^{\dagger}\left(\hat{H}_{in}
\otimes\hat{L}\right)\Psi-i\frac{X}{2N^{1/2}}\Psi^{\dagger}\left(\hat{H}^{(1)}_{in}
\otimes\hat{L}\right)\Psi\right\}}\right\rangle
\end{eqnarray}
where $\hat{U}^{-1}=\mbox{diag}
(Z_{J=0}^{(1)},Z_{J}^{(1)},Z_{J=0}^{(2)},Z_{J}^{(2)})=
\frac{z_1}{2}(\hat{I}_4+\hat{\Lambda})+
\frac{z_2}{2}(\hat{I}_4-\hat{\Lambda})+
\mbox{diag}(0,J_1,0,J_2)$
and notations for the supermatrices $\hat{L},\hat{\Lambda}$
and the supervector $\Psi$ are the same
as in Eqs.(\ref{suvec},\ref{matdiag}).

The subsequent procedure of dealing with the ensemble average in
Eq.(\ref{sm16}) is exactly the same as that presented in details
in Sec.III. The only difference is that the average is performed
only over the GUE matrix $\hat{H}_{in}$, whereas the matrix
$\hat{H}_{in}^{(1)}$ is considered to be arbitrary,
but {\it fixed} from
the same ensemble, see \cite{AlSi}. As a result, one has  
(cf.Eq.\ref{genfunav}):
\begin{equation}\label{sm17}
{\cal F}(J_1,J_2)=
\int[dR]\exp{\left\{-\frac{N}{2}\left[\mbox{Str}\hat{R}^2-\Omega\mbox{Str}
\hat{R}\hat{\Lambda}\right]-\mbox{Str}\ln{\hat{G}^{\cal  
F}}\right\}}
\end{equation}where
\begin{equation}\label{sm18}
\hat{G}^{\cal F}= \hat{G}^{\cal F}_1-\hat{\Gamma}\otimes\hat{U};
\quad \hat{G}^{\cal F}_1=\left[(E\hat{I}_4-
\hat{R})\otimes\hat{I}_N+\frac{X}{2N^{1/2}}\left(
\hat{H}_{in}^{(1)}\otimes\Lambda\right)\right]
\end{equation}
so that\begin{equation}\label{sm19}
\mbox{Str}\ln{\hat{G}^{\cal F}}=\mbox{Str}\ln{ \hat{G}^{\cal F}_1}
+\mbox{Str}\ln{\left\{\hat{I}_N-\left(\Gamma\otimes
\hat{U}\right)\left[\hat{G}^{\cal
F}_1\right]^{-1}\right\}}\end{equation}
In turn, one can expand $\hat{G}^{\cal F}_1$ in a series with respect
to $X$ :
\begin{equation}\label{sm20}
\mbox{Str}\ln{\hat{G}^{\cal F}_1}=N\mbox{Str}\ln{
\left(E\hat{I}_4-\hat{R}\right)}  
-\sum_{l=1}^{\infty}\frac{(-X/2N^{1/2})^l}{l}
\mbox{Tr}\left[H_{in}^{(1)}\right]^l\mbox{Str}\left[\hat{\Lambda}(E\hat{I}_4-
\hat{R})^{-1}\right]^l
\end{equation}
Now we use the fact that for any typical GUE matrix holds:
$\mbox{Tr}\left[H_{in}^{(1)}\right]^{2p}=O(N);\,\,\,
\mbox{Tr}\left[H_{in}^{(1)}\right]^{2p+1}=O(1);$ where $p\ge 0$ is an
integer. It is therefore evident, that in the limit $N\to \infty$
the only nonvanishing term in the expansion above is that with $l=2$.
We also can put effectively $\hat{G}^{\cal F}_1=
\left(E\hat{I}_4-\hat{R}\right)\otimes I_N$ in the second term in  
Eq.(\ref{sm19}) and represent it in a form of a sum over channels,
see Eq.(\ref{chanexp}).

 Collecting all the relevant terms, we obtain:
\begin{eqnarray}\label{sm21}
{\cal F}(J_1,J_2)=
\int[dR]\exp{\left\{-N\left[\frac{1}{2}\mbox{Str}\hat{R}^2+
\mbox{Str}\ln{(E\hat{I}_4-\hat{R})}\right]\right\}}\times\\  
\nonumber \exp{\left\{\frac{N\Omega}{2}\mbox{Str}
\hat{R}\hat{\Lambda}+\frac{X^2}{8}\mbox{Str}\left[\hat{\Lambda}
(E\hat{I}_4-\hat{R})^{-1}\right]^2\right\}}\prod_{a=1}^M\mbox{Sdet}^{-1}
\left[\hat{I}_4-\gamma_a\hat{U}(E\hat{I}_4-\hat{R})^{-1}\right]
\end{eqnarray}

This integral can be evaluated in the usual manner by saddle point
method in the limit $N\gg 1$. The saddle point manifold is
parametrized again by the matrices $\hat{R}=\frac{E}{2}\hat{I}_4-
\pi\nu_{sc}\hat{Q}\equiv
\left[E\hat{I}_4-\hat{R}\right]^{-1}$. Remembering also that
$\mbox{Sdet}\hat{U}=\left(\frac{Z_J^{(1)}Z_J^{(2)}}{Z_{J=0}^
{(1)}Z_{J=0}^{(2)}}\right)$, we obtain the following representation
for the correlation function $f(z_1,z_2)$ in terms of the
integral over the graded coset space:
\begin{eqnarray}\label{sm22}
f(z_1,z_2)=\lim_{J_1\to 0,J_2\to 0}\displaystyle{
\frac{\partial^2}{\partial J_1\partial
J_2}}\int [d\hat{Q}]\prod_{a=1}^M\mbox{Sdet}^{-1}
\left[\hat{U}^{-1}-\gamma_a\left(\frac{E}{2}\hat{I}_4-
\pi\nu_{sc}\hat{Q}\right)\right]\times
\\ \nonumber
\exp{\left\{-\frac{\omega}{2}\mbox{Str}
\hat{Q}\hat{\Lambda}+\frac{x^2}{8}\mbox{Str}\hat{Q}\hat{\Lambda}
\hat{Q}\hat{\Lambda}\right\}}
\end{eqnarray}
where we introduced scaled variables: $\omega=\pi\nu_{sc} N\Omega;\quad
x=\pi\nu_{sc} X$.

Remembering the definition of the supermatrix $\hat{U}^{-1}$ and  
performing the expansion of the superdeterminants up to the second
order with respect to $J_1,J_2$ one finds:
\begin{equation}\label{sm23}
\begin{array}{c}
\lim_{J_1\to 0,J_2\to 0}\displaystyle{
\frac{\partial^2}{\partial J_1\partial
J_2}}\prod_{a=1}^M\mbox{Sdet}^{-1}
\left[\hat{U}^{-1}-\gamma_a\left(\frac{E}{2}\hat{I}_4-
\pi\nu_{sc}\hat{Q}\right)\right]=\\
\left(\prod_{a=1}^M \mbox{Sdet}^{-1}\hat{B}_a\right)
\left[\sum_{a=1}^M\mbox{Str}\left(\hat{B}_a^{-1}
\hat{C}_1\hat{B}_a^{-1}\hat{C}_1
\right)+\sum_{a,b=1}^M\mbox{Str}
\left(\hat{B}_a^{-1}\hat{C}_1\right)\mbox{Str}\left(\hat{B}_b^{-1}\hat{C}_2
\right)\right]
\end{array}
\end{equation}
where
\begin{equation}\label{sm24}
\hat{B}_a=\frac{z_1}{2}(\hat{I}_4+\hat{\Lambda})+
\frac{z_2}{2}(\hat{I}_4-\hat{\Lambda})-\gamma_a\left(\frac{E}{2}\hat{I}_4-
\pi\nu_{sc}\hat{Q}\right);\quad \hat{C}_1=\mbox{diag}(0,1,0,0);\,\,  
\hat{C}_2=\mbox{diag}(0,0,0,1)
\end{equation}
In order to evaluate the integral over the coset space explicitly
we substitute the corresponding expressions (see Appendix B,  
Eqs.(\ref{asm4}-\ref{asm7})) in Eq.(\ref{sm23})
and perform the Grassmannian integrations remembering that in the  
chosen parametrization , see Appendix B, a nonvanishing
contribution comes (apart
from the terms proportional to the combination
$\alpha^*\alpha\beta^*\beta$) also from terms in the integrand  
containing no Grasmannians at all (the so-called
Parisi-Sourlas-Efetov-Wegner (PSEW)theorem, see \cite{Zirn,my}). We  
therefore find:
\begin{eqnarray}\label{sm25}\nonumber
f(z_1,z_2)=\sum_{a=1}^M\frac{1}{z_2(a)+i\tilde{\gamma}_a}
\sum_{b=1}^M\frac{1}{z_1(b)-i\tilde{\gamma}_b}+\\
\int_{1}^{\infty}\int_{-1}^1
\frac{d\lambda_1d\lambda_2}{(\lambda_1-\lambda_2)^2}
\exp{\left[-i\omega(\lambda_1-\lambda_2)-
\frac{x^2}{2}(\lambda_1^2-\lambda_2^2)\right]}\prod_{a=1}^M
\frac{{\cal D}_f(a)}{{\cal D}_b(a)}\times\\
\nonumber\left[\sum_{a=1}^M\left\{\frac{z_2(a)+i\tilde{\gamma}_a\lambda_1}
{{\cal D}_b(a)}-\frac{z_2(a)+i\tilde{\gamma}_a\lambda_2}
{{\cal D}_f(a)}\right\}
\sum_{b=1}^M\left\{\frac{z_1(b)-i\tilde{\gamma}_b\lambda_1}
{{\cal D}_b(b)}-\frac{z_1(b)-i\tilde{\gamma}_b\lambda_2}
{{\cal D}_f(b)}\right\}+\right.\\ \nonumber
\left.\sum_{a=1}^M\left(\frac{\tilde{\gamma}_a^2
\mid\mu_1\mid^2}{{\cal D}^2_b(a)}+\frac{\tilde{\gamma}_a^2
\mid\mu_2\mid^2}{{\cal D}^2_f(a)}\right)\right]
\end{eqnarray}
where the notations ${\cal D}_{f,b}(a),\tilde{\gamma}_a,z_{1,2}(a)$
are explained in Appendix B.

It is easy to see that the expression in the first line of  
Eq.(\ref{sm25}) is just the so-called "disconnected" part  
$f^{-}(z_1)f^+(z_2)$ of the
corresponding correlation function, which is given by:
$$
\sum_{a=1}^M\frac{1}{z_2(a)+i\tilde{\gamma}_a}
\sum_{b=1}^M\frac{1}{z_1(b)-i\tilde{\gamma}_b}=
\left\langle
\mbox{Tr}\frac{1}{z_1-i\epsilon-\hat{K}_{-X/2}(E-\Omega/2)}\right\rangle
\left\langle\mbox{Tr}\frac{1}
{z_2+i\epsilon-\hat{K}_{X/2}(E+\Omega/2)}\right\rangle
$$

The "connected" part of the correlation function $f_c(z_1,z_2)=
f(z_1,z_2)-f^{-}(z_1)f^+(z_2)$ can be written in the most elegant form
by noticing that:
\begin{eqnarray}\label{sm26}
\frac{z_2(a)+i\tilde{\gamma}_a\lambda_1}
{{\cal D}_b(a)}-\frac{z_2(a)+i\tilde{\gamma}_a\lambda_2}
{{\cal D}_f(a)}=-\frac{\partial}{\partial z_1}\ln{
\frac{{\cal D}_{f}(a)}{{\cal D}_{b}(a)}}
\\ \nonumber
\frac{z_1(a)-i\tilde{\gamma}_a\lambda_1}
{{\cal D}_b(a)}-\frac{z_1(a)-i\tilde{\gamma}_a\lambda_2}
{{\cal D}_f(a)}=-\frac{\partial}{\partial z_2}
\ln{\frac{{\cal D}_f(a)}{{\cal D}_b(a)}}
\end{eqnarray}
and
\begin{equation}\label{sm27}
\frac{\tilde{\gamma}_a^2
\mid\mu_1\mid^2}{{\cal D}^2_b(a)}+\frac{\tilde{\gamma}_a^2
\mid\mu_2\mid^2}{{\cal D}^2_f(a)}=\frac{\partial^2}
{\partial z_1\partial z_2}\ln{\frac{{\cal D}_f(a)}{{\cal D}_b(a)}}
\end{equation}

Taking these relations into account one finally obtains
the following compact expression:
\begin{eqnarray}\label{sm28}
f_c(z_1,z_2)=\int_{1}^{\infty}\int_{-1}^1
\frac{d\lambda_1d\lambda_2}{(\lambda_1-\lambda_2)^2}
\exp{\left[-i\omega(\lambda_1-\lambda_2)-
\frac{x^2}{2}(\lambda_1^2-\lambda_2^2)\right]}\times
\\ \nonumber
\frac{\partial^2}{\partial z_1\partial z_2}\prod_{a=1}^M\left[
\frac{z_1(a)z_2(a)+i\tilde{\gamma}_a
\lambda_2(z_1(a)-z_2(a))+\tilde{\gamma}_a^2}
{z_1(a)z_2(a)+i\tilde{\gamma}_a
\lambda_1(z_1(a)-z_2(a))+\tilde{\gamma}_a^2}\right],
\end{eqnarray}
with $\tilde{\gamma}_a=\pi\nu_{sc}\gamma_a$ and $z_p(a)=z_p-\gamma_aE/2$.
This expression
 constitutes one of the central results of the present paper.
In the rest of the present section we are going to use this  
relation intensively
 for extracting statistical properties of scattering
phase shifts and their derivatives.

\subsection{Correlations of phase shift densities at fixed values of
energy $E$ and external parameter $X$}

 The  general expression eq.(\ref{sm28}) can be further simplified
in the particular case $\omega=x=0$. Physically this means that
we are interested in studying correlation of phase shift densities at
two points $\theta_1$ and $\theta_2$, but at fixed values
 of the energy $E$ and the external parameter $X$.
Let us further assume that all channels are equivalent
$\gamma_a=\gamma; \quad a=1,2,...,M$ for the sake of simplicity.
Introducing notations: $A=\tilde{z}_1\tilde{z}_2+\tilde{\gamma}_a^2;\quad
 B=\tilde{\gamma}_a\left(z_1-z_2\right)$
where $\tilde{z}_{1,2}=z_{1,2}-E\gamma/2$ we can write:
\begin{equation}\label{sm29}
f_c^{\omega=x=0}(z_1,z_2)=\int_{1}^{\infty}
d\lambda_1\frac{\partial^2}{\partial z_1\partial z_2}
\frac{1}{(A+iB\lambda_1)^M}\int_{-1}^1\frac{d\lambda_2}
{(\lambda_1-\lambda_2)^2}(A+iB\lambda_2)^M
\end{equation}
The integration over $\lambda_2$ can be easily performed yielding:
\begin{eqnarray}\label{sm30}
f_c^{\omega=x=0}(z_1,z_2)=\frac{\partial^2}{\partial z_1\partial z_2}
\sum_{l=1}^M\left(\begin{array}{c}M\\l\end{array}\right)\frac{(-iB)^l}{(l-1)}
\int_{1}^{\infty}\frac{d\lambda_1}{[A+iB\lambda_1]^l}
\left\{(\lambda_1+1)^{l-1}-
(\lambda_1-1)^{l-1}\right\}=\\ \nonumber
\frac{\partial^2}{\partial z_1\partial z_2}
\sum_{l=1}^M\left(\begin{array}{c}M\\l\end{array}\right)\frac{(-1)^l}{(l-1)}
\sum_{q=0}^{l-1}\left(\begin{array}{c}l-1\\q\end{array}\right)\frac{(-1)^q}{q}
\left[\left(\frac{A-iB}{A+iB}\right)^q-1\right]
\end{eqnarray}
where in the second line we expanded the brackets $(\lambda_1\pm  
1)^l$ and performed the remaining integration over $\lambda_1$  
explicitly in each term.

After differentiation over $z_1,z_2$ with help of the relations
$$
A\frac{\partial B}{\partial z_{1,2}}-B\frac{\partial A}{\partial
z_{1,2}} =\tilde{\gamma}(\tilde{\gamma}^2+\tilde{z}_2^2);\quad
A^2+B^2=(\tilde{\gamma}^2+\tilde{z}_1^2)(\tilde{\gamma}^2+\tilde{z}_2^2)
$$
the sums over $l,q$ can be performed explicitly as well. As the
result, we obtain:
\begin{equation}\label{sm31}
f_c^{\omega=x=0}(z_1,z_2)=-\frac{\tilde{\gamma}^2}{B^2}\left(\left[
\frac{A-iB}{A+iB}\right]^M-1\right)
\end{equation}

In this point it is convenient to pass
from the variables $\tilde{z}_1,\tilde{z}_2$ to new "angular"  
variables $\tilde{\theta}_1,\tilde{\theta}_2$ defined as:  
\begin{equation}\label{sm32}
\tilde{z}_1=\tilde{\gamma}\tan{\tilde{\theta}_1},\quad
\tilde{z}_2=\tilde{\gamma}\tan{\tilde{\theta}_2}
\end{equation}
We obviously have:
$$
B=\tilde{\gamma}^2\frac{\sin{(\tilde{\theta}_1-\tilde{\theta}_2)}}
{\cos{\tilde{\theta}_1}\cos{\tilde{\theta}_2}};\quad
\frac{A-iB}{A+iB}=\exp{-2i(\tilde{\theta}_1-\tilde{\theta}_2)}
$$
Remembering the relation Eq.(\ref{sm14}) between the spectral correlation
function ${\cal K}_{E,\Omega,X}(z_1,z_2)$ and $f(z_1,z_2)$we notice  
that the pair spectral correlation function of the
densities of the "angles" $\tilde{\theta}$ defined as:
\begin{eqnarray}\label{sm33}
{\cal K}(\tilde{\theta}_1,\tilde{\theta}_2)= \langle
\rho_E(\tilde{\theta}_1)\rho_E(\tilde{\theta}_2)\rangle- \langle
\rho_E(\tilde{\theta}_1)\rangle\langle\rho_E(\tilde{\theta}_2)\rangle\\  
\nonumber
\rho_E(\tilde{\theta})=\frac{1}{M}\sum_{a=1}^M\delta\left(\tilde{\theta}-
\arctan{\frac{1}{\pi\nu_{sc}}\left[\frac{z_a(E)}{\gamma}-E/2\right]}\right)
\end{eqnarray}
can be written in a very simple form:
\begin{equation}\label{sm34}
{\cal K}(\tilde{\theta_1},\tilde{\theta_2})\mid_{\tilde{\theta_1}\ne
\tilde{\theta_2}}=
-\left(\frac{\sin{M(\tilde{\theta_1}-\tilde{\theta_2})}}{\pi
M\sin{(\tilde{\theta_1}-\tilde{\theta_2})}}\right)^2
\end{equation}

One immediately recognizes in Eq.(\ref{sm34}) the pair correlation
function of the Dyson Circular Unitary Ensemble ( see the book
\cite{Mehta} where this object is called "two-level cluster
function"). It corresponds to the following joint probability
density of $M$ variables $\tilde{\theta}_a=
\arctan{\frac{1}{\pi\nu_{sc}}\left[\frac{z_a(E)}{\gamma}-E/2\right]}
\quad a=1,2,...,M$:
\begin{equation}\label{sm35}
{\cal P}_M(\tilde{\theta}_1,...,\tilde{\theta}_M)=const\times
\prod_{a<b}\mid  
e^{2i\tilde{\theta}_a}-e^{2i\tilde{\theta}_b}\mid^2\propto
\prod_{a<b} \sin^2{\left(\tilde{\theta}_a-\tilde{\theta}_b\right)}
\end{equation}
with $-\pi/2\le\tilde{\theta}_a<\pi/2.$ Assuming this probability
density being proven, the joint probability density of phase shifts  
$\theta_a$ related to "angles"
$\tilde{\theta}_a$ as
$\tan{\theta_a/2}=-\gamma\left(\tan{\tilde{\theta}_a}+\tilde{E}/2\right)$
, where $\tilde{E}=E/(\pi\nu_{sc})$ is given by:   
\begin{equation}\label{sm36}
{\cal P}_M(\theta_1,...,\theta_M)={\cal
P}_M(\tilde{\theta}_1,...,\tilde{\theta}_M)\prod_{a=1}^M
\left|\frac{d\tilde{\theta}_a}{d\theta_a}\right|;\quad
\left|\frac{d\tilde{\theta}_a}{d\theta_a}\right|=
\frac{1}{2\tilde{\gamma}}\frac{\cos^2{\tilde{\theta}_a}}{\cos^2{\theta_a/2}}
\end{equation}
On the other hand one can write:
$$
\sin^2{(\tilde{\theta}_a-\tilde{\theta_b})}=\cos^2{\tilde{\theta}_a}
\cos^2{\tilde{\theta}_b}(\tan{\tilde{\theta}_a}-\tan{\tilde{\theta}_b})=
\frac{\cos^2{\tilde{\theta}_a}}{\tilde{\gamma}\cos^2{\theta_a/2}}
\frac{\cos^2{\tilde{\theta}_b}}{\tilde{\gamma}\cos^2{\theta_b/2}}
\sin^2{(\theta_a/2-\theta_b/2)}
$$
so that using the identity: $\prod_{a<b}u_au_b=(\prod_au_a)^{m-1}$
one obtains:
\begin{equation}\label{sm37}
{\cal P}_M(\theta_1,...,\theta_M)\propto \frac{1}{\tilde{\gamma}^{M^2}}
\prod_{a<b}\sin^2{(\theta_a/2-\theta_b/2)}\left(\prod_c
\frac{\cos^2{\tilde{\theta}_c}}{\cos^2{\theta_c/2}}\right)^M
\end{equation}
Using the relation between $\theta_c$ and $\tilde{\theta}_c$
and definitions of the quantities $\tilde{E},\tilde{\gamma}$
one finds after a simple algebra that:
$$
\frac{\cos^2{\tilde{\theta}_c}}{\cos^2{\theta_c/2}}=\frac{2}
{\left(1+\frac{\tilde{E}^2}{4}+\frac{1}{\tilde{\gamma}^{2}}\right)+
e^{i\theta_c}\left[\frac{\tilde{E}}{2i\tilde{\gamma}}+\frac{1}{2}
\left(1+\frac{\tilde{E}^2}{4}-\frac{1}{\tilde{\gamma}^2}\right)\right]
+e^{-i\theta_c}\left[-\frac{\tilde{E}}{2i\tilde{\gamma}}+\frac{1}{2}
\left(1+\frac{\tilde{E}^2}{4}-\frac{1}{\tilde{\gamma}^2}\right)\right]}=
$$
$$
\frac{\left(\gamma^2(4-E^2)/\left[1+\gamma^2+\gamma\sqrt{4-E^2}\right]\right)}
{\left(1-\left\langle S\right\rangle^*e^{i\theta}\right)
\left(1-\left\langle S\right\rangle e^{-i\theta}\right)}
$$
where we made use of Eq.(\ref{sm12}). Thus, we arrive finally at  
the following expression:
\begin{equation}\label{sm38}
{\cal P}_M(\theta_1,...,\theta_M)\propto\prod_{a<b}\mid e^{i\theta_a}-
e^{i\theta_b}\mid^2\prod_{c=1}^M\mid 1-\langle
S\rangle^*e^{i\theta_c}\mid^{-2M}\end{equation}
which is nothing other but the Poisson's kernel distribution,  
Eq.(\ref{Pua}).
Here the phase shifts $\theta_a$ may be restricted to an interval
$0\le\theta_a<2\pi$. Inverting the argumentation, we prove, that our
correlation function, Eq.(\ref{sm34}) follows from the Poisson's kernel,
Eq.(\ref{sm38}); this we have shown in the case of equivalent channels.
\subsection{Distribution of partial delay times and parametric
derivatives of phase shifts}
The knowledge of the spectral correlation function Eq.(\ref{sm1})
allows one to determine the distribution ${\cal P}_{\tau}(\tau)$of  
partial delay times
\begin{equation}\label{sm39}
\tau_a(E)=\frac{\partial \theta_a(E)}{\partial
E}=-\frac{2}{1+z_a(E)^2}\frac{\partial z_a(E)}{\partial E};\quad
a=1,...,M\end{equation}The distribution ${\cal P}_{\tau}(\tau)$ can  
be easily found if one knows
the joint probability density  ${\cal P}_E(z,v)$ defined as:
\begin{equation}\label{sm40}
{\cal  
P}_E(z,v)=\frac{1}{M}\left\langle\sum_{a=1}^M\delta(z-z_a)\delta\left(v-
\frac{\partial z_a(E)}{\partial E}\right)\right\rangle
\end{equation}
because of the relation:
\begin{equation}\label{sm41}
{\cal P}_{\tau}(\tau)=\frac{1}{M}\left\langle\sum_{a=1}^M\delta
\left(\tau-\tau_a\right)\right\rangle=\int_{-\infty}^{\infty}
\int_{-\infty}^{\infty}dzdv{\cal
P}_E(z,v)\delta\left(\tau+\frac{2v}{1+z^2} \right)
\end{equation}
where angular brackets stand for the ensemble average as  
before\cite{notr1}.

To determine the joint probability density ${\cal P}_E(z,v)$
we use its relation to the spectral correlation function
${\cal K}_{E,\Omega,X}(z_1,z_2)$ defined in Eq.(\ref{sm1}):
\begin{equation}\label{sm41a}
{\cal P}_E(z,v)=M\lim_{\Omega\to +0} \Omega{\cal  
K}_{E,\Omega,X=0}(z_1=z-v\Omega/2,z_2=z+v\Omega/2)
\end{equation}
relations of this kind were first used in \cite{KZ} and later on in
\cite{Aleur,prl94,curv}. To understand its origin we write for  
small positive $\Omega$:
\begin{eqnarray}\label{sm42}\nonumber
\Omega{\cal K}_{E,\Omega,X=0}(z_1=z-v\Omega/2,z_2=z+v\Omega/2)=\\
\Omega\frac{1}{M^2}\left\langle\sum_{a,b=1}^M\delta
\left[z-v\Omega/2-z_a(E-\Omega/2)\right]
\delta\left[z+v\Omega/2-z_b(E+\Omega/2)\right]
\right\rangle=\\ \nonumber
\Omega\frac{1}{M^2}\left\langle\sum_{a,b=1}^M\delta
\left[z-v\Omega/2+z_a(E-\Omega/2)\right]
\delta\left[v\Omega+z_a(E-\Omega/2)-z_b(E+\Omega/2)\right]\right\rangle
\end{eqnarray}
Expanding $z_a(E-\Omega/2)-z_b(E+\Omega/2)$
at small $\Omega$ as $z_a(E)-z_b(E)-\frac{\Omega}{2}
\left(\frac{\partial z_a(E)}{\partial E}+\frac{\partial
z_b(E)}{\partial E} \right)+...$ we immediately see that in the
limit $\Omega\to +0$ a nonvanishing contribution to Eq.(\ref{sm42})  
comes from the terms with equal channel indices
$a=b$ and the resulting expression is equivalent to  Eq.(\ref{sm41}).

To perform the limit $\Omega\to +0$ in the most economic manner
we pass from the variables $z_1,z_2$ to $z=(z_1+z_2)/2;\quad
v_s=(z_2-z_1)/2\omega$, with $v_s,\omega$ being the scaled
variables $v_s=v\Delta/2\pi;\quad \omega=\pi\Omega/\Delta$.
Correspondingly, the correlation function $f_c(z_1,z_2)$ acquires
the form:
\begin{eqnarray}\label{sm43}
f(z,v_s)=
\int_1^{\infty}d\lambda_1\int_{-1}^1 d\lambda_2
\exp{-i\omega(\lambda_1-\lambda_2)}\frac{1}{(\lambda_1-\lambda_2)^2}
\times
\\ \nonumber
\left[\frac{\partial^2}{\partial  
z^2}-\frac{1}{\omega^2}\frac{\partial^2}{\partial  
v_s^2}\right]\prod_{a=1}^M\displaystyle{
\left[\frac{z^2-\gamma_aEz+\gamma_a^2-2i\pi\nu_{sc}\gamma_a\omega
\lambda_2v_s-v_s^2\omega^2}
{z^2-\gamma_aEz+\gamma_a^2-2i\pi\nu_{sc}\gamma_a\omega
\lambda_1v_s-v_s^2\omega^2}\right]}
\end{eqnarray}

According to the general discussion presented above we should take
the real part of this expression and look for the term proportional
to $1/\omega$ at $\omega\to +0$. It is easy to understand that
such a singularity comes from that part of the integration region  
over the
"non-compact" variable $\lambda_1$, where
$\lambda_1\propto\omega^{-1}$. After a natural
rescaling $\lambda_1=t/\omega$ one can easily extract the
corresponding singular term, which turns out to come from the
term with the second derivative  
$\frac{1}{\omega^2}\frac{\partial^2}{\partial v_s^2}$ in the  
expression above.
Performing the calculation explicitly, we find the expression
for the joint probability density of variables $z$ and $v_s$
\begin{equation}\label{sm44}
{\cal P}_E(z,v_s)=-\frac{1}{2M\pi^2}
\displaystyle{\int_{-\infty}^{\infty}\frac{dt}{t^2}
e^{-it}\frac{\partial^2}{\partial v_s^2}\prod_{a=1}^M\frac{1}
{1-itv_s r_a^{-1}}}=-\frac{1}{2M\pi^2}\displaystyle{
\frac{1}{ v_s^3}
\int_{-\infty}^{\infty}dt
e^{itv_s^{-1}}\prod_{a=1}^M\frac{1}
{1-itr_a^{-1}}}
\end{equation}
where $r_a=\left[z^2-\gamma_aEz+\gamma_a^2\right]/2\pi\nu_{sc}\gamma_a$
and we made use of the identity:  
$\mbox{Re}\int_{0}^{\infty}dtf(it)=\frac{1}{2}
\int_{-\infty}^{\infty}dt f(it)$.

Performing the integration in the expression above (cf.  
Eq.(\ref{calf1})) one obtains:
\begin{eqnarray}\label{sm45}
{\cal P}_E(z,v_s)=\frac{(-1)^{M-1}}{\pi Mv_s^3}\prod_{a=1}^M\left[\frac
{z^2-\gamma_a
Ez+\gamma_a^2}{2\pi\nu_{sc}\gamma_a}\right]\theta(-v_s)\times
\\ \nonumber \sum_{b=1}^M
\exp{-\left[\frac
{z^2-\gamma_b Ez+\gamma_b^2}{2\pi\nu_{sc}\gamma_bv_s}\right]}
\prod_{c\ne b}\left[\frac
{z^2-\gamma_b Ez+\gamma_b^2}{2\pi\nu_{sc}\gamma_b}-\frac
{z^2-\gamma_c Ez+\gamma_c^2}{2\pi\nu_{sc}\gamma_c}\right]^{-1}
\end{eqnarray}
where $\theta(x)=1$ if $x\ge 0$ and $\theta(x)=0$ otherwise. We   
took into account that $z^2-\gamma_a Ez+\gamma_a^2\ge
 0$ as long as $\mid E\mid\le 2$, which is just the case
we are interested in ( we remind that the semicircle density
$\nu_{sc}(E)$ is non-vanishing for $\mid E\mid< 2$).

Substituting the expression Eq.(\ref{sm45}) into Eq.(\ref{sm41})
one trivially performs the integration over $v$ because of the  
$\delta-$ function and obtains for the distribution function
of scaled partial delay times $\tau_s=\frac{\tau\Delta}{2\pi}$
the following expression:
\begin{eqnarray}\label{sm46}\nonumber
{\cal P}_{\tau}(\tau_s)=\frac{(-1)^{M-1}}{2\pi  
M\tau_s^3}\int_{-\infty}^{\infty}
\frac{2dz}{(1+z^2)}\prod_{a=1}^M
\left[\frac
{z^2-\gamma_a Ez+\gamma_a^2}{\pi\nu_{sc}\gamma_a(1+z^2)}\right]
\sum_{b=1}^M\exp{-\left[\frac
{z^2-\gamma_b  
Ez+\gamma_b^2}{\pi\nu_{sc}\gamma_b\tau_s(1+z^2)}\right]}\times
\\ \prod_{c\ne b}\left[\frac
{z^2-\gamma_b Ez+\gamma_b^2}{2\pi\nu_{sc}\gamma_b(1+z^2)}-\frac
{z^2-\gamma_c
Ez+\gamma_c^2}{2\pi\nu_{sc}\gamma_c(1+z^2)}\right]^{-1}=
\\ \nonumber \frac{(-1)^{M-1}}{2\pi M\tau_s^3}\int_{-\pi}^{\pi}d\phi
\left[\prod_{a=1}^M R_a(\phi)\right]
\sum_{b=1}^M\exp{-\frac{R_b(\phi)}{\tau_s}}
\prod_{c\ne b}\left[R_b(\phi)-R_c(\phi)\right]^{-1}
\end{eqnarray}
Here$$R_a(\phi)=\displaystyle{\frac{(1+\gamma_a^2)-\gamma_a
E\sin{\phi}+(\gamma_a^2-1)\cos{\phi}}{2\pi\nu_{sc}\gamma_a}}
$$
and we changed the integration variable: $z=\tan{(\phi/2)}$.

The expression Eq.(\ref{sm46}) provides the distribution
of partial delay times for a general case of non-equivalent
channels and thus constitutes one of the most important results of
the present subsection.

Further simplifications are possible if we restrict our attention to the
particular case of equivalent channels $\gamma_a=\gamma;\quad  
a=1,2,...,M$:
\begin{eqnarray}\label{sm47}
{\cal P}_{\tau}(\tau_s)=\frac{1}{2\pi  
M!\tau_s^{M-1}}\int_{-\pi}^{\pi}d\phi
\left[\frac{(1+\gamma^2)-\gamma
E\sin{\phi}+(\gamma^2-1)\cos{\phi}}{2\pi\nu_{sc}\gamma}\right]^M\times\\
\nonumber\exp{-\frac{1}{2\pi\nu_{sc}\gamma\tau_s}\left[(1+\gamma^2)-\gamma
E\sin{\phi}+(\gamma^2-1)\cos{\phi}\right]}
\end{eqnarray}

The last expression can be put in a more elegant form upon using  
the identity:
$$
\int_{-\pi}^{\pi}f(p\cos{\phi}+q\sin{\phi})d\phi=2
\int_{0}^{\pi}f(\sqrt{p^2+q^2}\cos{\phi})d\phi
$$
and introducing the quantity $g=(\gamma+\gamma^{-1})/2\pi\nu_{sc}$
(related to the transmission coefficient as $g=2T^{-1}-1$, see  
Eq.(\ref{defg})),
so that $\frac{1}{2\pi\rho\gamma}\sqrt{(\gamma^2-1)^2+(\gamma E)^2}=
\sqrt{g^2-1}$. This finally gives:
\begin{eqnarray}\label{sm48}
{\cal P}_{\tau}(\tau_s)=\frac{1}{\pi M!\tau_s^{M+2}}
\int_0^{\pi}d\phi\left[g+\sqrt{g^2-1}\cos{\phi}\right]^M
\exp{\left\{-\frac{1}{\tau_s}\left[g+\sqrt{g^2-1}\cos{\phi}\right]\right\}}=\\
\nonumber
\frac{(-1)^M}{M!\tau_s^{M+2}}
\frac{\partial^M}{\partial\left(\tau_s^{-1}\right)^M}
\displaystyle{ \left[e^{-g\tau_s^{-1}}I_0
\left(\tau_s^{-1}\sqrt{g^2-1}\right)\right]}
\end{eqnarray}
where
$I_0(z)$ stands for the modified Bessel function.
This expression provides us with the explicit form of the
distribution of (scaled) partial delay times for the case
of equivalent channels. Below we
briefly analyze its most important features.

First of all, we notice that the distribution above assumes the  
simplest form for the
"critical" coupling $T=1$ (i.e. $g=1$)
 corresponding to the most strong overlap of
individual resonances allowed for the few-channel scattering, see
preceding sections. Under this condition one finds the following
distribution of scaled partial delay times:\begin{equation}\label{sm49}
{\cal P}_{\tau}(\tau_s)=\frac{1}{M!}\tau_s^{-M-2}e^{-1/\tau_s}
\end{equation}
The powerlaw tail $\tau_s^{-M-2}$ at $\tau_s\gg g$ which is
evident from the expression above for $g=1$ is actually a typical feature
of the time delay distribution for any values of the parameters
$g_a; a=1,...,M$ (see the discussion of Wigner-Smith
time delay in the next subsection).
  For the equivalent channels we obviously have:
\begin{equation}\label{sm50}
{\cal P}_\tau(\tau_s\gg g)=\frac{1}{M!}\tau_s^{-M-2}P_M(g)
\end{equation}
where $P_M(g)$ stands for the Legendre polynomial\cite{RG}. For the  
case of non-equivalent channels the asymptotic behavior  
$\tau_s^{-M-2}$ can be infered from Eq.(\ref{sm46}) upon noticing
that: $A_k=\sum_{b=1}^M R_b^k\prod_{c\ne b}(R_c-R_b)^{-1}\equiv 0$ for
$k=0,1,...,M-2$ and $A_k\ne 0$ for $k\ge M-1$, so that the integrand
is proportional to $1/\tau_s^{M-1}$. Combined with the factor
$\tau_s^{-3}$ in front of the integral it gives the desired behavior.

In Fig.3 we plotted the distribution Eq.(\ref{sm48}) for the  
opposite case of weakly open systems (the regime of isolated
resonances: $T\ll 1$, i.e. $g\gg 1$). Under this condition when   
$g\gg\tau_s$
 the modified Bessel function
can be replaced by its asymptotic expression valid for large
arguments. Taking into account also that $g-\sqrt{g^2-1}\approx
\frac{1}{2g}$ at $g\gg 1$, we find that the distribution function  
Eq.(\ref{sm48}) simplifies to the following form:
\begin{equation}\label{sm51}
{\cal P}_{\tau}(\tau_s)=\frac{(-1)^M}{M!}\tau_s^{-M-2}
\frac{\partial^M}{\partial\left(\tau_s^{-1}\right)^M}
\left[e^{-(2g\tau_s)^{-1}}\frac{1}{\sqrt{2\pi g\tau_s^{-1}}}\right]
\end{equation}
This expression is correctly normalized and plays the same role
for the distribution of partial delay times as that played
by the $\chi^2$ distribution in the issue of the resonance width
distribution. It is necessary to mention that Eq.(\ref{sm51})
is valid as long as $1,\tau_s\ll g$. At larger
values of $\tau_s$ the behavior changes to that given by
Eq.(\ref{sm50}).
It is interesting to note that in the parametrically large
region $(2g)^{-1}\ll 1, \tau_s\ll g$ one can neglect the exponential
term in Eq.(\ref{sm51}) and reduce this distribution to the
following form:
\begin{equation}\label{sm52}
 {\cal P}_\tau(\tau_s)=\frac{(2M-1)!!}{2^M M!}\frac{1}{\sqrt{2\pi g}}
\tau_s^{-3/2}
\end{equation}
This $\tau_s^{-3/2}$ behavior taking place irrespectively of the
number of open channels is therefore the most typical feature
of the partial delay times distribution for the
 regime of isolated resonances.
The origin of such a behavior can be understood analysing the
general expression for the Wigner-Smith delay times,
Eq.(\ref{wigtime2}) ( see a more detailed discussion after the
equation Eq.(\ref{w16})).

 At $\tau_s\sim (2g)^{-1}$
the distribution shows a maximum at a value ${\cal P}(\tau_s)\sim g$
and then is cut off exponentially at smaller $\tau_s$:
\begin{equation}\label{sm52a}
{\cal P}_\tau\left(\tau_s\ll (2g)^{-1}\right)=
\frac{2g}{\pi^{1/2}M!}\frac{1}
{(2g\tau_s)^{(M+3/2)}}
\exp{-\left[\frac{1}{2g\tau_s}\right]}
\end{equation}

All these features are evident from Fig.3.

Having at our disposal the exact
distribution eq.(\ref{sm48}) it is instructive to calculate the mean
value and the variance of the partial delay times. One
finds:
\begin{equation}\label{sm53}
\langle\tau\rangle=\frac{2\pi}{M\Delta};\quad \frac{\langle
\tau^2\rangle-\langle \tau\rangle^2}{\langle \tau\rangle^2}=\frac{2M
(T^{-1}-1)+1}{M-1}
\end{equation}

The first of these relations is
quite well known \cite{Gasparian,Harney,Lyub,Bauer,Lehmann}.
 It shows that the mean delay
time $\langle \tau\rangle$ is determined
by the mean level spacing
$\Delta$ of the closed system and the number $M$ of open channels. On
the other hand the magnitude of delay time fluctuations measured by
the relative variance of the partial delay times distribution, see
eq.(\ref{sm53}), is determined both by $M$ and $T$. Generically, the
fluctuations are the weaker the larger is the number of open channels
$M$ and the stronger is the coupling to continua: $1-T\ll T$. Let us
also mention as an interesting feature the divergency of the
time-delay variance at $M=1$, which is a consequence of the powerlaw
tail $\tau_s^{-M-2}$ typical for the distribution ${\cal
P}_{\tau}(\tau_s)$.

Here it is appropriate to mention that recently two other groups of  
authors \cite{Gopar,SZZ} addressed
the question of delay time distribution by different
approaches. Gopar, Mello and B\"{u}ttiker \cite{Gopar} verified  
numerically
 an old conjecture by Wigner \cite{Wigcon}
concerning invariance of poles and residues of the $K-$matrix under
a certain set of transformations, provided there is only one
 perfectly open channel: $M=1;\,T=1$ case in our notations.
When combined with the $\chi^2$ distribution of residues, this
conjecture was shown to produce the time
delay distribution for all three symmetry classes (
orthogonal, unitary and symplectic), the result for unitary class just
coinciding with Eq.(\ref{sm49}) for $M=1$.
In the paper \cite{SZZ} Seba, Zyczkowsky and Zakrzewsky arrived at  
Eq.(\ref{sm49}) for arbitrary $M$ after a set of shrewd,
but uncontrolled manipulations with eigenvalues and eigenfunctions
of $K-$matrix.
 Actually, these authors
suggested the following general expression for the scaled partial  
delay time
distribution claimed to be valid for arbitrary value of the  
coupling constant $\gamma$ and $E=0$:
\begin{equation}\label{szz}
{\cal P}_{szz}(\tau)=\frac{e^{-\gamma/\tau}}{\gamma^{M-1}M!
\tau^{M+2}}\,_1F_1\left[M,M+1,\left(\gamma-\gamma^{-1})/\tau\right)\right]
\end{equation}
where $_1F_1[M,M+1,z]$ is the confluent hypergeometric function.

This expression is quite different from ours given in  
Eq.(\ref{sm48}), the
two formulas coinciding in the limit $\gamma=1$ only.
However, one can
check that the expression Eq.(\ref{szz}) fails to fulfil the following
important condition: the mean delay time must be
independent of the degree of coupling to continua, measured by
$\gamma$. Instead, it should be determined by the mean level  
spacing $\Delta$.
This requirement is satisfied by our distribution Eq.(\ref{sm48}), see
Eq.(\ref{sm53}) and is known for a long time  
\cite{Gasparian,Harney,Lyub,Bauer,Lehmann}. It
 follows from the basic formula (\ref{wigtime2}):
\begin{equation}\label{test}
\langle \tau\rangle=\frac{2N}{M}\int d\Gamma\int d\omega
\rho(E-\omega;\Gamma)\frac{\Gamma/2}{\omega^2+\Gamma^2/4}
\end{equation}
where $\rho(E;\Gamma)\equiv\frac{1}{N}\langle\sum_{n=1}^N\delta(E-E_n)
\delta(\Gamma-\Gamma_n)\rangle$ is the density of the
$S-$ matrix poles. For few-channel case $M\ll N$ the typical scale
of the width $\Gamma$ is the mean
level spacing $\Delta=1/\left(N\rho(E)\right)$, see Sec.III,
so that the Lorentzian factor in the integrand of Eq.(\ref{test})
 can be replaced by $2\pi\delta(\omega)$ when evaluating the integral.
 This gives $\langle\tau\rangle=2\pi N\rho(E)/M$
in full agreement with Eq.(\ref{sm53}). At the same time,  for the  
particular case $M=1$ one can find that the first moment
corresponding to the distribution Eq.(\ref{szz}) is given by:
$\frac{\Delta}{2\pi}\langle\tau\rangle_{szz}=2\ln{\gamma}/\left(\gamma-\gamma^{-1}\right)$
in contradiction with the general discussion above.   This failure  
rules out the distributon Eq.(\ref{szz}) as the
 correct one and shows that the assumptions made in \cite{SZZ}
are justified only as long as $T=1$.

The distribution ${\cal P}_w(w)$ of parametric derivatives of phase  
shifts
$w_a=\frac{\partial \theta_a}{\partial X}$ can be found in a very
similar way. Proceeding in the same manner as in
Eqs.(\ref{sm40}-\ref{sm42}) one obtains:
\begin{equation}\label{sm54}
{\cal P}_w(w)=\frac{1}{M}\left\langle\sum_{a=1}^M\delta
\left(w-w_a\right)\right\rangle=\int_{-\infty}^{\infty}
\int_{-\infty}^{\infty}dzdu{\cal
P}_X(z,u)\delta\left(w+\frac{2u}{1+z^2} \right)
\end{equation} where
\begin{eqnarray}\label{sm55}
{\cal P}_X(z,u)=\frac{1}{M}\left\langle
\sum_{a=1}^M\delta(z-z_a)\delta\left(u-
\frac{\partial z_a(E,X)}{\partial X}\right)\right\rangle= \\ \nonumber
M\lim_{X\to 0} X
{\cal K}_{E,\Omega=0,X}(z_1=z-uX/2,z_2=z+uX/2)
\end{eqnarray}
Performing the limiting procedure $X\to 0$ in the same way as
$\Omega\to 0$ in Eqs.(\ref{sm41}-\ref{sm43}) we arrive at the
expression:
\begin{eqnarray}\label{sm56}
{\cal P}_X(z,u_s)=-\frac{1}{2M\pi^2}\displaystyle{
\int_{-\infty}^{\infty}\frac{dt}{t^2}
e^{-t^2/2}\frac{\partial^2}{\partial u_s^2}
\prod_{a=1}^M\frac{1}
{1-itu_s r_a^{-1}}}=\\ \nonumber  
\int_{-\infty}^{\infty}\frac{d\xi}{(2\pi)^{1/2}}\exp{-\xi^2/2}
\left[-\frac{1}{2M\pi^2}\displaystyle{
\int_{-\infty}^{\infty}\frac{dt}{t^2}
e^{-it\xi}\frac{\partial^2}{\partial u_s^2}\prod_{a=1}^M\frac{1}
{1-itu_s r_a^{-1}}}\right]
\end{eqnarray}
which can be written as:
\begin{equation}\label{sm56a}
{\cal P}_X(z,u_s)=
\int_{-\infty}^{\infty}\frac{d\xi}{(2\pi)^{1/2}\mid\xi\mid}
e^{-\xi^2/2}{\cal P}_E(z,u_s/\xi)
\end{equation}
where $u_s=u/(\pi\nu_{sc})$ and ${\cal P}_E(z,v_s)$ is the joint
probability density of $z$ and its derivative over the energy studied
earlier in this subsection. This fact means that a similar
relation Eq.(\ref{sm56a}) holds for the distribution function
${\cal P}_w(w_s)$
of scaled parametric derivatives  
$w_s=\frac{1}{2\pi\nu_{sc}}\frac{\partial \theta_a}{\partial X}$
and that of the scaled partial delay times ${\cal P}_{\tau}(\tau_s)$:
\begin{equation}\label{sm56b}
{\cal P}_w(w_s)=
\int_{0}^{\infty}\frac{d\xi}{(2\pi)^{1/2}\xi}
e^{-\xi^2/2}{\cal P}_{\tau}(\mid w_s\mid/\xi)
\end{equation}

The same relation was obtained in the paper \cite{SZZ} on a basis
of some plausible assumptions concerning parametric derivatives of
phase shifts.

\subsection{Parametric correlations of Wigner-Smith time delays}
 The expression for the correlation function
Eq.(\ref{sm28}) can be used to calculate the parametric correlations
of Wigner-Smith time delays $\tau_w(E,X)$ defined as
\begin{equation}\label{w1}
\tau_w(E,X)=-\frac{i}{M}\frac{\partial}{\partial E}
\ln{\mbox{det} S(E,X)}\equiv\frac{1}{M}\frac{\partial}{\partial E}
\sum_a\theta_a(E,X).
\end{equation}
To show this, we should remember
 the relation Eq.(\ref{totsh}) between the total phase shift
$\theta=\sum_a\theta_a$, the exact density of states for the {\it closed}
chaotic system $\nu_X(E)$ and the eigenvalues $z_a$ of the  
$K-$matrix. We see that
\begin{equation}\label{w2}
\tau_w(E,X)=\frac{2\pi N}{M}\nu_X(E)+\tau_z(E,X);\quad
\tau_z(E,X)=-2\frac{\partial}{\partial  
E}\frac{1}{M}\sum_{a=1}^M\arctan{z_a(E,X)}
\end{equation}Introducing the correlation function:
$$
C_W(\Omega,X)=\left\langle\tau_w(E-\Omega/2,-X/2)\tau_w(E+\Omega/2,X/2)
\right\rangle
$$
we see that it consists of three essentially different contributions:
\begin{equation}\label{w3}
C_W(\Omega,X)=C_{\tau\tau}(\Omega,X)+\frac{2\pi N}{M}
\left(C_{\nu\tau}(-\Omega,-X)+
C_{\nu\tau}(\Omega,X)\right)+(2\pi N/M)^2 C_{\nu\nu}(\Omega,X)
\end{equation}
where
\begin{equation}\label{w4a}
C_{\tau\tau}(\Omega,X)=
\left\langle\tau_z(E-\Omega/2,-X/2)\tau_z(E+\Omega/2,X/2)
\right\rangle
\end{equation}
\begin{equation}\label{w4b}
C_{\nu\tau}(\Omega,X)=
\left\langle\nu(E-\Omega/2;-X/2)\tau_z(E+\Omega/2,X/2)
\right\rangle
\end{equation}
\begin{equation}\label{w4c}
 C_{\nu\nu}(\Omega,X)=
\left\langle\nu(E-\Omega/2,-X/2)\nu(E+\Omega/2,X/2)
\right\rangle
\end{equation}
In what follows we are interested, as usual, in finding the "connected"
 part of all these correlation functions.This will be implicitly  
assumed below.
The correlation function $C_{\tau\tau}(\Omega,X)$ can be easily
related to that given by Eq.(\ref{sm28}) because of the relation:
$$
\tau_z(E,X)=-2\frac{\partial}{\partial E}\int_{-\infty}^{\infty}dz
\arctan{(z)}\,\rho_{E,X}(z)
$$
where $\rho_{E,X}(z)$ is the density of $K-$matrix eigenvalues
defined in the beginning of this section. As a result we have:
\begin{equation}\label{w5}
C_{\tau\tau}(\Omega,X)=\left[\frac{\partial^2}{\partial E^2}
-\frac{4\pi^2}{\Delta^2}\frac{\partial^2}{\partial \omega^2}\right]
\frac{1}{2\pi^2M^2}\mbox{Re}\int_{-\infty}^{\infty}dz_1
\arctan{z_1}\int_{-\infty}^{\infty}dz_2\arctan{z_2}
f_c(z_1,z_2)
\end{equation}
where we used the relation Eq.(\ref{sm14}) between the correlation  
function
${\cal K}_{x,\omega}(z_1,z_2)$ and that given by the Eq.(\ref{sm28})
and we used the scaled variable $\omega=\pi\Omega/\Delta$.
After such a rescaling it is obvious that the term containing the
second derivative $\frac{\partial^2}{\partial E^2}$ can be neglected
in comparison with the second one because of the large factor
$\Delta^{-2}$.
Substituting now the expression Eq.(\ref{sm28}) into eq.(\ref{w5})
one can easily perform the integration over $z_1,z_2$ by exploiting  
the presence of the second derivative $\partial^2/\partial z_1
\partial z_2$ in the function $f_c(z_1,z_2)$
 (this allows to convert factors
$\arctan{z_{1,2}}$ into $(1+z_{1,2}^2)^{-1}$ by partial
integrations) and noticing that
all poles of the expression
$$
\prod_{a=1}^M\left[
\frac{z_1(a)z_2(a)+i\tilde{\gamma}_a
\lambda_2(z_1(a)-z_2(a))+\tilde{\gamma}_a^2}
{z_1(a)z_2(a)+i\tilde{\gamma}_a
\lambda_1(z_1(a)-z_2(a))+\tilde{\gamma}_a^2}\right]
$$
lie in the {\it upper} half plane Im$z_1>0$ with respect to the variable
$z_1$ and in the {\it lower} half plane Im$z_1<0$ with respect to
the variable $z_2$. As a result, the integration can be performed
trivially by closing the integration contour over $z_1 (z_2)$
 in the lower (upper) half plane, correspondingly, and amounts to  
replacing $z_1=-i;z_2=i$ in the integrand. This gives:
\begin{equation}\label{w6}
C_{\tau\tau}(\omega,X)=2\left(\frac{\pi}{M\Delta}\right)^2\mbox{Re}
\int_{-1}^1d\lambda_2\int_{1}^{\infty}d\lambda_1
\exp\{i\omega(\lambda_1-\lambda_2)-\frac{x^2}{2}
(\lambda_1^2-\lambda_2^2)\}I(\lambda_1,\lambda_2)
\end{equation}
where
\begin{equation}\label{w7}
I(\lambda_1,\lambda_2)=1-\left\{\prod_{a=1}^M\left(\frac{1+\tilde{\gamma_a}
\lambda_2-i\gamma_aE/2}{1+\tilde{\gamma_a}
\lambda_1-i\gamma_aE/2}\right)+\prod_{a=1}^M
\left(\frac{1+\tilde{\gamma_a}
\lambda_2+i\gamma_aE/2}{1+\tilde{\gamma_a}
\lambda_1+i\gamma_aE/2}\right)\right\}+
\prod_{a=1}^M
\left(\frac{1+2\tilde{\gamma_a}
\lambda_2+\gamma_a^2}{1+2\tilde{\gamma_a}
\lambda_1+\gamma_a^2}\right)
\end{equation}
The first three terms are boundary contributions due to the partial
$z_1,z_2$ integrations.

Let us now show how to calculate the correlation function
$C_{\nu\tau}(\Omega,X)$ which is expressed to the leading order
in $N\gg 1$ as follows:
\begin{equation}\label{w8}
C_{\nu\tau}(\Omega,X)=\frac{\partial}{\partial\Omega}
\int_{-\infty}^{\infty}dz\arctan{z}\left\langle\nu_{-X/2}(E-\Omega/2)
\rho_{E+\Omega/2,X/2}(z)\right\rangle
\end{equation}

 To this end we represent the density
$\nu_X(E)$ in terms of the corresponding resolvent as  
$\nu_X(E)=\frac{1}{\pi N}\mbox{Im Tr}(E-i0^{+}-\hat{H}_{in})^{-1}$  
so that
\begin{eqnarray}\label{w9a}
{\cal K}_{\nu\tau}(\Omega,X;z)\equiv\left\langle\nu_{-X/2}(E-\Omega/2)
\rho_{E+\Omega/2,X/2}(z)\right\rangle=
\\ \nonumber
\frac{1}{2\pi^2MN}\mbox{Re}\left\langle
\mbox{Tr}\frac{1}{E-\Omega/2-\hat{H}(-X/2)-i0^{+}}
\mbox{Tr}\frac{1}{z-\hat{K}_(\Omega,X/2)+i0^{+}}\right\rangle
\end{eqnarray}
The calculation of the connected part of the averaged product of  
two traces of resolvents
in the preceding equation (we denote this quantity henceforth as
$f^{\nu}_c(z)$ omiting an explicit mentioning of the parameters
$\Omega,X$)
goes along exactly the same lines as the calculation of the
correlation function $f_c(z_1,z_2)$, see Eq.(\ref{sm13}).
Namely, one writes this function as:

\begin{equation}\label{w9}
\begin{array}{c}f^{\nu}_c(z)=\displaystyle{
\frac{\partial^2}{\partial J_1\partial
J_2}\left[\left(\frac{Z_J^{(2)}}{Z_{J=0}^
{(2)}}\right)^M
{\cal F}_{\nu}(J_1,J_2)\right]\mid_{J_1=J_2=0}} \\\\\displaystyle{ {\cal
F}_{\nu}(J_1,J_2)=\left\langle\frac{\mbox{
Det}[E-i0^{+}-\Omega/2-\hat{H}(-X/2)+J_1]
\mbox{Det}[E+\Omega/2-H_{\nu}(X/2;Z_J^{(2)})]}
{\mbox{Det}[E-i0^{+}-\Omega/2-\hat{H}(-X/2)]
\mbox{Det}[E+\Omega/2-H_{\nu}(X/2;Z_{J=0}^
{(2)})]}\right\rangle}
\end{array}
\end{equation}
 where we introduced the notations:
$Z_J^{(2)}=z+i0^{+}+J_2$ and
$H_{\nu}(X;Z_J^{(2)})=\hat{H}_{in}+\frac{X}{N^{1/2}}\hat{H}_{in}^{(1)}
+\frac{\pi}{Z_J^{(2)}}WW^{\dagger}$. The generating function
${\cal F}_{\nu}(J_1,J_2)$ is expressed in a standard way as a
Gaussian superintegral. Finally, the function $f^{\nu}_c(z)$
 is reduced (after exactly the same
manipulations as before, see Eqs.(\ref{sm16}-\ref{sm22})) to
the following representation in terms of the nonlinear $\sigma$-model:
\begin{eqnarray}\label{w10}
f^{\nu}_c(z)=\frac{\pi}{\Delta}\int[d\hat{Q}]\left[
E/2\hat{I}_4-\pi\nu\hat{Q}\right]_{ff}^{(11)}
\left(\prod_{a=1}^M\mbox{Sdet}^{-1}
\hat{B}_{\nu}(a)\right)\sum_{a=1}^M
\left(B_{\nu}^{-1}(a)\right)_{ff}^{(22)}\times
\\ \nonumber
\exp{\left\{-\frac{\omega}{2}\mbox{Str}
\hat{Q}\hat{\Lambda}+\frac{x^2}{8}\mbox{Str}\hat{Q}\hat{\Lambda}
\hat{Q}\hat{\Lambda}\right\}}
\end{eqnarray}
where the supermatrix $\hat{B}_{\nu}(a)$ is given by
$$
\hat{B}_{\nu}(a)=\displaystyle{\frac{1+\hat{\Lambda}}{2}+
\frac{1-\hat{\Lambda}}{2}\left(z(a)\hat{I}_4+\tilde{\gamma}_a\hat{Q}\right)}
$$
and we used the conventions of Appendix B for matrix elements
 as well as the notations: $z(a)=z-\gamma_a E/2;
\quad \tilde{\gamma}_a=\pi\nu_{sc}\gamma_a$. The explicit  
expressions for all matrix elements are given in Appendix B.  
Substituting them into the superintegral Eq.(\ref{w10})
we find again, that the term containing no Grassmannians at all gives
the contribution
$$
N(E/2+i\pi\nu)\sum_{a=1}^M\frac{1}{z(a)+i\tilde{\gamma}_a}
$$
which is exactly the "disconnected part" of the corresponding
correlation function. The connected part is given by:
\begin{equation}\label{w11}
f^{\nu}_c(z)=-i\frac{\pi^2}{\Delta^2}
\int_{-1}^1d\lambda_2\int_{1}^{\infty}d\lambda_1
\frac{1}{\lambda_1-\lambda_2}
\exp\{i\omega(\lambda_1-\lambda_2)-\frac{x^2}{2}(\lambda_1^2-\lambda_2^2)\}\frac{\partial}{\partial
z}\prod_{a=1}^M\frac{z(a)+i\tilde{\gamma}\lambda_2}
{z(a)+i\tilde{\gamma}\lambda_1}
\end{equation}
Substituting this expression into Eq.(\ref{w8}) one can again
trivially perform the integration over the variable $z$. As a  
result one obtains:
\begin{eqnarray}\label{w12}
C_{\nu\tau}(\omega,X)=-\frac{1}{2\pi MN}
\left(\frac{\pi}{\Delta}\right)^2\mbox{Re}
\int_{-1}^1d\lambda_2\int_{1}^{\infty}d\lambda_1
\exp\{i\omega(\lambda_1-\lambda_2)-\frac{x^2}{2}
(\lambda_1^2-\lambda_2^2)\}\times
\\ \nonumber
\left[1-\prod_{a=1}^M\left(\frac{1+\tilde{\gamma_a}
\lambda_2+i\gamma_aE/2}{1+\tilde{\gamma_a}
\lambda_1+i\gamma_aE/2}\right)\right]
\end{eqnarray}
Finally, the parametric correlation function  $C_{\nu\nu}(\Omega,X)$
 of the densities of states for a closed chaotic system
with broken time-reversal symmetry was found some time ago by  
Simons and Altshuler \cite{AlSi}:
\begin{equation}\label{w13}
 C_{\nu\nu}(\Omega,X)=\frac{1}{2\pi^2}
\left(\frac{\pi}{\Delta}\right)^2\mbox{Re}
\int_{-1}^1d\lambda_2\int_{1}^{\infty}d\lambda_1
\exp\{i\omega(\lambda_1-\lambda_2)-\frac{x^2}{2}
(\lambda_1^2-\lambda_2^2)\}
\end{equation}
Summing up all the contributions we find the desired expression for
the parametric correlation function of {\it scaled} Wigner-Smith  
time delays $\tilde{\tau}_s=\frac{\Delta}{2\pi}\tau_w$:
\begin{eqnarray}\label{w14}
C_W(\omega,x)\equiv\langle\delta\tilde{\tau}_{W}
(E-\Omega/2,-X/2)\delta\tilde{\tau}_{W}(E+\Omega/2,X/2)\rangle=
\\ \nonumber
\frac{1}{2M^2}
\int_{-1}^{1}d\lambda\int_{1}^{\infty}d\lambda_1
\cos{\left[\omega(\lambda_1-\lambda)\right]}
\exp\{-\frac{x^2}{2}(\lambda_1^2-\lambda^2)\}\prod_{a=1}^M
\left[\frac{1+\lambda\,g_a^{-1}}{1+\lambda_1\,g_a^{-1}}\right]
\end{eqnarray}
where we used the parameter $g_a=2T_a^{-1}-1$ introduced earlier.

It is interesting to mention that there exists an alternative way
to derive the pair correlation function of Wigner-Smith time delays
given in Eq.(\ref{w14}). The starting point in
that case is Eq.(\ref{wigtime2}). Then the
calculation of the correlation of fluctuations of Wigner-Smithtime  
delay $\delta\tau_{W}(E,X)=\tau_w-\langle\tau_w\rangle$ amounts to
evaluating the average product of the resolvents of the
non-Hermitian effective Hamiltonians ${\cal H}\pm i\pi WW^{+}$.
This can be done by exactly the same method as we use elsewhere in
the present paper. For the case of chaotic systems with preserved  
TRS and no external
parameter X such a calculation was
done earlier in \cite{Lehmann}.

Let us analyse the correlation function $C_W(x,\omega)$ in more
detail. For this purpose we find it convenient to
rewrite Eq.(\ref{w14}) in a slightly different form:
\begin{equation}\label{w15}
C_W(\omega,x)=\frac{1}{2}\left(R_M^{(1,c)}(\omega,x)
R_M^{(2,c)}(\omega,x)-R_M^{(1,s)}(\omega,x)R_M^{(2,s)}(\omega,x)\right)
\end{equation}
where\begin{equation}\label{w15a}
R_M^{(1,c)}(\omega,x)=\int_{0}^{\infty}dt_1\cos{(\omega
t_1)} \exp{\left[-x^2t_1-\frac{x^2t_1^2}{2}\right]}
\prod_{a=1}^M\displaystyle{\frac{1}{1+T_at_1/2}}
\end{equation}
\begin{equation}\label{w15b}
R_M^{(2,c)}(\omega,x)=\int_{0}^{2}dt_2\cos{(\omega
t_2)}\exp{\left[-x^2t_2+\frac{x^2t_2^2}{2}\right]}\prod_{a=1}^M  
(1-T_at_1/2)
\end{equation}
and the functions $R_M^{(p,s)}(\omega,x);\quad p=1,2$ are obtained
from the expressions for $R_M^{(p,c)}(\omega,x);\quad p=1,2$ by  
replacing $\cos{\omega t_p}$ by $\sin{\omega t_p}$.

First of all, the correlation function Eq.(\ref{w15}) taken at
$\omega=x=0$ gives the variance of the Wigner-Smith time delay  
distribution. In principle,
the corresponding integration can be performed for an arbitrary
set of transmission coefficients $T_a$. The resulting expressions
turn out to be quite cumbersome. They simplify in the case of all
equivalent channels $T_a=T;\quad a=1,...,M$ when we  
find:\begin{equation}\label{w16}
\frac{\langle \tau_w^2\rangle-\langle \tau_w\rangle^2}{\langle
\tau_w\rangle^2}=\frac{2}{T^2(M^2-1)}\left[1-
(1-T)^{M+1}\right]
\end{equation}
This expression shows the same qualitative
features ( divergencies at $M=1$ or $T\to 0$) as those following
from Eq.(\ref{sm53}).

For chaotic systems with only one open channel the Wigner-Smith  
time delay  $\tilde{\tau}_W$ just coincides with the partial
delay time $\tau_s$ and the corresponding distribution
is given by that in Eq.(\ref{sm48}).
Unfortunately, our methods give us no
possibility to find explicitly the distribution
${\cal P}_W(\tilde{\tau}_W)$ of Wigner-Smith time delay for an  
arbitrary number of open
channels $M>1$.  It is natural to put forward a conjecture, that the
divergence of the variance of Wigner-Smith time delay as long as
$M\to 1$ indicates that a (unknown) distribution ${\cal  
P}_W(\tau_w)$ possess the same powerlaw tail ${\cal
P}_W(\tilde{\tau}_W)\propto \tilde{\tau_w}^{-M-2}$ at large
$\tilde{\tau}_W$ as that typical for the distribution of partial
phase shift times. Another argument supporting this conjecture comes
from the general formula Eq.(\ref{wigtime2}). Taking the value $E$  
at random
it is evident that anomalously large time delay $\tau_w(E)\sim
\Gamma_n^{-1}$  corresponds to the
event when $E$ happens to be sufficiently close (at the distance  
$\delta E\lesssim\Gamma_n$) to a position $E_n$ of an anomalously  
narrow resonance $\Gamma_n\ll
\Delta$. The probability of such an event can be estimated as
$P_{\Gamma}\propto(\Gamma/\Delta)
\rho(\Gamma/\Delta\ll 1)\propto (\Gamma/\Delta)^M$, where we used  
the small width asymptotic $\rho(y\ll 1)\propto y^{M-1}$ of the
 resonance widths distribution Eq.(\ref{maineq}).
Then the asymptotic tail of the probability distribution of the  
time delay
can be estimated as ${\cal P}(\tau_w)\propto
\int d\Gamma\delta(\tau_w-\Gamma^{-1})P_{\Gamma}\propto \tau_w^{-M-2}$
in agreement with our conjecture. Here it is appropriate to mention  
that the same asymptotic behavior
is typical for the staying probability function $p(t)$, see
Eq.(\ref{uu1}).  The long-time asymptotic for $p(t)$ was found for  
the systems with preserved time-reversal symmetry in the  
papers\cite{Lew,Harney,ISS}.
 It is trivial to adjust the corresponding argumentation to the
present case and to recover the $\tau^{-(M+2)}$ behavior.

The expression Eq.(\ref{wigtime2}) allows one also to show that for  
weakly opened systems the distribution of the scaled delay times  
should demonstrate the universal behavior
${\cal P}(\tau)\propto \tau^{-3/2}$ in the parametrically large  
domain $g^{-1}\ll \tau\ll g$ , cf. Eq.(\ref{sm52}).
Indeed, for $g\gg 1$ the resonances do not overlap: $\Gamma_n\ll
\Delta$, and their widths $\Gamma_n$ follow the  
$\chi^2$distribution. It is therefore clear, that for any particular  
value of the
energy $E$ the sum in Eq.(\ref{wigtime2}) is dominated by a single
resonance whose position $E_n$ is the closest to $E$, that is
 $\tilde{\tau}_w\approx \frac{2}{M}\frac{y_n}{y_n^2+u_n^2}$
, where $y_n=\pi\Gamma/\Delta$ and $u_n=\frac{2\pi}{\Delta}(E-E_n)$.
 We can estimate the value of the  contribution coming from all  
neglected terms assuming that all resonances have
 the same widths $\langle \Gamma\rangle $ and are spaced equally  
with the mean spacing $\Delta$. This immediately gives the  
correction
to be of the order of $\delta\tilde{\tau}_w\propto
\langle\Gamma\rangle /M\Delta\sim g^{-1}$, where we used the  
formula Eq.(\ref{MS1}) in the limit $T\ll 1$. We conclude that the  
distribution of the scaled time delay
is correctly reproduced by the distribution of the"closest  
resonance term" as long as we are interested in the region  
$\tilde\tau_w\gg g^{-1}$. Assuming that the variable $u_n$ is
uniformly distributed in the interval $[-\pi,\pi]$ we find:
\begin{equation}\label{weak}
{\cal P}(\tilde{\tau}_w)=\int_0^{\infty}dy {\cal P}_{\chi^2}(y)
\int_{-\pi}^{\pi}du\delta
\left(\tilde{\tau}_w-\frac{2}{M}\frac{y}{y^2+u^2}
\right)=M^{-1/2}\tilde{\tau}_w^{-3/2}\int_{0}^{y_m(\tilde{\tau}_w)}
 dy \frac{y^{1/2}}{(2-M\tau y)^{1/2}}{\cal P}_{\chi^2}(y)
\end{equation} where  
$y_m(\tilde{\tau}_w)=\mbox{min}\left[\frac{2}{M\tilde{\tau}_w};
\pi^2\frac{M\tilde{\tau}_w}{2}\right]$.
Taking into account that the $\chi^2$ distribution ${\cal
P}_{\chi^2}(y)$ is cut exponentially at $y> 1/g$ we can safely  
neglect the term $M\tau y\ll 1$ and set the upper limit $y_m=\infty$  
 as long as $\tilde{\tau}_w\ll g$. This immmediately
 results in the anticipated $\tilde{\tau}_w^{-3/2}$ behavior.
On the other hand, we can put ${\cal P}_{\chi^2}(y)\propto g^M
y^{M-1}$ and $y_m=2/(M\tilde{\tau}_w)$ in the domain of extremely  
large time delays $\tilde{\tau}_w\gg g$, which results in the $g^M  
\tilde{\tau}_w^{-(M+2)}$ tail in full
agreement with the general discussion presented above.

The behavior of the delay time distribution in the domain of  
extremely small time delays $\tilde{\tau}_w<g^{-1}$ is
determined by contribution of many resonant terms in the expression
 Eq.(\ref{wigtime2}).  However, one can argue that the distribution
 should be exponentially cut: ${\cal P}(\tilde{\tau}_w
\propto\exp{-\left[const (\tau g)^{-1}\right]}$, as is indeed seen  
from the expression Eq.(\ref{sm52a}).
This behavior is a typical one for a sum of random variables of the  
form $\sum_n y_n/u_n^{-2}$, with $\langle y_n\rangle\sim g^{-1}$  
(the so-called stable Levy distribution, see similar arguments in  
\cite{Hacken}).

The correlation function Eq.(\ref{w15}) acquires quite a simple  
form for the case of many weakly open channels: $T_a\ll 1,
a=1,2,..,M$ but $ \Gamma=\sum_a T_a\gg 1$. Then we can put effectively:
$\prod_a(1- t_2T_a/2)\approx\exp{-\Gamma  
t_2/2};\quad\prod_a(1+t_1T_a/2)^{-1}\approx \exp{-\Gamma t_1/2}$ and  
 also
 neglect the terms $\pm \frac{x^2t_{1,2}^2}{2}$ in the exponents
of the integrands in Eqs.(\ref{w15a},\ref{w15b}). The corresponding
integrals can be calculated exactly giving:
\begin{equation}\label{w17}
C_W(\omega,x)=\frac{(\Gamma_X^2-\omega^2)\left[1-e^{-2\Gamma_X}
\cos{2\omega}\right]
+2\Gamma_X\omega  
e^{-2\Gamma_X}\sin{2\omega}}{(\omega^2+\Gamma_X^2)^2}     
\end{equation}
where $\Gamma_X=\Gamma+x^2$.
Let us note that for $x=0$ this
expression is actually valid for arbitrary $\Gamma$. For arbitrary  
value of $x$ the condition of validity is $\Gamma_X\gg 1$.
Neglecting the exponentially small terms we arrive at the simple
expression:
\begin{equation}\label{w18}
C_W(\omega,x)=\frac{(\Gamma_X^2-\omega^2)}{(\omega^2+\Gamma_X^2)^2}  
   \end{equation}

As will be shown in the next subsection, this expression is
nothing other but the semiclassical formula for parametric
time delay correlations in systems with broken time-reversal  
invariance. For the case of preserved time-reversal symmetry and no  
external
parameters the Eq.(\ref{w18}) was derived in \cite{Lehmann}.

\subsection{ Semiclassical theory for parametric correlations
of time delays.}

A general semiclassical expression for the Wigner-Smith time delay
in terms of a periodic orbit expansion has been given by Balian
and Bloch \cite{Balbl}. It is formally identical to Gutzwiller's
trace formula. The corresponding expression for the pair correlation
function of time delay (without taking into account a parametric
dependence) for chaotic scattering was derived by
Eckhardt\cite{Eckhardt}. In parallel,
Berry and Keating \cite{Keating} developed a method allowing to  
take parametric correlations into account
for the case of a {\it closed} chaotic system pierced by a magnetic  
flux serving to  break down the time-reversal
symmmetry. Below we show briefly how to
 combine both approaches to arrive at the semiclassic expression
for the parametric correlation function of time  delay in that case;
 see also related discussion in the papers \cite{Dosmifr,Pluhar,Bruus}.
The semiclassical periodic orbit expansion for the "fluctuating
 part" of a time delay of a quantum particle
with an energy $E+\Omega/2;\quad \Omega\ll E$ moving in a  
systempierced by a magnetic flux line with flux $\phi$ (measured in  
units of flux quanta
$\phi_0=2\pi c/e$) is:
\begin{equation}\label{sc1}
\delta\tau_w(E+\Omega/2,\phi)=\sum_j A_je^{\frac{i}{\hbar}\left[S_j(E)+
\frac{\Omega}{2}T_j\right]}e^{2\pi iw_j\phi}
\end{equation}
where the summation goes over all periodic orbits with period
$T_j=\partial S_j/\partial E$, with $S_j(E)$ being the corresponding
action, $A_j=\frac{e^{i\mu_j}T_j}{2\pi\sqrt{\mbox{det}(M_j-1)}}$
being the amplitude and $\mu_j,M_j$
being the Maslov phase and stability matrix
 corresponding to the given periodic orbit. The winding number $w_j$
counts the number of times the orbit winds around the flux line.

Thus, for the parametric correlation function one finds:
\begin{eqnarray}\label{sc2}
C_W(\Omega,X)\equiv\langle\delta\tilde{\tau}_{W}
(E-\Omega/2,\phi-X/2)\delta\tilde{\tau}_{W}(E+\Omega/2,\phi+X/2)\rangle=
\\ \nonumber
\left\langle\sum_{j,k}\mid
A_jA_k\mid\exp{\left\{\frac{i}{\hbar}\left[S_j(E)-S_k(E)\right]+\frac{i\Omega}{2\hbar}(T_j+T_k)+2\pi  
i(w_j-w_k)\phi+\pi
iX(w_j+w_k)\right\}}\right\rangle\end{eqnarray}
where the averaging goes over the energy spectrum.
According to standard argumentation\cite{Eckhardt,Keating} one can  
restrict
oneself to the so-called "diagonal approximation" taking into
account only contributions with coinciding indices $j=k$:
\begin{equation}\label{sc3}
C_W^{diag}(\Omega,X)=\left\langle\sum_j\mid A_j\mid^2\exp
{\left[i\frac{\Omega}{\hbar}T_j+2\pi iXw_j\right]}\right\rangle
\end{equation}
The next important step uses the fact that
winding numbers for orbits in any narrow window of periods are
essentially irregular and Gaussian distributed \cite{Berob,Keating},
see also discussion in \cite{Pluhar,Bruus}:
$$
{\cal P}(w_j)=\frac{1}{2\pi\sigma(T_j)}\exp{-w_j^2/2\sigma(T_j)^2}
$$
where the variance $\sigma(T_j)$ increases linearly with period:
$\sigma(T)=\beta T$. The constant $\beta$ is system dependent;
 for a particle with mass $m$ moving in a billiard of the area $A$  
it is proportional to $\left(2E/mA\right)^{1/2}$  
\cite{Bruus,Pluhar}.
Taking the discrete nature of the winding numbers into account one  
can write:
\begin{equation}\label{sc4}
\overline{\exp{2\pi iX w_j}}=\displaystyle{\frac{\sum_
{k=-\infty}^{\infty}e^{-(2\pi)^2\sigma(T_j)(X-k)^2}}  
{\sum_{k=-\infty}^{\infty}e^{-(2\pi)^2\sigma(T_j)k^2}}\approx
e^{-4\pi^2\beta X^2 T_j}}
\end{equation}
where we used $X\ll 1$ (i.e. change of the magnetic flux is much
smaller than $\phi_0$) and neglected exponentially small terms
$O\left(\exp{-4\pi^2\sigma(T)}\right)$.
Substituting this average in the correlation function of time delay
it is convenient to consider its Fourier transform $C(t,X)=\int  
d\Omega e^{-i\Omega t/\hbar}
C_W^{diag}(\Omega,X)$. We have:\begin{equation}\label{sc5}
C_W(t,X)=\sum_j\mid A_j\mid^2\delta(t-T_j)e^{-4\pi^2\beta X^2 T_j}=
e^{-\tilde{x}^2 t}\sum_j\mid A_j\mid^2\delta(t-T_j)
\end{equation}
where we denoted $\tilde{x}=2\pi\beta^{1/2}X$.

For {\it closed} chaotic systems the sum in the preceding equation is
known to be proportinal to the time $t$ (this is the famous Hannay-Ozorio
 de Almeida sum rule \cite{Ozorio}). For open chaotic systems
Eckhardt \cite{Eckhardt} gave some arguments in favor of
replacing this sum by $te^{-\Gamma_{cl} t}$, with $\Gamma_{cl}$  
being the classical escape rate from the chaotic region. Using this  
fact we see that (under
the assumptions we made) the semiclassical expression for the
Fourier transform of the correlation
function of time delays is given by:
\begin{equation}
C_W(t,X)=te^{-(\Gamma_{cl}+\tilde{x}^2)t}
\end{equation}After Fourier-transforming this expression back we  
see that the result
turns out to be identical to that given in Eq.(\ref{w18}) upon  
identification
 $\Gamma_{cl}\to\Gamma;\quad \tilde{x}\to x$.

Another interesting point to be mentioned is that the form of the  
time delay correlation function given in the
eq.(\ref{w17}) (which contains Eq.(\ref{w18}) as a limiting case)
was obtained by Shushin and Wardlaw \cite{Shushin} in the model
of chaotic scattering on a leaky surface of constant negative
curvature. At the first glance such a correspondence is quite a  
surprising fact since the model considered in \cite{Shushin}
corresponds formally to {\it
one-channel} scattering, but the result Eq.(\ref{w17}) was derived
under the assumption of {\it many weak} chanels.
In order to understand that fact one
should remember that the model considered in \cite{Shushin}  
possesses quite a peculiar property: all its resonance poles turned  
out to have
{\it exactly the same} widths. It is at variance with the known  
form of the resonance widths distribution for one-channel scattering  
in a
generic chaotic system, see eq.(\ref{maineq}), where resonance
widths  fluctuate strongly. At the same time, if we consider the
limiting case of many weak channels: $M\gg 1, g\gg 1$ and  
$M/g=\Gamma_{ef}$
fixed , the distribution of resonance widths  tends to the
delta-functional one $\rho(\Gamma)=\delta(\Gamma-\Gamma_{ef})$.
This fact can be easily infered from the Eq.(\ref{gap}).
We see that effectively it is just the limiting case of many weak
channels that corresponds to non-fluctuating resonance widths.
Under these conditions the correlations of the time delays are
determined by the statistics of the positions of resonances. For  
the model of scattering on a leaky surface of negative curvature the  
positions of resonances are given by the zeroes of Riemann  
zeta-function on the so-called critical line in the complex plane.  
According
to the celebrated Montgomery conjecture (verified
numerically\cite{Oldyzko} and supported by sound analytical results
\cite{Bogomol})
statistical properties
of these zeroes are identical to those of eigenvalues of large
random GUE matrices. All these facts taken into account it is no  
more a surprise that the correlations of time delays for both models  
coincide in the considered region of parameters.
\section{Summary and Conclusions}

In the present paper we analyzed in much detail the universal  
features of statistics of resonances, phase shifts and delay times  
for a generic open chaotic
quantum system with broken time-reversal invariance. This was  
achieved by replacing the Hamiltonian of the chaotic region
by a large Random Matrix taken from the Gaussian Unitary Ensemble.
Employing the well-developed method of mapping the problem to
the so-called supersymmetric nonlinear $\sigma-$model we succeeded
in deriving explicit analytical expressions for various
distributions and correlations functions , see  
Eqs.(\ref{mainres}-\ref{calf2}),(\ref{maineq}),(\ref{sm28}),(\ref{sm46}),(\ref{sm48}),
(\ref{sm56b}) and (\ref{w14}), characterizing the above mentioned  
quantities for arbitrary finite number of open
channels and arbitrary strength of coupling to continua.

The best candidates for checking the validity of the expressions
obtained are realistic models of mesoscopic ballistic devices  
subject to applied magnetic field.
Closed \cite{magbil} as well as open\cite{Alh,Bruus,magscat}  
systems of this kind
were intensively investigated recently and the statistics of  
S-matrix elements
and related quantities was available among other characteristics.  
Very recently, the issue of dwell times
inside the chaotic region attracted some research interest as well
\cite{Wang}. All these facts allow us to expect that our results
can be verified independently in the numerical experiments.
It is  interesting to mention that recently another type of chaotic  
systems with broken time-reversal
invariance became available experimentally\cite{TRScav}. The authors
used microwave resonators of billiard shape with a "handle" which
allows only unidirectional propagation of radiation in it thus
breaking the symmetry between the wave and its time reversal counterpart.

It is important to mention that our results are also of potential
experimental relevance. Indeed, the issue of time-delay  
fluctuations turns out to be intimately related to the statistical  
properties of mesoscopic capacitors
 \cite{But2,Gopar}. Considering the case of a mesoscopic cavity  
coupled by a $M$-channel lead to one electronic reservoir and  
capacitively to another reservoir
Gopar, B\"{u}ttiker and Mello \cite{Gopar} suggested the following  
expression  for the low frequency AC admittance of such a structure:
\begin{equation}\label{GMB}
G^{I}(\omega)=-i\omega C_e\alpha\quad ;\quad  
\alpha=\frac{\tau_w}{\eta+\tau_w}
\end{equation}
where $C_e$ stands for a geometric capacitance relating the charge  
$Q$ on the plate to the voltage $U$ across the capacitor, $\tau_w$  
stands for the dimensionless Wigner time delay, and  
$\eta=\frac{C_e}{Me^2/\Delta}$, with
$\Delta$ being the mean level spacing for the cavity.

For macroscopic cavities $\eta\to 0$ and  the dimensionless capacitance
$\alpha$ is equal to unity resulting in the classical expression  
for the capacitive response:
$G^{I}(\omega)=-i\omega C_e$. In contrast, for small enough cavities
 $\eta$ has to be taken into account and the fluctuating delay time  
$\tau_w$
results in a fluctuating admittance.

As it was discussed above, for one open channel $M=1$ the  
distribution of Wigner time delay is identical to the distribution  
of partial delay times
 ${\cal P}_{\tau}(\tau)$
and is given by Eq.(\ref{sm48}). This fact immediately allows one  
to write down the
distribution of the dimensionless capacitance $\alpha$ as:
\begin{equation}\label{GBM1}
{\cal P}_{\alpha}(\alpha)=\frac{\eta}{(1-\alpha)^2}{\cal P}_{\tau}
\left[\tau=\frac{\eta\alpha}{1-\alpha}\right]
\end{equation}
For the perfect coupling case $T=1$ the corresponding distribution  
was analyzed in \cite{Gopar}. In the opposite limiting case of weak  
coupling $T\ll 1$ the universal $\tau^{-3/2}$ time delay  
distribution, see Eq.(\ref{sm52}) results in the following  
expression:
\begin{equation}\label{weak1}
{\cal P}_{\alpha}(\alpha)\propto \left(\frac{\eta}{T}\right)^{-1/2}
\alpha^{-3/2}(1-\alpha)^{-1/2}
\end{equation}
as long as $\alpha\gg T/\eta$ and $1-\alpha\gg \eta T$. As it  
follows from our previous discussion after Eq.(\ref{w16}), this form  
of the distribution should be valid for arbitrary number of weakly  
open channels.

Actually, our knowledge of
the general
 expression for Wigner time delay variance , see Eq.(\ref{w16}),
provides us with the possibility to determine the variance of the  
low-frequency admittance $G^I(\omega)$ in the limit of many open  
channels.

Indeed, in the limit $M\gg 1, T\sim 1$ our expression just says  
that the variance of time delay is of the order $1/M^2\ll 1$ as  
compared with the squared mean value $\langle \tau_w\rangle ^2$.  
Thus, we can represent the fluctuating time delay in a form  
$\tau_w=\langle \tau\rangle+\delta\tau_w$, where typical scale of  
the fluctuating part is of the order of $\delta\tau_w\sim \langle  
\tau_w\rangle/M$.  Substituting this expression to Eq.(\ref{GMB})
and expanding with respect to the fluctuating part $\delta \tau_w$  
one obtains  to the first nontrivial order:
$$
G^{I}(\omega)=\langle G^{I}(\omega)\rangle\left(1+\delta\tau_w\frac{\eta}
{\langle \tau_w\rangle \left(\eta+\langle  
\tau_w\rangle\right)}+...\right)
$$
and immediately extracts the variance of the admittance :
$$
\frac{\langle \left(G^{I}(\omega)\right)^2\rangle-\langle  
G^{I}(\omega)\rangle^2}{\langle G^{I}(\omega)\rangle^2}=
\frac{2}{T^2M^2}\frac{\tau_{RC}^2}{\langle \tau_w\rangle ^2}
$$
where $\tau_{RC}^{-1}=\langle \tau_w\rangle^{-1}+\eta^{-1}$
is the so-called RC-time and we substituted the
expression Eq.(\ref{w16})  for the time delay variance taking into  
account that $M\gg 1$.
Such an expression for the particular case $T=1$ was very recently  
derived by Brouwer and B\"{u}ttiker by a different method \cite{BB}.

 The majority of the numerical data concerning various statistical  
properties of the
scattering matrix for  open chaotic
systems corresponds to the case of preserved time-reversal invariance
see e.g.\cite{Smicol,Doron,Ishio,Lin}. This case is not only  
simplier from numerical point of view ( opposite to the situation  
with
the analytical calculations) , but also the most
relevant experimentally. As a consequence,
numerical studies on resonance width
statistics\cite{Del,Rydb,Burg,Seba2,chemf2,Weaver}
 as well as on
properties of scattering phase shifts and their derivatives  
\cite{JPich,Dietz,Eduardo}
were restricted to the systems of that symmetry class.
  It is necessary to note that some analytical results for poles  
and time delays for time-reversal invariant scattering
 are already available in the
literature for some time. In particular, for only one open channel
the joint probability distribution of all N complex resonance poles
is known\cite{Sok} ( however, not the density of these poles
in complex plane)
  as well as the distribution of Wigner time delay
for perfect coupling to continuum\cite{Gopar}. Essential progress
was achieved by Lehmann et al.\cite{Lehmann} who calculated
the correlation function of time delays for two different values of
energy and any number of open channels.
Actually,  the calculation similar to that done in the present
paper can be successfully carried out for the whole crossover region
between the orthogonal and unitary symmetry classes. The results  
will be published elsewhere
\cite{FSS}

 Let us also mention that recent numerical results
\cite{Seba2}
show that the resonance widths distribution derived in the present
paper can be applied for the systems with preserved time reversal
invariance quite satisfactorily after replacing the number of
channels $M$ by $M/2$. This fact is not so surprising, taking into
account that such feature as the powerlaw tail $1/y^2$ of that
distribution is actually a generic property following from the chaotic
classical dynamics only, see the discussion in Sec.III.

For a majority of models in atomic and molecular physics
 parameters of all the resonances can be determined even without  
expensive
calculations of $S-$matrix elements. The most effective method
is the so-called complex scaling (or complex rotation) method \cite{cosc}
successfully used for the systems  exhibiting chaotic behavior  
\cite{Del,Rydb,Burg}. It is interesting to mention that a crossover  
from isolated to
overlapping resonance regime was detected recently for the dissociation
reaction $HO_2\to H+O_2$  in one open channel
case \cite{chemf2}. One can hope that applying
the complex rotation method to this sytem one could extract the
widths of resonances with sufficient accuracy and to observe
a transition from the $\chi^2$ distribution towards that with
the $1/y^2$ tail.
 We would like to point out that
the whole $S$-matrix as a function of energy of incoming waves was  
measured in real experiments\cite{cav1,Dosmifr}, and even used
to calculate the average time delay\cite{Dosmifr}. In principle,
the positions of resonances in the complex plane can be extracted
if one knows $\hat{S}(E)$ with sufficient accuracy. For example,
one can use that fact that for any number of open channels
the determinant $\mbox{det}\hat{S}(E)$ as a function of energy has its
singularities (which are just resonance poles $E_n-i\Gamma_n/2$)
only in the lower half plane $Im{\cal E}<0$. As a result it can be
written as:
\begin{equation}\label{detsm}
\mbox{det}\hat{S}({\cal E})=e^{i\delta}\prod_{n}
\frac{{\cal E}-E_n-i\Gamma_n/2}{{\cal E}-E_n+i\Gamma_n/2}
\end{equation}
where $\delta$ is the phase of potential scattering irrelevant for
our discussion. Provided the values of $\mbox{det}\hat{S}(E)$ for
real $E$  are known,
one can restore the determinant of $S-$matrix in the upper  
half-plane $\mbox{Im}E>0$
by the relation:
\begin{equation}\label{ext}
\mbox{det}\hat{S}(E+iI)=\frac{I}{\pi}\int_{-\infty}^{\infty}
\frac{d\omega}{(\omega-E)^2+I^2}\mbox{det}\hat{S}(E)\end{equation}

It is easy to see that the two relations (\ref{detsm}) and
(\ref{ext}) allow to determine all the resonance parameters
$E_n,\Gamma_n$ from zeroes of the $S-$matrix determinant in the upper
half-plane $\mbox{Im}{\cal E}>0$. Of course, the practical
implementation of this procedure requires highly accurate data
 for $\hat{S}(E)$ which is not the case in the mentioned experiments
due to noise and damping in resonator walls. However, one can hope
that the progress in the experimental set-up could make such a
measurement feasible in future.

Finally, as an interesting perspective for future research we would  
like to mention the issue of $S-$matrix statistics for systems
exhibiting the Anderson localization phenomenon. This issue
attracts research attention for some period\cite{phaseloc} and  
increasing amount of numerical results are already available
\cite{JPich,Shep,Eduardo,Borgo} requiring a systematic
analytical insight into the problem.

\section{Acknowledgement}

The authors are grateful to
E.Akkermans, B.Altshuler, O.Bohigas, M.B\"{u}ttiker,F.-M.Dittes, V.  
Gasparian, I.Guarneri, G.Hackenbroich,
V.Falko, K.Frahm, F.Izrailev,B.Khoruzhenko, V.Kravtsov, K.Makarov ,  
A.Mirlin, J.-L.Pichard, D.Shepelyansky,
H.-J.St\"{o}ckmann, R.Weaver, and K.Zyczkowsky for useful comments  
and suggestions.
We are mostly obliged to  N.Lehmann, E.Mucciolo, P.Seba and  
V.Sokolov for numerous illuminating discussions on different issues  
of chaotic scattering and $S-$matrix statistics, to P.Seba for  
making his unpublished numerical data on resonance widths  
distribution
available to us, to D.Braun and N.Lehmann for their help in
preparing figures and to the authors of the
papers\cite{Hacken,Seba2,Weaver,Eduardo,SZZ} for sending their papers
to us prior to publication. We also appreciate very much the kind  
invitation by Prof. Jean Bellissard to write a contribution
to the present volume which stimulated us to write the paper
in its present form.

The financial support by the project "Quantum Chaos" (grant  
No.INTAS-94-2058)
and by
SFB 237 "Unordnung und gro{\ss}e Fluktuationen
" is acknowledged with thanks.

\appendix
\section{Regularization of the eigenvalue density for nonnormal
random matrices}\label{nonnormal}

Let us consider a random, but fixed non-Hermitian $N\times N$ random
matrix ${\cal H}$, for which we only assume that generically its
complex eigenvalues are non-degenerate. Taking the second derivative
of the potential Eq.(\ref{logpot}) (apart from the factor $1/N$)
 with respect to the energy ${\cal
E}$ and its complex conjugate ${\cal E}^*$, we obtain Poisson's
equation \cite{fnotel}:
\begin{equation}\label{N1}
-\frac{\partial^2}{\partial {\cal E}\partial {\cal E}^*}\Phi=
\mbox{Tr}\frac{1}{({\cal H}-{\cal E})^{\dagger}({\cal H}-{\cal
E})+\kappa^2} \kappa^2\frac{1}{({\cal H}-{\cal E})({\cal H}-{\cal
E})^{\dagger}+\kappa^2}=\pi\rho_{\kappa}(E,Y)
\end{equation}
with a density $\rho_{\kappa}$ which is always positive, because the
operators appearing are both positive. We will show below that
$\rho_{\kappa}$ goes to a sum of two-dimensional $\delta-$functions
in the complex energy plane:
\begin{equation}\label{N2}
\lim_{\kappa\to 0}\rho_{\kappa}(E,Y)=\sum_{j=1}^N\delta^2({\cal
E}-{\cal E}_j)
\end{equation}
where ${\cal E}_{j}$ are the eigenvalues of ${\cal H}$. The weight
for each $\delta-$function is one. Here the integral over
$\rho(E,Y)dEdY$ is normalized to $N$: $\int
\rho_{\kappa}(E,Y)dEdY=N$, which can be kept finite. This is true for
any $\kappa>0$ and can simply be shown by Stokes theorem.

Now let us consider the Hermitian eigenvalue problem
\begin{equation}\label{N3}
\left[({\cal H}-{\cal E})^{\dagger}({\cal H}-{\cal
E})+\kappa^2\right]\psi_i=\lambda_i\psi_i
\end{equation}
The eigenvalues $\lambda_i$ we again assume to be generically
non-degenerate, and the $\psi_i$ form a complete orthonormalised set.
It follows:
\begin{equation}\label{N4}
\left[({\cal H}-{\cal E})({\cal H}-{\cal E})^
{\dagger}+\kappa^2\right]({\cal H}-{\cal E})\psi_{i}=
\lambda_i({\cal H}-{\cal E})\psi_i
\end{equation}
so that $\phi_i=({\cal H}-{\cal E})\psi_i/\sqrt{\lambda_i-\kappa^2}$
is a normalised eigenvector of a second Hermitian eigenvalue problem
for the operator $\left[({\cal H}-{\cal E})({\cal H}-{\cal
E})^{\dagger}+\kappa^2\right]$. Such an eigenvector corresponds to
the same eigenvalue $\lambda_i$, provided $\lambda_i\ne \kappa^2$,
and ${\cal E}$ is not an eigenvalue of ${\cal H}$. If $({\cal
H}-{\cal E})\psi_0=0$, we nevertheless can find a normalised
eigenvector $\phi_0$, orthogonal to all $\phi_i;\quad i\ne0$, with
$({\cal H}-{\cal E})({\cal H}-{\cal
E})^{\dagger}\phi_0=0$, i.e. $\phi_0$ is an eigenvector of ${\cal
H}^{\dagger}$ with the eigenvalue ${\cal E}^*$.

Now we may expand $\rho_{\kappa}$ in terms of these  
eigenfunctions:\begin{equation}\label{N5}
\rho_{\kappa}(E,Y)=\frac{\kappa^2}{N}\sum_{ik}\frac{1}{\lambda_i}
\mid(\psi_i,\phi_k)\mid^2\frac{1}{\lambda_k}
\end{equation}
from which one sees explicitly that $\rho_{\kappa}$ is positive.
Here $(\psi_i,\phi_k)$ stands for the complex scalar product in
Hilbert space.

If ${\cal E}$ is not an eigenvalue of ${\cal H}$ (and ${\cal E}^*$
is not one of ${\cal H}^{\dagger}$), then $\rho_{\kappa}$ goes to
zero proportionally to $\kappa^2$ for $\kappa\to 0$.
Therefore the weight to the normalization of $\rho_{\kappa}$
comes only from the neighborhoods of the eigenvalues ${\cal E}_j$
of ${\cal H}$. If ${\cal E}_0$ is exactly an eigenvalue of ${\cal  
H}$, then we know the lowest eigenvalue $\lambda_0$ of $({\cal  
H}-{\cal
E}_0)^{\dagger} ({\cal H}-{\cal E}_0)+\kappa^2$ ( which is obviously
$\lambda_0=\kappa^2$) and all other eigenvalues are higher by amounts
independent of $\kappa$. That means:
\begin{equation}\label{N6}
\rho_{\kappa}(E,Y)\approx\frac{1}{\pi\kappa^2}\mid(\psi_0,\phi_0)\mid^2
\end{equation}
for ${\cal E}={\cal E}_0$ and $\kappa\to 0$, which diverges as it must
for a $\delta-$function at an eigenvalue ${\cal E}_0$.

In order to see how $\rho_{\kappa}$ varies with energy near an
eigenvalue ${\cal E}_0$, we may set ${\cal E}={\cal E}_0+\delta{\cal
E}$ and calculate $\lambda_0$ by perturbation theory. Only the second
order perturbation contributes and the surprisingly simple result is:
\begin{equation}\label{N7}
\lambda_0\approx\kappa^2+\mid\delta E\mid^2\mid(\phi_0,\psi_0)\mid^2
\end{equation}
This means that in the neighborhood of an eigenvalue ${\cal E}_0$
the function $\rho_{\kappa}$ has the form:
\begin{equation}\label{N8}
\rho_{\kappa}(E,Y)\approx\frac{\kappa^2}{\pi}\frac{\mid(\phi_0,\psi_0)\mid^2}
{(\kappa^2+\mid\delta E\mid^2\mid(\phi_0,\psi_0)\mid^2)^2}
\end{equation}
We will not consider rare cases, in which the vector $\psi_0$ is
occasionally orthogonal to $\phi_0$. Then we see that
$\rho_{\kappa}(E,Y)$ has in the limit $\kappa\to 0$ the form of a  
two-dimensional $\delta-$ function with weight one. Its width goes  
to zero like
$\kappa/\mid(\phi_0,\psi_0)\mid$. This is valid in the neighborhood
of one isolated eigenvalue, which is however arbitrary. This proves
that the Eq.(\ref{N2}) is indeed correct.

\section{The parametrization of the matrices $\hat{Q}$}\label{param}
The supermatrices $\hat{Q}$ belonging to the graded coset space
$U(1,1/2)/U(1,1)\times U(1,1)$ can be parametrized as
 $\hat{Q}=\hat{U}^{-1}\hat{M}\hat{U}$ where
\begin{equation}\label{efe}\begin{array}{ccc}
\hat{U}=\left(\begin{array}{cc} u&0\\0&v\end{array}\right)&
 \hat{M}=\left(\begin{array}{cc} -iM_{1}& M_{12}\\ M_{21} &  
iM_{1}\end{array}\right)&
\hat{u}^{\dagger}=\hat{u}^{-1};\quad \hat{v}^{\dagger}=
\hat{k}\hat{v}^{-1}\hat{k}\\
\hat{u}=\left(\begin{array}{cc} 1-\frac{\alpha^*\alpha}{2}&-\alpha^*
\\ \alpha&1+\frac{\alpha^*\alpha}{2}\end{array}\right)&
\hat{v}=\left(\begin{array}{cc} 1+\frac{\beta^*\beta}{2}&-i \beta^*
\\ i\beta &1-\frac{\beta^*\beta}{2}\end{array}\right)&
M_{1}=\left(\begin{array}{cc}\lambda_1& 0\\
0&\lambda_2\end{array}\right)\\
M_{12}=\left(\begin{array}{cc}
\mid\mu_1\mid e^{i\phi_{1}}&0\\
0& i\mid\mu_2\mid e^{-i\phi_{2}}\end{array}\right)&
M_{21}=\left(\begin{array}{cc}\mid \mu_1\mid e^{-i\phi_{1}}&
0\\ 0&i\mid \mu_2\mid e^{i\phi_{2}}\end{array}\right)\end{array}.
\end{equation}
where $\alpha,\alpha^*,\beta,\beta^*$ are Grassmann variables,
$\lambda_1\in(1,\infty);\lambda_2\in(-1,1);\phi_1,\phi_2\in(0,2\pi)$
and $\lambda_{1,2}$ are related to $\mid\mu_{1,2}\mid$ via
$\lambda_1^2-\mid\mu_1\mid^2=1; \quad \lambda_2^2+\mid\mu_2\mid^2=1$.

It is convenient to have also the explicit expressions for the matrix
elements of
$\hat{Q}=\left(\begin{array}{cc}
Q_{11}&Q_{12}\\Q_{21}&Q_{22}\end{array}\right)$. We have
\begin{equation}\label{Efe}
\quad Q_{11}=-iu^{-1}\left(\begin{array}{cc}
\lambda_{1}&0\\0&\lambda_{2}\end{array}\right)u\equiv
\left(\begin{array}{cc}-i\left[\lambda_{1}-\alpha^{*}\alpha
(\lambda_{1}-\lambda_{2})\right]& i\alpha^{*}(\lambda_{1}-\lambda_{2})\\
i\alpha(\lambda_{1}-\lambda_{2})&
-i\left[\lambda_{2}-\alpha^{*}\alpha(\lambda_{1}-
\lambda_{2})\right]\end{array}\right)
\end{equation}
\begin{equation}\begin{array}{c}
Q_{22}=iv^{-1}\left(\begin{array}{cc}
\lambda_{1}&0\\0&\lambda_{2}\end{array}
\right)v\equiv
\left(\begin{array}{cc} i\left[\lambda_{1}+\beta^{*}
\beta(\lambda_{1}-\lambda_{2})\right]&  
\beta^{*}(\lambda_{1}-\lambda_{2})\\
\beta(\lambda_{1}-\lambda_{2})&i\left[\lambda_{2}+\beta^{*}\beta(\lambda_{1}-
\lambda_{2})\right]\end{array}\right)
\end{array}\end{equation}
\begin{equation}\begin{array}{c}
Q_{12}=u^{-1}\left(\begin{array}{cc}
\mu_{1}&0\\0&i\mu_{2}^{*}\end{array}\right)v=\\
\left(\begin{array}{cc}\mu_{1}(1-\alpha^{*}\alpha/2)(1+\beta^{*}\beta/2)-
\alpha^*\beta\mu_2^*&-i\beta^*(1-\alpha^{*}\alpha/2)\mu_1+i\alpha^*
(1-\beta^*\beta/2)\mu_2^*\\
-\alpha(1+\beta^*\beta/2)\mu_1-\beta(1+\alpha^*\alpha/2)\mu_2^*&
i\alpha\beta^*\mu_1+i(1+\alpha^*\alpha/2)(1-\beta^*\beta/2)
\mu_2^*\end{array}\right)
\end{array}\end{equation}
\begin{equation}\begin{array}{c}
Q_{21}=v^{-1}\left(\begin{array}{cc}
\mu_{1}^{*}&0\\0&i\mu_{2}\end{array}\right)u=\\
\left(\begin{array}{cc}\mu_{1}^*(1-\alpha^{*}\alpha/2)(1+\beta^{*}\beta/2)+
\alpha\beta^*\mu_2&-\beta^*(1+\alpha^{*}\alpha/2)\mu_2-\alpha^*
(1+\beta^*\beta/2)\mu_1^*\\
+i\alpha(1-\beta^*\beta/2)\mu_2-i\beta(1-\alpha^*\alpha/2)\mu_1^*&
-i \alpha^*\beta\mu_1^*+i(1+\alpha^*\alpha/2)(1-\beta^*\beta/2)
\mu_2\end{array}\right)
\end{array}\end{equation}
where we introduced the notations:
$\mu_{1}=\mid\mu_1\mid e^{i\phi_{1}};\quad \mu_{2}=\mid\mu_2\mid
e^{i\phi_{2}}.$
The expressions above are frequently referred to as the
"Efetov parametrization" for the matrices $\hat{Q}$.

 We denote
the corresponding measure  as $d\hat{Q}$. Straightforward, but
lengthy calculation gives:
\begin{equation}\label{measure}
d\hat{Q}=\frac{d\lambda_{1}d\lambda_{2}}{(\lambda_{1}-\lambda_{2})^{2}}
\frac{d\phi_{1}d\phi_{2}}{(2\pi)^2}d\alpha^{*}d\beta^{*}d\alpha d\beta
\end{equation}

In the rest of this Appendix we present the explicit expressions for
supertraces, superdeterminants, and matrix elements, entering
different expressions in the main text,
see the Eqs. (\ref{densus},\ref{sm23},\ref{w10}).
\begin{enumerate}
\item{\bf For resonance widths calculation.}\\
We have:
$$
\mbox{Str}\hat{Q}\hat{\Lambda}=\left(Q_{bb}^{(11)}-Q_{bb}^{(22)}\right)
-\left(Q_{ff}^{(11)}-Q_{ff}^{(22)}\right)=-2i(\lambda_1-\lambda_2)
$$
$$
\mbox{Str}\hat{Q}(\hat{\Lambda}-\hat{K}_b\hat{L})=
Q_{ff}^{(22)}-Q_{ff}^{(11)}=2i\lambda_2-i(\lambda_1-\lambda_2)
(\alpha^*\alpha-\beta^*\beta)
$$
$$
\mbox{Str}\hat{Q}\hat{\Sigma}_L=i\left(Q_{bb}^{(12)}-Q_{bb}^{(21)}\right)
-\left(Q_{ff}^{(12)}+Q_{ff}^{(21)}\right)=
$$
$$
=i(\mu_1-\mu_1^*-\mu_2-\mu_2^*)-i\frac{\alpha^*\alpha-\beta^*\beta}{2}
(\mu_1-\mu_1^*+\mu_2+\mu_2^*)-i\alpha^{*}\beta(\mu_2^*-\mu_1^*)-
$$
$$
i\alpha\beta^*(\mu_1+\mu_2)-\frac{i}{4}\alpha^*\alpha\beta^*\beta
(\mu_1-\mu_1^*-\mu_2-\mu_2^*)
$$
The superdeterminant $\mbox{Sdet}^{-1}\left[I+i\frac{1}{2}
\gamma_a E\hat{\Lambda}+
i\pi\nu_{sc}\gamma_a\hat{Q}\hat{\Lambda}\right]$ can be easily
evaluated because the supermatrix $\hat{U}$, Eq.(\ref{efe}),
commutes with $\hat{\Lambda}$  and therefore can be omitted under the
sign of the superdeterminant:
\begin{equation}\label{sude1}
\mbox{Sdet}^{-1}\left[I+i\frac{1}{2}
\gamma_aE\hat{\Lambda}+
i\pi\nu_{sc}\gamma_a\hat{Q}\hat{\Lambda}\right]=\mbox{Sdet}^{-1}\left[I+i\frac{1}{2}
\gamma_aE\hat{\Lambda}+
i\pi\nu_{sc}\gamma_a\hat{M}\hat{\Lambda}\right]\end{equation}
The supermatrix $\hat{A}=\left[I+i\frac{1}{2}
\gamma_aE\hat{\Lambda}+
i\pi\nu_{sc}\gamma_a\hat{Q}\hat{\Lambda}\right]$ is however block
-diagonal in the fermion-boson arrangement:
$\hat{A}_{bf}=\hat{A}_{fb}=0$, see Eq.(\ref{efe}), and therefore
$\mbox{Sdet}^{-1}\hat{A}=\mbox{Det}\hat{A}_{ff}/\mbox{Det}\hat{A}_{bb}$.
Trivial calculation gives:
$$
\mbox{Sdet}^{-1}\left[I+i\frac{1}{2}
\gamma_aE\hat{\Lambda}+
i\pi\nu_{sc}\gamma_a\hat{Q}\hat{\Lambda}\right]=\frac
{1+2\pi\nu_{sc}(E)\gamma_a\lambda_2+\gamma_a^2}
{1+2\pi\nu_{sc}(E)\gamma_a\lambda_1+\gamma_a^2}
$$
which is reduced to the form used in the text of the paper, see
 
e.g. Eq.(\ref{densus}) upon introducing the transmission  
coefficients
$T_a$, Eq.(\ref{trans}) and the parameters $g_a=2/T_a-1$.

\item{\bf For scattering phase shifts statistics.}\\
First of all, using $\hat{U}\hat{\Lambda}=\hat{\Lambda}\hat{U}$
one has
$$
\mbox{Str}(\hat{Q}\hat{\Lambda}\hat{Q}\hat{\Lambda})=
2\mbox{Str}(M_{11}^2-M_{12}M_{21})=-4(\lambda_1^2-\lambda_2^2)
$$
The main object entering the calculation of eigenphases
correlation function is the supermatrix
$\left(\hat{B}_a\right)^{-1}$, where:
\begin{equation}\label{asm1}\begin{array}{c}
\hat{B}_a=\frac{1}{2}z_1(\hat{I}_4+\hat{\Lambda})+\frac{1}{2}z_2(\hat{I}_4-
\hat{\Lambda})-\gamma_a\left(\frac{E}{2}I_4-\pi\nu_{sc}\hat{Q}\right)\equiv
\hat{U}^{-1}\hat{b}(a)\hat{U}\\
\hat{b}(a)=\mbox{diag}\left(z_1(a)\hat{I}_2,
z_2(a)\hat{I}_2\right)+\tilde{{\gamma}}_a\hat{M}
\end{array}\end{equation}
where we used the notations: $z_p(a)=z_p-\gamma_a\frac{E}{2};\,p=1,2$
and $\tilde{\gamma}_a=\pi\nu_{sc}\gamma_a$, the supermatrix
$\hat{M}$ being defined in Eq.(\ref{Efe}). One can invert
$\hat{b}_a$ easily noticing that in the {\it boson-fermion}
arrangement this matrix is block-diagonal:
$\hat{b}(a)
=\mbox{diag}\left(\hat{b}_{bb}(a),\hat{b}_{ff}(a)\right)$, where
$$
\hat{b}_{bb}(a)=\left(\begin{array}{cc}z_1(a)-i\tilde{\gamma}_a\lambda_1&
\mu_1\tilde{\gamma}_a\\ \mu_1^*\tilde{\gamma}_a& z_2(a)+
i\tilde{\gamma}_a\lambda_1\end{array}\right)\quad\mbox{and}\quad
\hat{b}_{ff}(a)=\left(\begin{array}{cc}z_1(a)-i\tilde{\gamma}_a\lambda_2&
\mu_2^*\tilde{\gamma}_a\\ \mu_2\tilde{\gamma}_a& z_2(a)+
i\tilde{\gamma}_a\lambda_2\end{array}\right)$$
so that
$\hat{b}^{-1}(a)=\mbox{diag}\left(\left[\hat{b}^{-1}(a)\right]_{bb},\left[\hat{b}^{-1}(a)\right]_{ff}\right)$,  
where\begin{equation}\label{asm2}
\left[\hat{b}(a)\right]^{-1}_{bb}=\frac{1}{{\cal D}_b(a)}
\left(\begin{array}{cc}z_2(a)+i\tilde{\gamma}_a\lambda_1&
-\mu_1\tilde{\gamma}_a\\ -\mu_1^*\tilde{\gamma}_a& z_1(a)-
i\tilde{\gamma}_a\lambda_1\end{array}\right)
\end{equation}
\begin{equation}\label{asm2a}
\left[\hat{b}(a)\right]^{-1}_{ff}=\frac{1}{{\cal D}_f(a)}
\left(\begin{array}{cc}z_2(a)+i\tilde{\gamma}_a\lambda_2&
-i\mu_2^*\tilde{\gamma}_a\\ -i\mu_2\tilde{\gamma}_a& z_1(a)-
i\tilde{\gamma}_a\lambda_2\end{array}\right)
\end{equation}
and we introduced notations:
\begin{equation}\label{asm3}\begin{array}{c}
{\cal D}_b(a)\equiv \mbox{det}\hat{b}_{bb}(a)
=z_1(a)z_2(a)+i\tilde{\gamma}_a\left(z_1(a)-z_2(a)\right)
\lambda_1+\tilde{\gamma}_a^2\\
{\cal D}_f(a)\equiv \mbox{det}\hat{b}_{ff}(a)=
z_1(a)z_2(a)+i\tilde{\gamma}_a\left(z_1(a)-z_2(a)\right)
\lambda_2+\tilde{\gamma}_a^2
\end{array}\end{equation}
so that $\mbox{Sdet}\left(\hat{B}_a^{-1}\right)=
\mbox{Sdet}\left(\hat{b}^{-1}(a)\right)=\displaystyle{\frac{{\cal  
D}_f(a)}
{{\cal D}_b(a)}}$.

Rearranging the supermatrix $\hat{b}^{-1}(a)$ in {\it
advanced-retarded} order
we can find easily the supermatrix $\hat{B}^{-1}_a=
\hat{U}^{-1}\hat{b}^{-1}(a)\hat{U}$. Actually, we need only its
elements in the {\it fermion-fermion} block (see Eq.(\ref{sm23})):
$$
\mbox{Str}\left(\hat{B}^{-1}_a\hat{C}_1\right)=
-\left(B^{-1}_a\right)^{11}_{ff};\quad
\mbox{Str}\left(\hat{B}^{-1}_a\hat{C}_2\right)=
-\left(B^{-1}_a\right)^{22}_{ff}
$$
and  
$\mbox{Str}\left(\hat{B}^{-1}_a\hat{C}_1\hat{B}^{-1}_a\hat{C}_2\right)=
-\left(B^{-1}_a\right)^{21}_{ff}\left(B^{-1}_a\right)^{12}_{ff}$.\\

We find, correspondingly:
\begin{equation}\label{asm4}
\left(B^{-1}_a\right)^{11}_{ff}=\frac{z_2(a)+i\tilde{\gamma}_a\lambda_2}{{\cal
D}_f(a)}-\alpha^*\alpha\left[\frac{z_2(a)+i\tilde{\gamma}_a\lambda_1}{{\cal
D}_b(a)}-\frac{z_2(a)+i\tilde{\gamma}_a\lambda_2}{{\cal
D}_f(a)}\right]
\end{equation}
\begin{equation}\label{asm5}
\left(B^{-1}_a\right)^{22}_{ff}=\frac{z_1(a)-
i\tilde{\gamma}_a\lambda_2}{{\cal
D}_f(a)}+\beta^*\beta\left[\frac{z_1(a)-i\tilde{\gamma}_a\lambda_1}{{\cal
D}_b(a)}-\frac{z_1(a)-i\tilde{\gamma}_a\lambda_2}{{\cal
D}_f(a)}\right]
\end{equation}
\begin{equation}\label{asm6}
\left(B^{-1}_a\right)^{12}_{ff}=-\frac{i\tilde{\gamma}_a\mu_1}{
{\cal D}_b(a)}\alpha\beta^* -\frac{i\tilde{\gamma}_a\mu_2^*}{
{\cal D}_f(a)}\left(1+\alpha^*\alpha/2\right)
\left(1-\beta^*\beta/2\right)\end{equation}
\begin{equation}\label{asm7}
\left(B^{-1}_a\right)^{21}_{ff}=\frac{i\tilde{\gamma}_a\mu_1^*}{
{\cal D}_b(a)}\alpha^*\beta -\frac{i\tilde{\gamma}_a\mu_2}{
{\cal D}_f(a)}\left(1+\alpha^*\alpha/2\right)
\left(1-\beta^*\beta/2\right)\end{equation}
\item{\bf For time-delay correlations.}\\
The main new object here is the supermatrix (see Eq.(\ref{w10}):
\begin{equation}\label{asm8}
\hat{B}_{\nu}(a)=\frac{1+\hat{\Lambda}}{2}+\frac{1-\hat{\Lambda}}{2}
\left[z(a)\hat{I}_4+\tilde{\gamma}_a\hat{Q}\right]=
\hat{U}\left(\begin{array}{cc}\hat{I}_2& 0\\
\tilde{\gamma}_a\hat{M}_{21} &z(a)\hat{I}_2+i\tilde{\gamma}_a\hat{M}_1
\end{array}\right)\hat{U}^{-1}
\end{equation}
where we again used that matrices $\hat{U}$ and $\hat{\Lambda}$
commute. The matrix $\hat{B}_{\nu}(a)$ is simple to invert.
 Performing the calculation we find:
\begin{equation}\label{asm9}
\left(\hat{B}_{\nu}(a)^{-1}\right)_{ff}^{22}=\frac{1}{z(a)+
i\tilde{\gamma}_a\lambda_2}-\beta^*\beta\left(
\frac{1}{z(a)+i\tilde{\gamma}_a\lambda_1}-\frac{1}{z(a)+
i\tilde{\gamma}_a\lambda_2}\right)
\end{equation}
and the corresponding superdeterminant is given by:
\begin{equation}\label{asm9a}
\mbox{Sdet}^{-1}\hat{B}_{\nu}(a)=\frac{z(a)+
i\tilde{\gamma}_a\lambda_2}{z(a)+
i\tilde{\gamma}_a\lambda_1}
\end{equation}
\end{enumerate}

\newpage
\section*{FIGURE CAPTION}
\begin{itemize}
\item
Fig 1. A generic model for chaotic scattering : an irregular shaped  
cavity
 attached to the  perfect lead. The Hamiltonian associated with the  
cavity region is simulated by a random matrix $\hat{H}_{in}$.
\item
Fig.2 The distribution of scaled resonance widths $\rho(y)$ for  
$M=1$ (solid)
$M=2$ (dash-dotted) and $M=3$ (dotted line) equivalent open
channles. The effective coupling is maximal: $g=1$. As the result,  
the distributions demonstrate $M/(2y^2)$ asymptotic behavior at  
large $y$.
\item
Fig.3 The distribution of scaled  partial delay times ${\cal P}(\tau_s)$
for $M=1$ (solid)
$M=2$ (dash-dotted) and $M=3$ (dotted line) equivalent open
channles. The effective coupling $g=10$ corresponds to weakly open
systems.

\end{itemize}


\begin{references}
\bibitem[*]{leave}on leave from:
Petersburg Nuclear Physics Institute,
Gatchina 188350, St.Petersburg District, Russia
\bibitem{Smilansky} U.Smilansky,
 in "Chaos and Quantum Physics". Proceedings of
the  Les-Houches Summer School. Session LII, ed. by M.J.Giannoni  
et.al, 372 (North Holland, Amsterdam, 1991)
\bibitem{Gasp} P.Gaspard in:, "Quantum Chaos". Proceedings of E.  
Fermi Summer School,
 ed.by G.Casati et.al., 307 (1991)
\bibitem{Baldo} M.Baldo,E.G.Lanza, A.Rapisarda, Chaos 3, 691(1993)
\bibitem{chaonan} D.Stone in:" Mesoscopic Quantum Physics" , Les  
Houches Summer School, Session LXI, 1994; edited by E.Akkermans et  
al., p.325
(Elsever 1995)
\bibitem{chaonan1} W.A.Lin, J.B.Delos and R.Jensen , Chaos 3, 655 (1993)
\bibitem{Marcus} C.M.Marcus,A.J.Rimberg,R.M.Westervelt et al.
,Phys.Rev.Lett. 69 ,506 (1992) and Phys.Rev.B 48, 2460
(1993); J.A.Folk,S.R.Patel,S.F.Godijn et al., Phys.Rev. Lett.  
76,1699 (1996)
\bibitem{TRS}M.J.Berry, J.A.Katine,R.M.Westervelt and A.C.Gossard
,Phys.Rev.B 50, 17721 (1994)
\bibitem{cav} H.J.St\"{o}ckmann and J.Stein, Phys.Rev.Lett.
64, 2215(1990); S.Sridhar , Phys.Rev.Lett. 67 , 785 (1991)
\bibitem{cav1} J.Stein,H.-J.St\"{o}ckmann and U.Stoffregen,
 Phys.Rev.Lett. 75, 53 (1995)
\bibitem{Dosmifr} E.Doron,U.Smilansky and A.Frenkel, Phys.Rev.Lett.  
65,3072 (1990) and in \cite{Gasp}, p.399.
\bibitem{TRScav} P.So, S.Anlage, E.Ott and R.Oerter , Phys.Rev.Lett.
74, 2662 (1995); U.Stoffregen,J.Stein,
H.-J.St\"{o}ckmann et al. ,ibid., 2667
\bibitem{Alt} H.Alt, H.-D.Graf, Harney H.L. et al.,
Phys.Rev.Lett. 74 , 62(1995) ; H.Alt, P.von Brentano, H.-D. Gr\"{a}f
 ,Nucl.Phys.A 560,293 (1993)
\bibitem{chaonuc}  C.H.Dasso, M.Pollarolo, M.Saraceno , Nuclear
Physics A 602, 77(1996) and references therein.
\bibitem{Rot} I.Rotter , Rep.Prog.Phys. 54, 635 (1991)
\bibitem{chaoat} Atomic Spectra and Collisions in External
Fields, edited by M.H.Nayfeh et al. (Plenum, N.Y., 1989), vol.2

\bibitem{Del} B.Gremaud, D.Delande and J.C.Gay , Phys.Rev.Lett.
70, 1615 (1993); K.Dupret, J.Zakrzewsky and D.Delande
, Europh.Lett. 31 , 251 (1995)
\bibitem{Rydb} J.Main and G.Wunner, J.Phys.B:At.Mol.Opt.Phys 27,  
2835 (1994)
\bibitem{Schinke}  R.Schinke, H.-M.Keller, M.Stumpf and A.J.Dobbyn
, J.Phys.B:At.Mol.Opt.Phys 28,  2928 (1995)
\bibitem{Lomb}M.Lombardi, T.H.Seligman , Phys.Rev.A 47 , 3571(1993)
\bibitem{Barr}H.U.Baranger and P.Mello , Phys.Rev.Lett.
73,142 (1994); Europh.Lett. 33 , 465 (1996);
 P.Mello and H.Baranger , Physica A 220
, 15
(1995) and references therein.
\bibitem{BS} R.Bl\"{u}mel, U.Smilansky , Phys.Rev.Lett. 64, 241
(1989); E.Doron and U.Smilansky , Nonlinearity 5 ,  
1055(1992)\bibitem{Eckhardt} B.Eckhardt, Chaos 3 , 613(1993)
\bibitem{Weidrev} H.A.Weidenm\"{u}ller in:"Chaos and Quantum Chaos"  
, Ed. by W.D.Heiss, (Springer, 1992)
\bibitem{Melrev} P.A.Mello in: "Mesoscopic Quantum Physics", Les
Houches Summer School, Session LXI, 1994; edited by E.Akkermans et
al.,435 (Elsever 1995)
\bibitem{Lew} C.H.Lewenkopf and H.A.Weidenm\"{u}ller, Ann.Phys.
212, 53 (1991)
\bibitem{Gutzwiller} M.Gutzwiller , "Chaos in Classical and Quantum
Mechanics" (Springer Verlag,1991)
\bibitem{Balbl} R.Balian and C.Bloch , Ann.Phys. 85, 514 (1974)
\bibitem{AlSi} B.L.Altshuler and B.D.Simons in:
 "Mesoscopic Quantum Physics", ed. by E.Akkermans et al,
 Les Houches Summer School,Session LXI, 5 (Elsever,1995)
\bibitem{Bohigas} O.Bohigas, in : "Chaos and Quantum Physics".
 Proceedings of
the  Les-Houches Summer School. Session LII, ed. by M.J.Giannoni  
et.al,91 (North Holland, Amsterdam, 1991)

\bibitem{Mehta} M.Mehta  " Random Matrices" (Academic Press, N.Y.1990)
\bibitem{RMT} T.A.Brody,J.Flores,J.B.French,P.A.Mello,A.Pandey
and S.S.M.Wong , Rev.Mod.Phys. 53 , 385(1981)
\bibitem{Berry} M.Berry ,Proc.Roy.Soc.Lon.A 400,229 (1985)  
\bibitem{Khm} B.A. Muzykantsky, D.E.Khmelnitsky
,JETP Lett 62, 76 (1995)
\bibitem{AAA}A.Andreev, O.Agam,B.D.Simons and B.Altshuler
,Phys.Rev.Lett 76, 3947  (1996)
\bibitem{Bogkeat} E.Bogomolny, J.Keating , Phys.Rev.Lett. 77, 1472
(1996)
\bibitem{MW} C.Mahaux and H.A.Weidenm\"{u}ller,
 "Shell Model Approach in Nuclear Reactions"(North-Holland,  
Amsterdam,1969)
\bibitem{Levine} R.D.Levine , "Quantum Mechanics of Molecular
Rate Processes", (Oxford, 1969)
\bibitem{Livsic} M.S.$\mbox{Liv\v{s}i\^{c}}$ , Sov.Phys.JETP 4, 91  
(1957) and
 the book : "Operators, oscillations, waves: open
systems". Amer.Math.Soc.Transl. 34 (Providence, R.I.  
1973)\bibitem{KNO} I.Yu.Kobzarev, N.N.Nikolaev, L.B.Okun ,Yad.Phys.
10 , 864 (1969) [In Russian]
\bibitem{Pavlov} B.S.Pavlov , Teor. Mat.Fiz.  59 , 345 (1984)  
[Engl.transl:
Theor.Math.Phys. 59, 544 (1984)] ;
Russian Math. Surveys 42,  127 (1987) ; Yu.A. Kuperin, K.Makarov,  
S.P. Merkuriev, A.K. Motovilov, B.S. Pavlov , Theor. Math. Phys. 75,  
 630 (1988) and Sov. Nucl. Phys. 48, 224 (1988)
\bibitem{Mak} K.Makarov , St.Petersburg Math. J. 4,  967 (1993)
\bibitem{VWZ} J.J.M.Verbaarschot,H.A.Weidenm\"uller and
M.R.Zirnbauer, Phys.Rep 129, 367 (1985);\bibitem{Fesh} H.Feshbach ,  
Ann.Phys. 5, 357 (1958) and "Topics in the theory of Nuclear
Reactions ", in: Reaction dynamics (Gordon and Breach,N.Y.,1973)
\bibitem{Mello} P.Mello, P.Pereyra and T.H.Seligman, Ann.Phys.
(NY) 161, 254 (1985)
\bibitem{Smicol}R.Bl\"{u}mel and U.Smilansky , Phys.Rev.Lett.
60, 477 (1988) and
Physica D  36 , 111(1989);E.Doron and  
U.Smilansky , Nucl.Phys.A 545, 455
(1992); R.Bl\"{u}mel,B.Ditz,C.Jung and U.Smilansky
, J.Phys.A:Math.Gen. 25, 1483 (1992)
\bibitem{Brouwer} P.W.Brouwer , Phys.Rev.B 51 , 16875 (1995)
\bibitem{FPich} K.Frahm, J.-L.Pichard , J. Phys.I France 5, 847
(1995); J.Rau , Phys.Rev.B 51, 7734(1995)
\bibitem{Macedo} A.M.S. Macedo ,Phys.Rev. E 50,R659  
(1994)\bibitem{Kukulin} I.Kukulin, M.Krasnopolsky and J.Horacek:
"Theory of resonances: Principles and Applications",
(Kluwer,Dordrecht 1989).
\bibitem{Zworski} M.Zworski, "Counting Scattering Poles.",
in: " Spectral and Scattering Theory", ed. M.Ikawa, NY (1994)
\bibitem{cosc} W.Reinhardt , Ann.Rev.Phys.Chem 33, 223 (1982);
J.K.Ho , Phys.Rep. 99,1 (1983)
\bibitem{Burg} J.Burgd\"{o}rfer,X.Yang and J.Muller , Chaos,Solitons and
Fractals 5,1235 (1995)

\bibitem{Sok} V.Sokolov and G.Zelevinsky , Phys.Lett. 202B, 10
(1988); Nucl.Phys.A 504,562 (1989); S.Mizutori,
G.Zelevinsky ,Z.Phys.A 346 , 1(1993)
\bibitem{John} W.John, B.Milek, H.Schanz, P.Seba, Phys.Rev.Lett.  
67, 1949 (1991); R.Gawlista, P.Seba , Phys.Rev.A 46 , 6056 (1992)

\bibitem{Haake} F.Haake, F.Izrailev,N.Lehmann et al., Z.Phys.B 88  
,359 (1992)
\bibitem{Deso} M.Desouter-Lecomte, F.Culot, J.Chem.Phys.98,7819  
(1993) and references therein.
\bibitem{Dallwig} S.Dallwig,I.Weese,Th.Weiss and Ch.Schlier
, J.Chem.Phys. 104, 4347 (1996)
\bibitem{Lehm2}N.Lehmann, D.Saher,V. Sokolov and H.-J.Sommers
 , Nucl.Phys.A 582, 223 (1995)
\bibitem{Dittes} E.Sobeslavsky, F.M.Dittes, and I.Rotter,
J.Phys.A: Math.Gen. 28 , 2963 (1995);
M.M\"{u}ller, F.M.Dittes, W.Iskra
and I.Rotter, Phys.Rev.E 52 ,  5961(1995)
\bibitem{Remacle} F.Remacle, M.Munster,B.Pavlov-Verevkin and
M.Desouter-Lecomte ,Physics Letters A 145, 265 (1990);
F.Remacle and L.D.Levine , Physics Letters A 173, 284 (1993)
\bibitem{chemf1} K.Someda, H.Nakamura and F.Mies, Chem.Phys. 187,  
195 (1994)
\bibitem{chemf2} A.Dobbin, M.Stumpf, H.-M.Keller and R.Schinke ,  
J.Chem.Phys. 104, 8357 (1996); S.Mahapatra
,J.Chem.Phys. 105 ,344 (1996)
\bibitem{Drozdzh} S.Drozdz, A.Trellakis and J.Wambach,
Phys.Rev.Lett. 76, 4891 (1996)\bibitem{Weaver}J.Burkhardt and  
R.Weaver , J.Acoust.Soc.Am. 100, 320
 (1996)
\bibitem{Flam} V.V. Flambaum, A.A.Gribakina, G.F.Gribakin , Phys.Rev.A
 54 , 2066 (1996)
\bibitem{Hacken} G.Hackenbroich and J.U.N\"{o}ckel "Dynamical
tunneling in open systems" (1996), unpublished
\bibitem{Porter} C.E.Porter , "Statistical theory of
spectra: Fluctuations" (Academic, N.Y., 1965)
\bibitem{NO2} Th.Zimmermann, H.K\"{o}pel, L.S.Cederbaum et al.
 , Phys.Rev.Lett. 61 , 3 (1988)
\bibitem{Alh} R.A.Jalabert, A.D.Stone and Y.Alhassid,  
Phys.Rev.Lett. 68 , 3468 (1992)
\bibitem{Mucc} E.R.Mucciolo, N.Prigodin and B.L.Altshuler,  
Phys.Rev.B 51,1714 (1995)
\bibitem{FSres} Y.V.Fyodorov and H.-J.Sommers ,Pis'ma  
Zh.Eksp.Teor.Fiz. 63, 970 (1996); [JETP Letters,
63, 1026 (1996)]
\bibitem{Doron} E.Doron and U.Smilansky , Nonlinearity 5, 1055
(1992)
\bibitem{Selig} C.Jung, T.H.Seligman , J.Phys.A:Math.Gen. 28, 1507
(1995) ;  B.Dietz, M.Lombardi, T.H.Seligman,
J.Phys.A:Math.Gen 29, L95 (1996)
\bibitem{Dietz} B.Dietz, M.Lombardi, T.H.Seligman ,  
Phys.Lett.A215,181 (1996)
\bibitem{JPich} R.A.Jalabert, J.-L.Pichard , J. Phys.I France
 5, 287 (1995)\bibitem{Eduardo} E.Mucciolo, R.A.Jalabert,  
J.-L.Pichard and
B.L.Altshuler , unpublished
\bibitem{Butt} M.B\"{u}ttiker in: "Electronic Properties of
Multilayers and Low-Dimensional Semiconductors Structures", ed.by  
J.M.Chamberlain et al., p.297  (Plenum Press,N.Y.  
1990)\bibitem{time} R.Landauer and Th.Martin , Rev.Mod.Phys. 66,217
(1994)\bibitem{delayop} W.O.Amrein and M.B.Cibils, Helv. Phys.Acta  
60, 481
(1987); C.Bracher and M.Kleber , Annalen der Physik 4, 696
(1995)
\bibitem{Gasparian} V.Gasparian et al., Phys.Rev.Lett. 75,2312  
(1995); Phys.Rev.B 51, 6743 (1995);
G.Ionnacone , Phys.Rev.B 51 ,4727 (1995); P.J.Price,
Phys.Rev.B 48, 17301 (1993)  \bibitem{Wigner} E.P.Wigner ,  
Phys.Rev. 98,145 (1955);
F.Smith , Phys.Rev. 118 (1960),349
\bibitem{Harney} H.L.Harney, F.M.Dittes and A.M\"{u}ller
 ,Ann.Phys. 220, 159 (1992)
\bibitem{Lyub} L.Lyuboshits , Physics Letters B 72,41 (1977);
Yad.Fiz.27, 948  (1978) [Sov.J.Nucl.Phys. 27,502
(1978)]; Pis'ma ZHETF 28, 32 (1978) [JETP Lett.
28,30 (1978)]
\bibitem{Lyubver} L.Lyuboshits , Yad.Fiz. 37, 292 (1983)
[Sov.J.Nucl.Phys. 37,174 (1983)]
\bibitem{Bauer} M.Bauer, P.A.Mello and K.W.McVoy , Z.Physik A
293,151 (1979)
\bibitem{ISS} F.Izrailev, D.Saher, V.Sokolov , Phys.Rev. E 49,130 (1994)
\bibitem{Lehmann} N.Lehmann,D.Savin, V. Sokolov and H.-J.Sommers
 , Physica D 86, 572 (1995)
\bibitem{But2} M.B\"{u}ttiker, A.Pretre and H.Thomas,Phys.Rev.Lett.  
70, 4114 (1993) and Z.f.Physik B 94,133
 (1994)
\bibitem{Gopar} V.Gopar, P.Mello and M.B\"{u}ttiker , Phys.Rev.Lett. 77 ,
3005 (1996)
\bibitem{BB} P.Brouwer and M.B\"{u}ttiker
"Charge relaxation and dwell time in the fluctuating admittance of  
a chaotic
cavity.", preprint cond-mat 9610144 (1996)
\bibitem{Akk} E.Akkermans,A.Auerbach,J.Avron and B.Shapiro,  
Phys.Rev.Lett. 66, 76 (1991); E.Akkermans, Europh.Lett.15 ,709(1991)
\bibitem{FSS} Y.V.Fyodorov, D.Savin and H.-J.Sommers , in preparation
\bibitem{FStime} Y.V.Fyodorov and H.-J.Sommers , Phys.Rev.Lett.
76,4709 (1996)
\bibitem{Seba1} P.Seba , Phys.Rev.B 53, 13024 (1996)
\bibitem{Seba2} S.Albeverio, F.Haake, P.Kurasov,M.Kus and P.Seba ,
J.Math.Phys. 37 , 4888 (1996)

\bibitem{fnote}  Actually, this condition
taken alone is sufficient to ensure that the corresponding operator  
${\cal H}$ is a symmetric one rather than
self-adjoint.For the full proof
of a self-adjoint nature of the construction see
e.g. \cite{Seba2}

\bibitem{fnote1}{Close to thresholds the expression Eq.(\ref{uu})  
for time delay should be modified  by taking into account the  
explicit energy dependence of wavevectors $k_a(E)$ more accurately.
As a result, new terms may arise, see e.g. \cite{delayop,Gasparian}}
\bibitem{Som}H.-J. Sommers, A.Crisanti, H.Sompolinsky and
Y.Stein  , Phys.Rev.Lett. 60, 1895 (1988)

\bibitem{VZ} J.J.M.Verbaarschot and  
M.R.Zirnbauer,J.Phys.A:Math.Gen. 17, 1093 (1985)
\bibitem{Efrev} K.B.Efetov, Adv.Phys. 32, 53 (1983)

\bibitem{my} Y.V.Fyodorov in: "Mesoscopic Quantum Physics", Les
Houches Summer School, Session LXI, 1994; edited by E.Akkermans et
al., 493 ( Elsever 1995)
\bibitem{fnote2} Note that this auxilliary function does not in general
have the property of positivity.
\bibitem{fnote3} Strictly speaking, absolute convergence is obtained
only for $\hat{\Gamma}$ being positive. This can be achieved formally
by adding further very weakly coupled $N-M$ channels, the couplings of
which can be put to zero after the calculation is done. The
corresponding coupling constants should be chosen so small that they
do not modify our saddle-point considerations, see below. This is
always possible at any , whatever large, but fixed, value of $N$.
Having such regularization in mind, the whole procedure is well-defined.

\bibitem{Khor} Y.V.Fyodorov, B.Khoruzhenko
and H.-J.Sommers "Almost Hermitian Random Matrices:
eigenvalue density in the complex plane" preprint cond-mat 9606173 (1996)
  Phys.Lett. A, submitted

\bibitem{Zirn} M.R.Zirnbauer, Nucl.Phys.B 265,375 (1986);
Phys.Rev.B 34, 6394 (1986)

\bibitem{fnote4} Already here one sees that contributions of $N-M$
additional very weakly coupled channels may be neglected if the
corresponding couplings $\gamma_a$ approach zero faster than $1/N$.

\bibitem{Molgap} P.A.Moldauer , Phys.Rev.C 11,426 (1975)
\bibitem{Mol} P.A.Moldauer ,Phys.Rev. 157 , 907(1967);
M.Simonius , Phys.Lett. 52B, 279 (1974)
\bibitem{French} J.B.French, P.A.Mello and A.Pandey , Phys.Lett.B
80,17 (1978)\bibitem{Shep} F.Borgonovi, I.Guarnery, D.L.Shepelyansky,
Phys.Rev.A 43, 4517 (1991)

\bibitem{RG} I.S.Gradshtein, I.M.Ryzhik "Table of integrals,series
and products", (Academic Press, 1980) formulae:4.292.4; 6.646.1
\bibitem{unpub} N.Lehmann, H.-J.Sommers , in preparation

\bibitem{notr1} The distribution ${\cal P}_{\tau}(\tau)$
is actually a {\it density} of partial delay times and should not  
be confused
with the joint probability density of {\it all} $M$ partial delay times.
In general, it is also different from the distribution of  
Wigner-Smith time delay, being equivalent to the latter only for one  
open channel $M=1$.

\bibitem{KZ} V.E.Kravtsov, M.R.Zirnbauer, Phys.Rev.B
46, 4332 (1992)
\bibitem{Aleur} A.Taniguchi, A.Hashimoto, B.D.Simons and  
B.Altshuler Europh.Lett 27, 335 (1994)
\bibitem{prl94} Y.V.Fyodorov , Phys.Rev.Lett. 73, 2688  
(1994)\bibitem{curv} Y.V.Fyodorov and H.-J.Sommers , Phys.Rev.E  
51,2719
(1995) ; and Z. Physik B 99,123 (1995)
\bibitem{SZZ} P.Seba, K.Zyczkowski, J.Zakrzewski , Phys.Rev.E 54,  
324 (1996)
\bibitem{Wigcon} E.P.Wigner, Ann.Math. 53,36 (1951);ibid 55,7 (1952)
\bibitem{Keating} M.Berry and J.P.Keating, J.Phys.A:Math.Gen 27,  
6167 (1994)
\bibitem{Pluhar}Z.Pluhar,H.A.Weidenm\"{u}ller,J.A.Zuk et al.
 ,Ann.Phys.(NY) 243,1 (1995)
\bibitem{Bruus} H.Bruus, C.Lewenkopf, E.Mucciolo, Phys.Rev.B 53 ,  
9968 (1996)

\bibitem{Berob} M.Berry, M.Robnik, J.Phys.A.19, 649 (1986)

\bibitem{Ozorio}J.H. Hannay and A.M. Ozorio de Almeida, J.Phys.A:
Math.Gen.17,  3429 (1984)

\bibitem{Shushin} A.Shushin and D.W.Wardlaw, J.Phys.A:Math.Gen.
 25, 1503 (1992)

\bibitem{Oldyzko} A.M.Odlyzko: "The $10^{20}$ {th} zero of the  
Riemann zeta-function and $70$ million of its neighbors", preprint  
ATT Bell
Laboratories (1989)

\bibitem{Bogomol} E.Bogomolny, J.P.Keating, Nonlinearity 8,1115 (1995)

\bibitem{magbil}
 Z.Yan ,R.Harris , Europh.Lett. 32, 437 (1995);
 O.Bohigas,M.J.Giannoni,A.M.Ozorio de Almeida and C.Schmit
   , Nonlinearity 8, 203 (1995)
\bibitem{magscat} A.D.Stone and H.Bruus , Physica B 189, 43 (1993);  
Phys.Rev.B 50, 18275 (1994);
Y.Wang J.Wang and H.Guo , Phys. Rev.B 49, 1928 (1994) ;
 X.Yang, H.Ishio and J.Burgd\"{o}rfer , Phys.Rev.B  52, 8219 (1995);

\bibitem{Wang} Y.Wang,N.Zhu,J.Wang and H.Guo , Phys.Rev.B
53, 16408 (1996)
\bibitem{Ishio}H.Ishio and K.Nakamura , J.Phys.Soc.Jpn 61,2649
(1992) and ibid, 3939; J.Stat.Phys. 83, 203 (1996);
H.Ishio and J.Burgd\"{o}rfer , Phys.Rev.B 51, 2013 (1995)

\bibitem{Lin} W.A.Lin and R.Jensen ,Phys.Rev.B 53,3638 (1996)

\bibitem{phaseloc} P.W.Anderson,P.A.Lee, Suppl.Prog.Theor.Phys. 69,  
212 (1980); A.M.Jayannavar,G.Vijayagovindan and N.Kumar,
 Z.Phys.B 75,77 (1989); J.Heinrichs, J.Phys.Cond.Mat.
7 , 6291 (1995)

\bibitem{Borgo} F.Borgonovi,I.Guarnery, Phys.Rev. E 48, R2347 (1993)
\bibitem{fnotel} We have not indicated unit matrices $\hat{I}$ here  
just treating them as numbers.
\end{references}
\end{document}